\documentclass[12pt]{article}

\usepackage{newtxtext,newtxmath}

\usepackage{graphicx}
\usepackage{makecell}
\usepackage{rotating}
\usepackage[letterpaper,margin=1in]{geometry}

\renewenvironment{abstract}
	{\quotation}
	{\endquotation}

\date{}

\makeatletter
\renewcommand{\fnum@figure}{\textbf{Figure \thefigure}}
\renewcommand{\fnum@table}{\textbf{Table \thetable}}
\makeatother

\usepackage{scicite}
\usepackage{url}

\newcommand\eureka{{\texttt{Eureka!}}}

\def\scititle{
	Cloudy mornings and clear evenings on a gas giant exoplanet
}
\title{\bfseries \boldmath \scititle}

\author{
	Sagnick Mukherjee$^{1,2,3\ast}$,
	David K. Sing$^{1,4}$,
	Guangwei Fu$^{1}$,
        Kevin B. Stevenson$^{5}$,\and
        Stephen P. Schmidt$^{1}$,
        Harry Baskett$^{6}$, 
        Mei Ting Mak$^{6,7}$,
        Patrick McCreery$^{1}$,\and
        Natalie H. Allen$^{1}$,
        Katherine A. Bennett$^{4}$,
        Duncan A. Christie$^{8}$,
        Carlos Gascón$^{9,10}$,\and
        Jayesh Goyal$^{11,12}$,
        \'Eric H\'ebrard$^{6}$,
        Joshua D. Lothringer$^{13}$,\and
        Mercedes L\'opez-Morales$^{13}$, 
        Jacob Lustig-Yaeger$^{5}$,
        Erin M. May$^{5}$,
        L. C. Mayorga$^{5}$,\and
        Nathan Mayne$^{6}$,
        Lakeisha M. Ramos Rosado$^{1}$,
        Henrique Reggiani$^{14}$,\and
        Zafar Rustamkulov$^{4}$,
        Kevin C.\ Schlaufman$^{1}$,
        Kristin S. Sotzen$^{5}$,
        Daniel Thorngren$^{1}$,\and
        Le-Chris Wang$^{1}$,
        Maria Zamyatina$^{6}$\and     
    \small$^{1}$William H.\ Miller III Department of Physics and Astronomy, Johns Hopkins University, Baltimore, USA.\and
	\small$^{2}$Department of Astronomy and Astrophysics, University of California, Santa Cruz, USA.\and 
    \small$^{3}$School of Earth and Space Exploration, Arizona State University, Tempe, USA.\and
    \small$^{4}$Department of Earth \& Planetary Sciences, Johns Hopkins University, Baltimore, USA \and
    \small$^{5}$Johns Hopkins Applied Physics Laboratory, Laurel, USA.\and
    \small$^{6}$Department of Physics and Astronomy, University of Exeter, Exeter, UK\and
    \small$^{7}$Atmospheric, Oceanic, and Planetary Physics Department, University of Oxford, Oxfordshire, UK\and
    \small$^{8}$ Max Planck Institute for Astronomy, Heidelberg, Germany \and
    \small$^{9}$ Center for Astrophysics, Harvard $\&$ Smithsonian, Cambridge, USA  \and
    \small$^{10}$ Institut d'Estudis Espacials de Catalunya, Barcelona, Spain \and
    \small$^{11}$School of Earth \& Planetary Sciences, National Institute of Science Education and Research, Jatni, India \and
    \small$^{12}$Homi Bhabha National Institute, Training School Complex, Anushaktinagar, Mumbai, India \and
    \small$^{13}$Space Telescope Science Institute, Baltimore, USA \and
    \small$^{14}$ Gemini South, Gemini Observatory, National Science Foundation's \\
    \small National Optical-Infrared Astronomy Research Laboratory, La Serena, Chile \and
	\small$^\ast$ Corresponding author. Email: smukhe50@asu.edu\and
}

\begin{document} 

\maketitle

\begin{abstract} \bfseries \boldmath

The spectra of exoplanet atmospheres are affected by aerosols (clouds and hazes) of uncertain origin. Proposed aerosol formation mechanisms include gas condensation or photochemical reactions. We measure the transmission spectrum of the tidally locked gas giant exoplanet WASP-94A~b and identify asymmetry in its atmosphere. The morning limb is cooler and cloudy, while the evening limb is hotter and exhibits gaseous H$_2$O absorption features. We interpret this difference as due to the formation of cloud droplets near the morning limb, which evaporate during circulation to the evening limb. The dominant aerosols are clouds cycling between the day and night sides of the atmosphere, not photochemical hazes. The resulting asymmetry can severely bias chemical abundance measurements, unless limb-resolved spectroscopy is available.

\end{abstract}

\noindent

Aerosols are ubiquitous in the atmospheres of planets in the Solar System \cite{baines09} and have been observed in exoplanet atmospheres\cite{sing16,inglis24,ehrenreich20}. Aerosols affect the observed spectra of exoplanet atmospheres by muting gaseous absorption features, introducing additional spectral features, and changing the continuum slope \cite{inglis24,kempton23,gao23,wakeford15}. Physically, aerosols can cause atmospheric heating or cooling, and alter atmospheric chemistry. There is limited information about the nature of these aerosols, their three dimensional (3D) atmospheric distribution, and the physical processes that determine their properties. It is debated
whether aerosols in the atmospheres of highly irradiated giant exoplanets (known as hot Jupiters) are predominantly clouds formed through condensation of minerals from atmospheric gases \cite{visscher2010apj,wakeford15} or hazes produced by photochemical reactions driven by ultraviolet (UV) radiation from the host star\cite{porco05,lavega20,kempton17}.  

The physical processes that govern aerosols in hot Jupiters -- 3D atmospheric circulation, aerosol response to temperature, and UV irradiation, are uncertain inputs to atmospheric models of these planets \cite{steinrueck23,parmentier13,lines19,christie21}. Simulations predict that atmospheric circulation can transport aerosols across a large fraction of the planet's circumference or depth of its atmosphere \cite{parmentier13,lines19,steinrueck23}, and that aerosols respond to 3D temperature gradients via evaporation, condensation, and chemical reactions \cite{parmentier16,Arfaux22,arfaux24}. Theoretical estimates of the circulation timescales and aerosol particle sizes are uncertain by orders of magnitude \cite{steinrueck21,komacek19}. 

Several observational techniques have been employed to investigate aerosols, including phase-curve observations that probe various phases of the planet as it orbits the host star\cite{coulombe25,kempton23}. Infrared phase-curves constrain the 3D structure of exoplanet atmospheres, but if the targeted planet has abundant aerosols, this technique leads to degeneracies between the inferred thermal and cloud properties \cite{birmingham17,molliere20}. Spectroscopic techniques at high spectral resolution that spatially separate the leading and trailing terminators (or limbs) of the planet (known as limb-resolved spectroscopy) have shown chemical asymmetry between the limbs of some hot Jupiters \cite{louden15,ehrenreich20}, which might be due to condensation of gases in the colder night-side of the planet. However, that technique does not directly probe aerosols, so alternative physical mechanisms have been proposed to explain these asymmetries \cite{savel23,savel22}.

Limb-resolved transmission spectroscopy at low spectral resolution \cite{espinoza24,murphy24limb,fortney10} uses transit light curves (the stellar flux as a function of time as the exoplanet passes between the star and the observer) to separately measure the transit depths for the leading morning limb and the trailing evening limb of the planet \cite{Jones2022,espinoza21}. This technique demonstrated limb asymmetry in two gas giant exoplanets, driven by temperature difference between the limbs \cite{murphy24limb,espinoza24}. Those studies did not find signatures of scattering driven by aerosols in either limb.

Uncertainties in aerosol physics limit our understanding of the formation and evolution of a wide variety of exoplanetary systems. The chemical compositions of exoplanets are typically inferred from measurements of their atmospheres\cite{oberg11,welbanks19}, but key uncertainties in the aerosol physics can bias these measurements \cite{line16,madhu2019,mai19}.

\section*{Observations of WASP-94A~b}
The hot Jupiter exoplanet WASP-94A~b has a mass of 0.456 Jupiter masses (M$_{\rm Jupiter}$) and a radius of 1.72 Jupiter radii (R$_{\rm Jupiter}$) \cite{neveu14}. We observed a single transit of WASP-94A~b using the James Webb Space Telescope (JWST). The Near Infrared Imager and Slitless Spectrograph (NIRISS) instrument was used to obtain low-resolution time-resolved spectra between 0.8 to 2.8~{$\upmu$}m \cite{methods}. We construct white light curves of the transit in two broadband wavelength ranges covering strong H$_2$O absorption features, 0.9 to 1.2~{$\upmu$}m and 1.35 to 1.5~{$\upmu$}m, using the Fast InfraRed Exoplant Fitting Lyghtcurve Suite (FIREFLy) analysis pipeline \cite{rustamkulov22,methods}. 
We then fitted the data using two models with different assumed planet shapes: a spherically symmetric model and a model with limb-to-limb asymmetry \cite{methods,espinoza21,Jones2022}. We initially fitted only the 0.9 to 1.2~{$\upmu$}m band to constrain the planetary system parameters, including the mid-transit time ($T_0$) in each model. We then used the best-fitting $T_0$ values from the 0.9 to 1.2~{$\upmu$}m light curve as fixed parameters in the model fitting of the 1.35 to 1.5~{$\upmu$}m band, in both models. 

Fig.~\ref{wlcfit}A shows the observed light curve and both best-fitting models for the 1.35 to 1.5~{$\upmu$}m band. The residuals for the asymmetric limb model (Fig.~\ref{wlcfit}B) are consistent with zero, whereas those of the spherical planet model (Fig.~\ref{wlcfit}C) show substantial deviations during ingress and egress. We calculate that the data prefer the asymmetric limb model over the spherical limb model with a statistical significance of 6$\sigma$ \cite{methods} , indicating asymmetric transit depths for the two limbs.

\begin{figure}[h]
\centering
\includegraphics[width=1\textwidth]{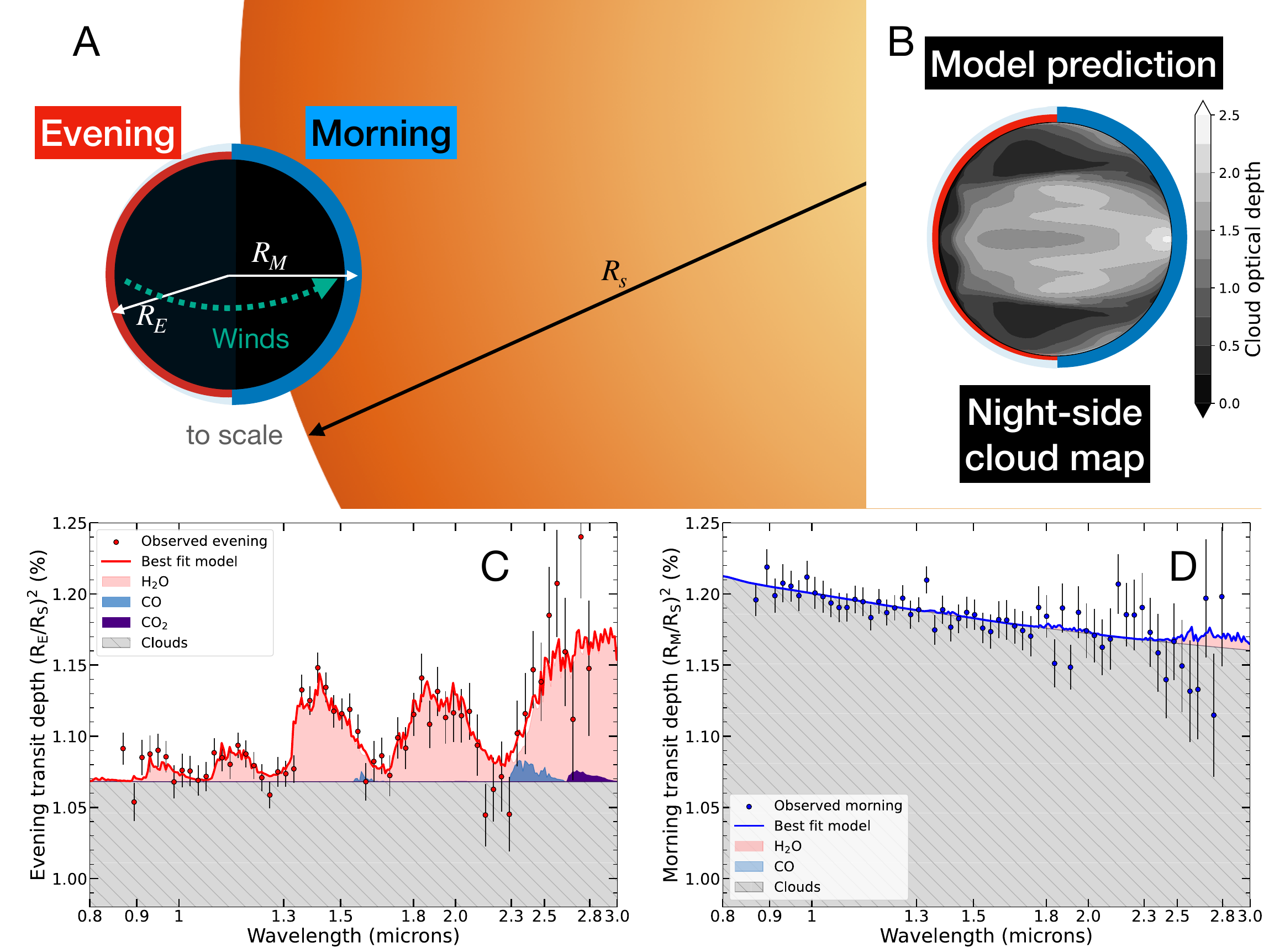}
\caption{{\bf Observing geometry, model predictions, and transmission spectrum of the morning and evening limbs of WASP-94A\,b}. ({\bf A}) The observing geometry of WASP-94A\,b and its host star (to scale). The red and blue regions depict the evening and morning limbs of the planet, respectively. The radii of the evening limb ($R_{\rm E}$), morning limb ($R_{\rm M}$), and the star ($R_{\rm S}$) are indicated with white (for planet) and black (for star) arrows. The dotted cyan arrow depicts the direction of the equatorial jet expected from theoretical models of the planet.  ({\bf B}) The night-side cloud map of WASP-94A~b predicted by the GCM. The predicted optical depth of clouds in the night-side is shown with the gray colorscale, and the evening and morning limbs are depicted with the same colors as (A). ({\bf C}) Points are the observed transmission spectrum of the evening limb, on a non-linear wavelength scale. The red line is the best-fitting model from atmospheric retrieval analysis. The red, blue, indigo, and gray shaded regions are the contributions of H$_2$O, CO, CO$_2$, and clouds to the model, respectively. ({\bf D}) Same as (C) but for the morning limb. All error bars show 1$\sigma$ uncertainties.}\label{figgascont}
\end{figure}

The NIRISS spectroscopic light curves were binned to a spectral resolving power R$\sim$50. We fitted the light curve at each wavelength using the asymmetric limb transit model \cite{methods}. The light curve models at each wavelength were used to derive a transmission spectrum for each limb separately. Because hot Jupiters like WASP-94A~b are expected to be tidally locked to their host stars \cite{showman02}, we identify the leading morning limb and the trailing evening limb during the ingress and egress (Fig.~\ref{figgascont}A). The resulting transmission spectra of the two limbs are shown in Fig.~\ref{figgascont}C-D. In addition to the FIREFLy analysis pipeline, we also use two independent data analysis pipelines, \texttt{Eureka!} and the Fu pipeline, to measure the morning and evening limb transmission spectra \cite{methods}. The spectra extracted from the three pipelines are consistent with each other \cite{methods}.

The offset between the morning and evening limb spectra depends on the precision on $T_0$ \cite{murphy24}. We used observations of another transit of WASP-94A~b measured with JWST's Near Infrared Spectrograph (NIRSpec) in combination with the NIRISS observations to improve the precision on $T_0$ \cite{ahrer25,methods}. 
We measured the extent by which the offset between the morning and evening limb spectra might change due to our measured $T_0$ and account for it within our modeling analysis \cite{methods}. Star spots or faculae can also contaminate the transmission spectra of exoplanets \cite{sing11}. We used observations of the host star WASP-94A obtained with the Transiting Exoplanet Survey Satellite (TESS) to find that such spots or faculae do not contaminate our observed transmission spectra  \cite{methods}.

\section*{Asymmetry caused by clouds not hazes}

The morning limb spectrum has no prominent gas absorption features and a sloped continuum, rising at shorter wavelengths, which is a signature of high-altitude aerosols \cite{lecavelier08}. The evening limb spectrum does exhibit absorption features, consistent with gaseous H$_2$O \cite{methods}, but has no substantial evidence of aerosols. This difference implies that the dominant aerosols on WASP-94A~b are clouds, not hazes.

Hazes are theoretically expected to be preferentially produced on the UV-irradiated permanent day-side (because WASP-94A~b is tidally locked), then transported to other parts of the atmosphere by planet-wide atmospheric circulation \cite{kempton17,steinrueck23}. Large haze particles (with particle radius $r>$ 30 nm) would therefore be more abundant on the evening limb than the morning limb \cite{steinrueck21}. That is the opposite to what we observe. Small haze particles ($r<$ 30 nm) can get concentrated more on the morning limb \cite{steinrueck21,mak2025}. However, theoretical modeling of several hot Jupiters shows that their over-concentration in the morning limb do not cause a complete suppression of gas absorption features in the morning limb while maintaining prominent absorption features on the evening limb at the wavelengths of our observations \cite{mak2025}. 

Clouds are predicted to form in the cooler night-side of the planet then circulate downwind to the morning limb \cite{lines19,parmentier16}. Further transport to the much hotter day-side is predicted to cause a large fraction of the cloud droplets to evaporate \cite{lines19}. This would cause the evening limb to be clearer than the morning limb, as we observed. We use a 3D general circulation model (GCM) to theoretically predict the atmospheric structure of WASP-94A~b \cite{methods} without fitting the observed spectra. The GCM predicts cloud formation on the night-side, with a clear evening and a cloudy morning limb (Fig.~\ref{figgascont}B). By fitting the morning limb spectrum with atmosphere models that include various types of aerosols, we find that clouds provide better fit to the morning limb spectrum than several haze species \cite{methods}. 

\section*{Temperature variations and diurnal cycles}

We constructed an atmospheric model featuring two separate atmospheric columns with distinct temperature and cloud structures, each representing the morning and evening limbs \cite{methods}. The temperature profiles of the two columns converge together in the deep atmosphere based on theoretical predictions \cite{parmentier13,lines19}. We use this atmosphere model to fit the transmission spectrum of both limbs using a Bayesian atmospheric retrieval framework \cite{methods}. Fig.~\ref{figgascont}C-D also shows the best-fitting retrieved theoretical model. We find the evening spectrum is dominated by H$_2$O absorption, which is detected with 10$\sigma$ significance \cite{methods}. The morning limb spectrum is dominated by cloud absorption, which is detected with 9$\sigma$ significance \cite{methods}.

\begin{figure}[h]
\centering
\includegraphics[width=1\textwidth]{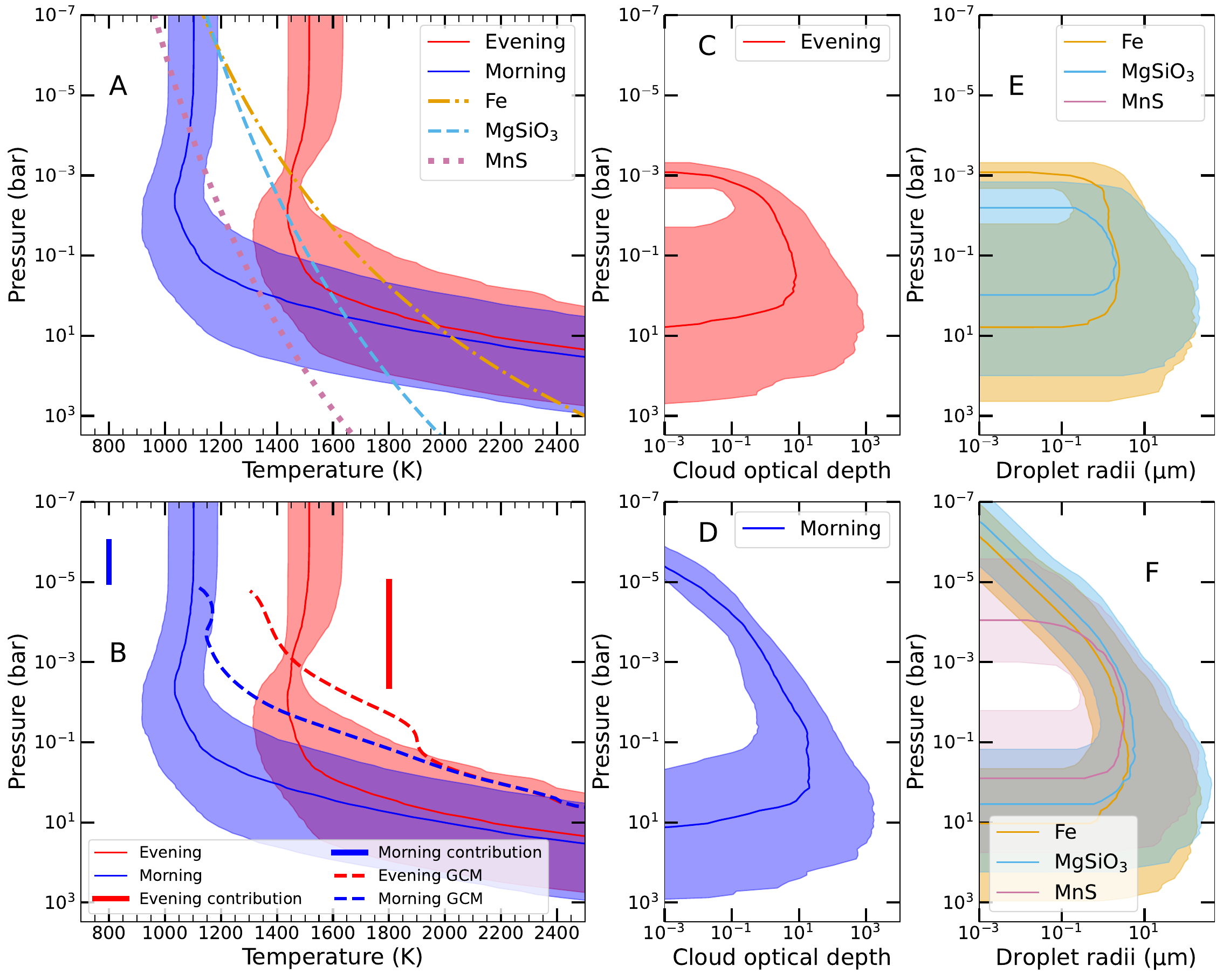}
\caption{{\bf Retrieved temperature and cloud structure of the morning and evening limbs.} ({\bf A}) Retrieved temperature-pressure ($T(P)$) profiles for the morning (blue) and evening (red) limbs, with lines indicating the median values and shading the $1\sigma$ uncertainty. Dashed, dot-dashed, and dotted lines are theoretical condensation curves for MgSiO$_3$ (cyan), Fe (orange), and MnS (purple) clouds, respectively. ({\bf B}) $T(P)$ profiles of the morning (blue) and evening (red) limbs predicted by the GCM are shown using dashed lines and compared with the retrieved $T(P)$ profiles. The vertical red and blue bars show the pressures probed by the evening and morning limb spectra, respectively. These were calculated from the contribution function of the best-fitting retrieved model \cite{methods}. ({\bf C})-({\bf D}) The retrieved cloud optical depth profiles, at 1.5~{$\upmu$}m wavelength, for the evening and morning limbs. ({\bf E})-({\bf F}) Retrieved mean droplet radius profiles for different cloud species for the evening (E) and morning limbs (F). The solid lines are the median profiles and the shaded regions show the $1\sigma$ uncertainty for MgSiO$_3$ (cyan), Fe (orange), and MnS (purple) clouds. }\label{figtpclouds}
\end{figure}

Fig.~\ref{figtpclouds}A shows the retrieved temperature-pressure ($T(P)$) profiles of each limb, compared to the theoretically calculated condensation curves \cite{visscher2010apj} of three potential cloud species:Fe, MgSiO$_3$, and MnS at the temperatures of WASP-94A~b. The $T(P)$ profile of the evening limb is hotter than the morning limb, with a temperature difference of 449$\pm$83\,K, which crosses the condensation curves at low pressure (high altitude). We therefore expect any cloud droplets at high altitudes on the morning limb would be evaporated at the same altitudes on the evening limb. We calculate the pressure range that contribute to the morning and evening limb spectrum with contribution functions for transmission spectroscopy \cite{methods, rustamkulov23}. These pressure ranges are shown with the red and blue vertical bars in Fig.~\ref{figtpclouds}B. The morning limb spectrum probes pressures $\le$0.01\,mbar whereas the evening limb spectrum probes  0.01 to 4\,mbar. Fig.~\ref{figtpclouds}B also shows $T(P)$ profiles for each limb of WASP-94A~b predicted by our 3D GCM, which matches the measured temperature difference between limbs \cite{methods}.

Fig.~\ref{figtpclouds} also shows the retrieved cloud properties of each limb. Fig.~\ref{figtpclouds}C-D shows the cloud optical depth profiles, which indicate that the evening limb becomes transparent at pressures $<1$ mbar but the morning limb remains optically thick until $\sim$0.01\,mbar. The retrieved temperature difference between the planet's limbs is sufficient to drive the day-night cloud cycling \cite{methods}, leading to the difference in the cloud optical depth between the limbs and hence the H$_2$O absorption features. 
Fig.~\ref{figtpclouds}E-F shows the mean cloud droplet sizes for the three cloud species at each limb; we find that the morning limb has droplets with mean radius between 0.1 to 1~{$\upmu$}m in the 0.01 to 1\,mbar pressure range.

The distribution of cloud droplets is thought to be connected to atmospheric circulation timescales, including the vertical mixing and the horizontal advection timescales \cite{ackerman2001cloud,gao18}. We consider the balance between vertical lifting of cloud droplets and their gravitational settling to constrain the strength of vertical mixing, assuming steady state \cite{methods}. This is quantified by the one dimensional (1D) vertical eddy diffusion parameter $K_{\rm zz}$; we derive $\log_{10}(K_{\rm zz}/{\rm cm}^2 {\rm s}^{-1} )=11.73\pm0.87$. This is consistent with the average altitude-dependent $K_{\rm zz}$ profile predicted by the GCM \cite{methods}. The vertical mixing timescale is $\sim$0.06 to 4~days. If the horizontal advection timescale at 0.01\,mbar is much slower than the vertical mixing timescale, then cloud droplets are stable at these high altitudes. However, if the horizontal advection timescale is faster, then cloud droplets are not replenished quickly enough to be maintained at high altitude in the morning limb. 

The GCM predicts an equatorial jet that extends across the day- and night-sides of WASP-94A~b \cite{methods}. The fastest advection timescale (which governs horizontal transport of the cloud droplets from the morning to the evening limb) is within this equatorial jet and typically $\sim$ 1.15 to 2.2~days. The jet in the GCM spans latitudes $\pm$40$^\circ$ for pressures $<1$ bar. If the cloud droplets trace the wind velocities of the jet at higher altitudes in the GCM, then the vertical mixing timescale must be closer to 0.06 days than 4 days.

\section*{Compositional measurement effects}
\begin{figure}[h]
\centering
\includegraphics[width=1\textwidth]{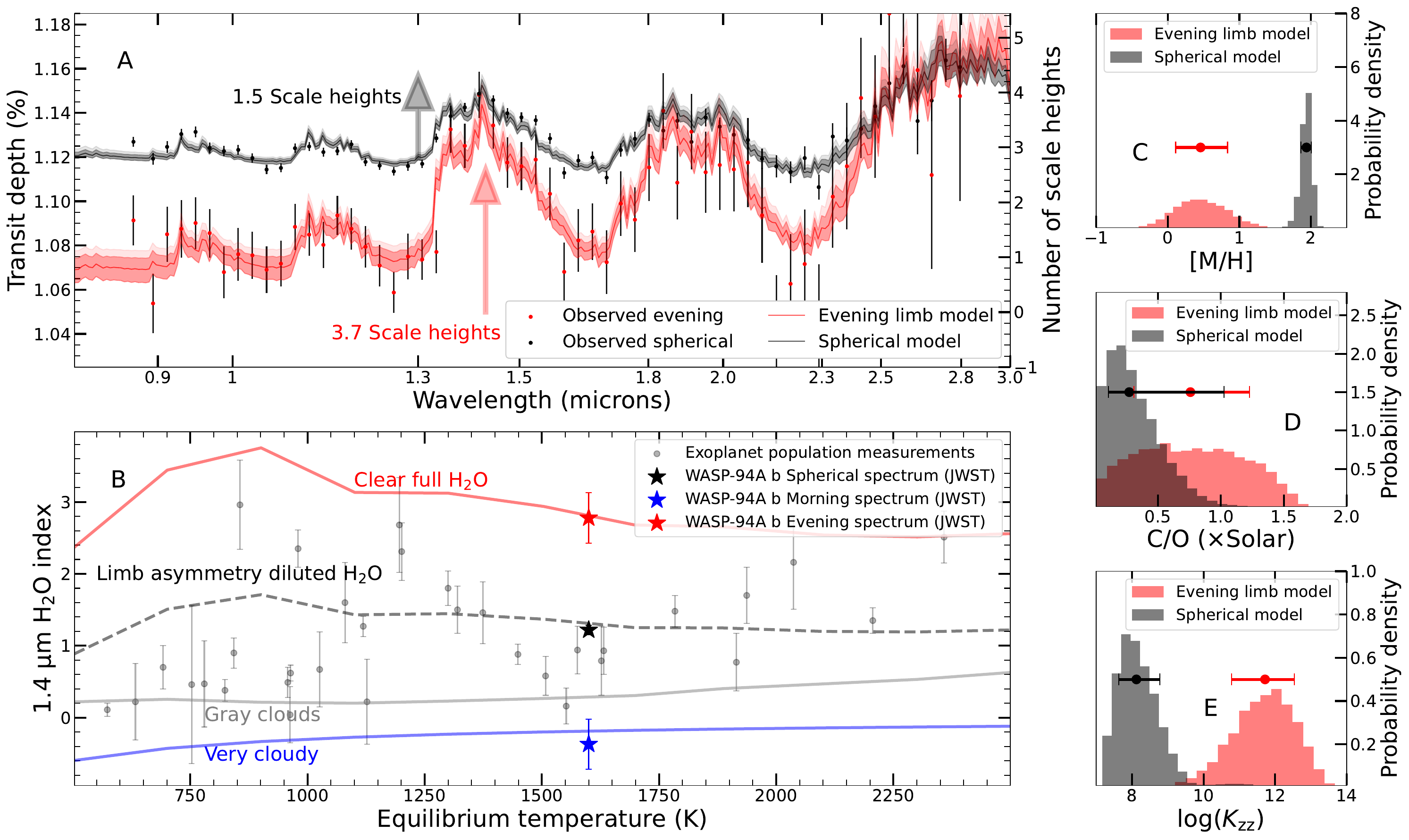}
\caption{{\bf Inferred composition of WASP-94A~b and the effect of limb asymmetry.} ({\bf A}) The observed transmission spectrum of the evening limb of WASP-94A~b (red points) compared to the average spectrum extracted assuming the planet is spherically symmetric (black points). The solid lines show the median retrieved model spectra, and the shaded regions show the 1$\sigma$ uncertainties for each. The y-axis on the right side shows the number of atmospheric pressure scale heights probed by the observations, which has been calculated using the equilibrium temperature, gravity, and retrieved atmospheric mean molecular weight of WASP-94A~b. ({\bf B}) Amplitude of the 1.4~{$\upmu$}m H$_2$O absorption feature for previous measurements of exoplanets \cite{fu17} (light gray points). The black point shows the H$_2$O feature amplitude measured from the spherical spectrum of WASP-94A~b, while the red and blue points show the H$_2$O amplitude measured separately for the evening and morning limbs, respectively. Overlaid are the H$_2$O amplitude predicted by self-consistent atmospheric models assuming a clear atmosphere (red), a very cloudy atmosphere (blue), an atmosphere with clouds that have wavelength-independent absorption (gray), and an atmosphere with limb-asymmetry (gray dashed). ({\bf C}) The inferred probability distribution for atmospheric metallicity determined from the two spectra in (A). The inferred values differ by 4$\sigma$. ({\bf D}) and ({\bf E}) are the same as (C) but for the C/O ratio and $K_{\rm zz}$, respectively. All error bars show 1$\sigma$ uncertainties. }\label{figmhcto}
\end{figure}

To approximate the spectrum that would have been observed if we had not resolved the two limbs, we derive a planet-wide spectrum using the the spherically symmetric model (Fig.~\ref{figmhcto}A) \cite{methods}. In addition to H$_2$O, we identified an additional absorption feature that could not be explained with our atmospheric model as it is due to outflowing metastable helium in the unbinned spherical spectrum at 1.083~{$\upmu$}m \cite{methods}. This absorption feature indicates rapid atmospheric mass-loss from WASP-94A~b \cite{methods}. 

We bin the spherical spectrum to the resolving power of the limb asymmetry spectrum for further analysis. The absorption features in the binned spherical spectrum are muted by a factor of $\sim$2, compared to the evening limb spectrum. We then applied a retrieval method, that models the spectrum with only one atmospheric column, to the spherical spectrum for comparison with the retrieval method with two atmospheric columns that was used for the limb-resolved spectra. The two modeling frameworks are otherwise identical. Fig.~\ref{figmhcto}C-E shows the resulting constraints on the chemical composition and cloud properties from the limb-resolved and spherical spectra. The resulting constraints on the logarithm of metallicity, which is the abundance of heavier elements relative to Hydrogen, of WASP-94A~b are +0.46$\pm$0.36 from the limb-resolved observations, and +1.937$\pm$0.073 from the spherical spectrum. These values are inconsistent at greater than 4$\sigma$ statistical significance. We conclude that not accounting for the limb asymmetry biases the inferred metallicity. This bias was theoretically predicted \cite{line16}; it arises from the dilution of gas absorption features in the spherical spectrum. The inferred carbon-to-oxygen ratios (Fig.~\ref{figmhcto}D) remain consistent but have different probability distributions for the same reasons. Fig.~\ref{figmhcto}E shows that $K_{\rm zz}$ is also affected by this bias, with the inferred values differing by 3$\sigma$. We also apply another retrieval method to the spherical spectrum, which models the planet-wide spectrum by linearly combining spectra from fully cloudy and fully clear atmosphere columns using a cloud-coverage fraction parameter \cite{methods}. We find that this approach doesn't alleviate these biases \cite{methods}. 

We caution that this bias is unlikely to be limited to this planet or similar hot Jupiters. Fig.~\ref{figmhcto}B shows the amplitude of the 1.4~{$\upmu$}m H$_2$O absorption features measured from previous observations of transiting exoplanets with a variety of sizes, masses, and temperatures \cite{fu17}. These H$_2$O amplitudes were measured from the 1D spectrum without accounting for limb-asymmetry. The H$_2$O amplitude we measure from the spherical spectrum of WASP-94A~b follows the same trend as other hot Jupiters (1000-2000 K). However, our separate measurements for the two limbs of WASP-94A~b are at the extremes of the range observed in other hot Jupiters. We calculate the spherical planet H$_2$O feature amplitude from a grid of fully clear and very cloudy one-dimensional atmospheric models, and a grid of models with one clear and one very cloudy limb (as we inferred for WASP-94A~b) \cite{methods}. If other hot Jupiters also have asymmetric limbs due to day/night cloud cycling, these models imply that their H$_2$O feature would be diluted to varying levels, potentially biasing the inferred composition and cloud properties. We expect similar issues to apply to smaller exoplanets, such as sub-Neptunes, which are theoretically predicted to have substantial aerosol cover \cite{kempton23,coulombe25}.



\begin{figure}[h]
\centering
\includegraphics[width=0.7\textwidth]{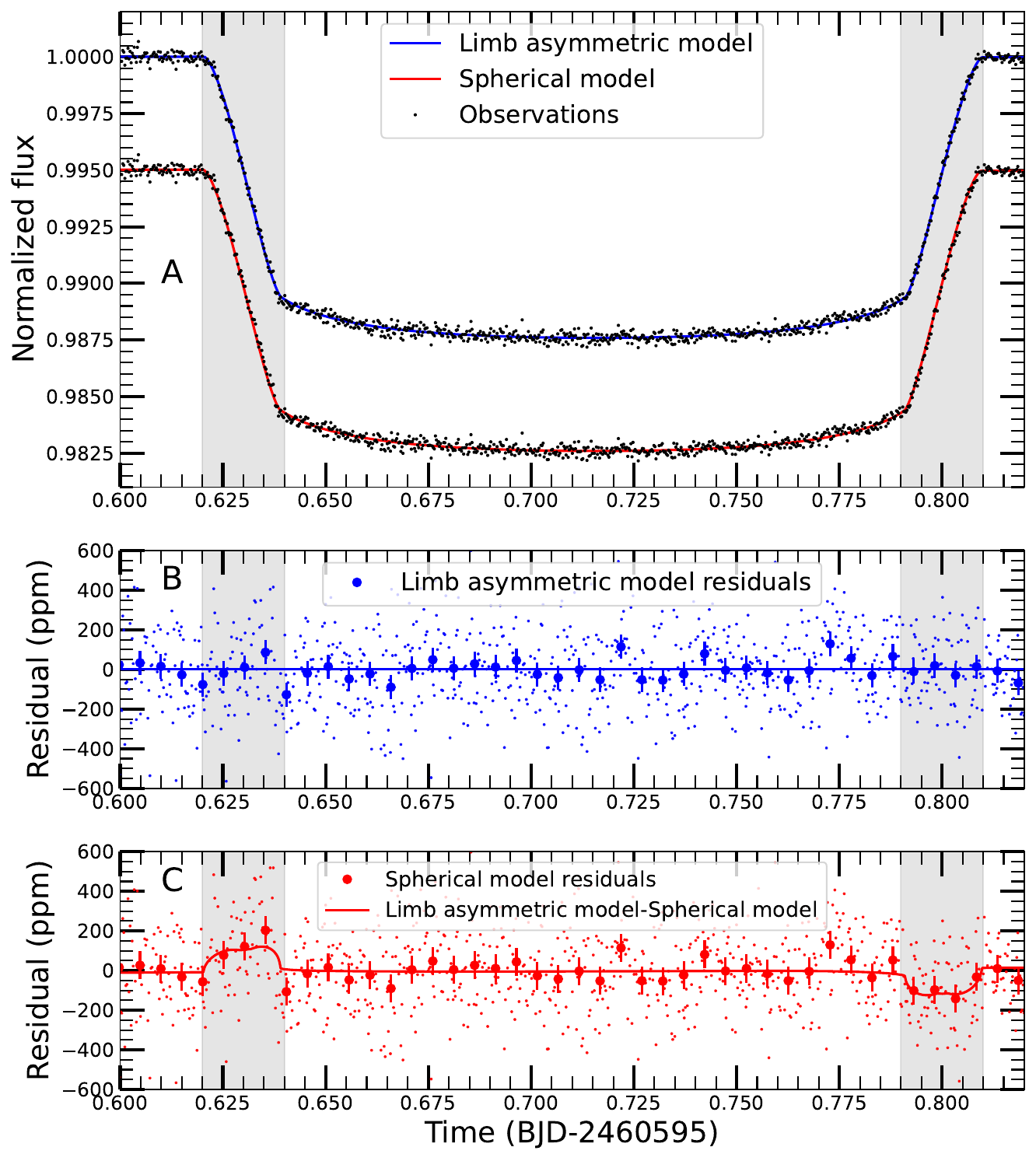}
\caption{{\bf Transit light curves of WASP-94A~b and fitted models.} ({\bf A}) Black points are the broadband 1.35 to 1.5~{$\upmu$}m light curve overlain with the fitted models assuming a spherical planet (red line) or allowing for asymmetric limbs (blue line). The same observations are plotted twice, with a constant vertical offset applied to separate the two model fits for clarity. The time in Barycentric Julian Date (BJD) is shown in the x-axis and the gray shaded region marks the times of ingress and egress. ({\bf B}) Small blue points are the residuals between the asymmetric limb model and the data. These residuals are binned along the time axis and shown with the large blue points along with their 1$\sigma$ uncertainty. The blue line marks zero and the residuals are all consistent with zero. ({\bf C}) Same as (B) but for the spherical planet model, which is inconsistent with zero during ingress and egress. The red line shows the difference between the asymmetric limb and the spherical model.}\label{wlcfit}
\end{figure}

\clearpage 

%
\bibliography{science_template} 
\bibliographystyle{sciencemag}

%
%
%
%
%
%


\section*{Acknowledgments}
SM thanks Jonathan Fortney, Kazumasa Ohno, and Namrata Roy for discussion and feedback. We thank the anonymous reviewers whose feedback helped improve this manuscript. SM also acknowledges use of the lux supercomputer at University of California Santa Cruz, funded by National Science Foundation Major Research Instrumentation grant AST 1828315. HB, NM, EH and MZ acknowledge use of the Monsoon2 system, a collaborative facility supplied under the Joint Weather and Climate Research Programme, a strategic partnership between the Met Office and the Natural Environment Research Council. This work is based on observations made with the National Aeronautics and Space Administration/European Space Agency/Canadian Space Agency James Webb Space Telescope. The data were obtained from the Mikulski Archive for Space Telescopes at the Space Telescope Science Institute, which is operated by the Association of Universities for Research in Astronomy, Inc., under National Aeronautics and Space Administration contract NAS 5-03127 for JWST.

\paragraph*{Funding:}
SM was supported by the Templeton Theory-Experiment (TEX) Cross Training Fellowship from the John Templeton Foundation and the 51 Pegasi b postdoctoral fellowship from the Heising-Simons Foundation. SPS was supported by the National Science Foundation Graduate Research Fellowship Program under Grant No. DGE2139757. 
HB, MTM, NM, EH and MZ acknowledge support from a Science and Technology Facilities Council astronomy observation and theory small award [ST/Y00261X/1]. NM was also supported by a United Kingdom Research and Innovation Future Leaders Fellowship [grant number MR/T040866/1]. MTM acknowledges funding from the Bell Burnell Graduate Scholarship Fund [grant number BB005], administered and managed by the Institute of Physics. PM and KCS were supported by the National Aeronautics and Space Administration under Grant No. 80NSSC23K0266 issued through the Exoplanets Research Program (XRP). DKS received support from JWST GO program number 5924 through a grant from the Space Telescope Science Institute, which is operated by the Association of Universities for Research in Astronomy, Inc., under National Aeronautics and Space Administration contract NAS 5-03127.



\paragraph*{Author contributions:}


SM performed the \texttt{FIREFLy} data reduction of the observations. SM led the development of asymmetric limb atmospheric retrievals, performed the aerosol species retrievals, and led manuscript preparation. DKS was the principal investigator of the JWST program and oversaw new developments within the FIREFLy data reduction pipeline and its associated data analysis. DKS also contributed to atmospheric modeling, theoretical interpretations, and manuscript preparation. GF and KBS performed independent data reductions. SPS performed the patchy cloud atmospheric retrievals on the spherical planet spectrum. HB and MTM performed the GCM simulation, with contributions from NM, MZ, EH and DAC. PM modeled the helium ouflow signature in the planet's spherical spectrum, with contribution from KCS and JL. NHA contributed to the stellar contamination analysis. MLM, JLY, LCM, and JG contributed to the theoretical design and implementation of the two column Bayesian atmospheric retrieval for simultaneous analysis of the morning and evening limb spectrum. KBS, GF, NHA, KB, CG, LMRR, CW, EM, DT, ZR, and HR contributed to new developments in data reduction and analysis pipelines required for this work. KBS, GF, SPS, HB, PM, NHA, KB, DAC, CG, JG, EH, MLM, EM, LCM, NM, LMRR, HR, ZR, DT, CW, MZ, and MTM contributed to manuscript preparation. MTM, KBS, NHA, JL, JLY, EM, LCM, LMRR, KCS, KSS, and DT contributed to substantial revisions of the manuscript. NHA, KB, CG, LMRR, CW, EM, ZR, and KSS contributed to observation planning. GF, KBS, NHA, KB, JG, JL, MLM, NM, LMRR, HR, ZR, and KCS contributed to preparation of the JWST proposal.



\paragraph*{Competing interests:}
There are no competing interests to declare.
\paragraph*{Data and materials availability:}

The raw JWST observations obtained with the NIRISS instrument are available on the Mikulski Archive for Space Telescopes (MAST) through program number 5924 \cite{Mukherjee2025-mastdoi}. Our reduced transmission spectra of the morning and evening limbs are archived at Zenodo \cite{mukherjee_2025_15085825}. The best-fitting models, uncertainties, and retrievals and GCM outputs are available in the same Zenodo repository \cite{mukherjee_2025_15085825}. The raw JWST observations obtained with the NIRSpec instrument are available on MAST through program number 3154 \cite{ahrer2025-mastdoi}. The TESS observations are also available on MAST \cite{tessmukherjee2025-mastdoi}. The FIREFly data analysis pipeline is available at {https://github.com/sagnickm/FIREFLy} and released on Zenodo \cite{firefly_zenodo}. The \texttt{Eureka!} pipeline is available at {https://github.com/kevin218/Eureka} and released on Zenodo \cite{the_eureka_team_2025_19076001}. The Fu data analysis pipeline is available at {https://github.com/guangweifu/Tswift/tree/main} and released on Zenodo \cite{guangweifu_2026_tswift}.
\subsection*{Supplementary materials}
Materials and Methods\\
Figs. S1 to 24\\
Tables S1 to S7\\
References \textit{(57-\arabic{enumiv})}\\ 


\newpage


\renewcommand{\thefigure}{S\arabic{figure}}
\renewcommand{\thetable}{S\arabic{table}}
\renewcommand{\theequation}{S\arabic{equation}}
\renewcommand{\thepage}{S\arabic{page}}
\setcounter{figure}{0}
\setcounter{table}{0}
\setcounter{equation}{0}
\setcounter{page}{1} 


\begin{center}
\section*{Supplementary Materials for\\ \scititle}


Sagnick Mukherjee$^{\ast}$,
David K. Sing,
Guangwei Fu,
Kevin B. Stevenson,\\
Stephen P. Schmidt,
Harry Baskett, 
Mei Ting Mak,
Patrick McCreery,
Natalie H. Allen,\\
Katherine A. Bennett,
Duncan A. Christie,
Carlos Gascón,
Jayesh Goyal,\\
\'Eric H\'ebrard,
Joshua D. Lothringer,
Mercedes L\'opez-Morales, \\ 
Jacob Lustig-Yaeger,
Erin M. May,
L. C. Mayorga,
Nathan Mayne,\\
Lakeisha M. Ramos Rosado,
Henrique Reggiani,
Zafar Rustamkulov,\\
Kevin C.\ Schlaufman,
Kristin S. Sotzen,
Daniel Thorngren,
Le-Chris Wang,\\
Maria Zamyatina\\
\small$^\ast$Corresponding author. Email: smukhe50@asu.edu\\

\end{center}

\subsubsection*{This PDF file includes:}
Materials and Methods\\
Figures S1 to 24\\
Tables S1 to S7\\


\newpage


\subsection*{Materials and Methods}




\section{Data Reduction}\label{datareduc}
We observed WASP-94A~b using NIRISS in the single-object slitless-spectroscopy (SOSS) mode on 11 to 12 October 2024. The observations started at 22:36:23 on 11 October Universal Time (UT) and continued for 11.51 hours, ending at 09:12:22 on 12 October UT. The observations were obtained with the SUBSTRIP256 subarray and included 3 groups per integration. There were 1458 integrations for a total exposure time of 8.9 hours. The ``NISRAPID" readout setting, which reads out all frames,  was used for the NIRISS detector. Because WASP-94 is a binary system, the on-sky orientation of the detector was set to near the maximum the cross-dispersion separation between the two stars, which was 15$''$ (229 pixels). The binary companion (WASP-94B) was located beyond the 256 pixel detector subarray, being separated by $\sim$229 pixels at the near optimal orientation. The point spread function (PSF) wing of WASP-94B is partly visible at the top of the 2D images, but did not contaminate the WASP-94A spectrum.

We use three independent data reduction pipelines to analyze these observations -- FIREFLy, \texttt{Eureka!}, and Fu. Below, we briefly describe the data reduction steps performed within each of these pipelines.

\subsection{FIREFLy}
The \texttt{FIREFLy} data reduction is described elsewhere \cite{rustamkulov22,rustamkulov23,sing24}. We use v1.0.0 which was optimized to handle NIRISS/SOSS observations \cite{schmidt25}. For the NIRISS/SOSS observations of WASP-94A~b, we implemented several additional steps within the analysis pipeline, described below.

The \texttt{FIREFLy} routines analyze stage zero NIRISS/SOSS data by modifying of the JWST calibration pipeline \cite{Bushouse2024} and starting from the uncalibrated images. FIREFLy modifies the method of flicker noise removal, which has a power spectral density inversely proportional to frequency ($f$) (or $1/f$ noise). This modification is necessary due to the slitless nature of the SOSS data. \texttt{FIREFLy} constructs a custom PSF mask for each observation using a detector count threshold. We used a mask count threshold of 12 analog-to-digital unit (ADU). This mask is then applied to exclude the PSF and any background sources across the detector. The background was removed from each group by scaling a background template to a region with minimal contamination from the PSF or other sources. With the background subtracted from each group, we used the unmasked regions to subtract the $1/f$ noise in a column-by-column basis at the group level. This step substantially reduces the $1/f$ noise in the data. The background is then re-added before the linearity step is applied. We find the noise is reduced by $\sim$5\% using this process compared to only applying a $1/f$ noise subtraction at the integration stage.

To further clean the data, \texttt{FIREFLy} employs a static bad pixel cleaning routine by constructing a filter-smoothed median detector image, then cleaning pixels from each integration if they are 7$\sigma$ outliers from the filter smoothed median image. Cosmic ray hits on the detector are cleaned from each integration using the Laplacian edge cosmic ray detection technique \cite{vandokkum01}. We also perform an additional cleaning of the cosmic rays using a temporal cleaning method, where a rolling median detector image is created with 11 integrations at a time and 7$\sigma$ outliers of this rolling median image are cleaned from the respective integrations. At this stage, we perform another $1/f$ noise and background removal at the integration level by producing another custom PSF mask with a 15$\sigma$ count threshold. To extract 1D spectrum from the cleaned detector images, we use a full aperture width of 31 pixels. We vary this aperture width through higher and lower values to select this optimum aperture width on the basis of the lowest median absolute deviation (MAD) value for the extracted white light curve of the transit. For these data, we found the optimal MAD is 76~ppm. The extracted white light curve between 0.85 to 2.75~{$\upmu$}m along with the best-fitting asymmetric planet limb model are shown in Fig.~\ref{wlcwhole}. Fig.~\ref{LCcolorplot}A shows the wavelength vs. time 2D map of the normalized extracted spectroscopic light curves with \texttt{FIREFLy}.

To determine the statistical significance of preference of the asymmetric planet model over the spherical planet model, we construct two broadband light curves between 0.9 to 1.2~{$\upmu$}m and 1.35 to 1.5~{$\upmu$}m. We use a spherical planet model (Batman \cite{batman15}) and an asymmetric planet model (Catwoman \cite{espinoza21,Jones2022}) for fitting the light curve between 0.9 to 1.2~{$\upmu$}m, allowing the mid-transit time ($T_0$) to be a fitted parameter. Then, we fit the light curve between 1.35 to 1.5~{$\upmu$}m with both models after fixing $T_0$ to the value obtained from fitting the light curve between 0.9 to 1.2~{$\upmu$}m with each model (Fig.~\ref{wlcfit}). We use two bands to prevent the spherical planet model from compensating for limb asymmetry by varying $T_0$, because $T_0$ should be same at all wavelengths (Fig.~\ref{T0vscatwoman}). We find that the asymmetric planet model fit is preferred over the spherical planet model fit with a difference in the Bayesian information criteria (BIC) of $\Delta{\rm BIC}=31.23$. We also find that the asymmetric planet model has a lower reduced-${\chi}^2$ metric ($\chi_{\rm \nu}^2$) than the spherical planet model with a difference $\Delta{\chi_{\rm \nu}^2}=0.027$. This corresponds to a 6$\sigma$ preference of the asymmetric planet model over the spherical planet model.

\subsubsection{Constraining the mid-transit time}




We use the Catwoman routine \cite{Jones2022,espinoza21} to fit the white light curve with the asymmetric limb model to measure the transit depth of each limb of the planet and to constrain the system parameters of the planet. We trim the first 10 integrations in all the NIRISS/SOSS light curves to allow for detector settling. While fitting the white light curve, we fix the period of the planet to 3.9502 days \cite{kokori2023}, because the impact of the $1\sigma$ period uncertainty on the transit light curve is $\sim$ $10^{-4}$~ppm. We vary the transit depths of the morning and evening limbs (($R_{\rm M}$/$R_{\rm S}$)$^2$ and ($R_{\rm E}$/$R_{\rm S}$)$^2$), where $R_{\rm M}$ and $R_{\rm E}$ are the radii of the morning and evening limb of the planet, respectively. $R_{\rm S}$ is the stellar radius. We also vary the mid-transit time ($T_0$), the semi-major axis ($a$)-to-stellar radius ratio ($a/R_{\rm S}$), impact parameter ($b$), and the limb darkening coefficients ($u_+$ and $u_-$). We assume a quadratic limb darkening law in our transit model \cite{brown2001}. The limb darkening coefficient $u_+$ is defined as $u_1+u_2$ and $u_-$ is $u_1-u_2$, where $u_1$ and $u_2$ are the coefficients of the linear and quadratic terms in the limb darkening law \cite{brown2001}, respectively. Free parameters were $a/R_{\rm S}$, impact parameter ($b$), the limb darkening coefficients, and the morning/evening transit depths without any applied priors. Initial values for those system parameters were taken from \cite{ahrer24}. We detrend the white light curve with a model that varies linearly with time. We use the \texttt{emcee} sampler \cite{emcee13} with 100 walkers and 1000 iterations to fit the white light curve. Table \ref{tab:wlcfits} summarizes the free parameters and the resulting constraints on each from the white light curve. Fig.~\ref{wlcwhole} shows the FIREFly white light curve and the best-fitting model from Catwoman.

WASP-94A~b has also been observed with JWST's NIRSpec instrument, with the G395H grating \cite{ahrer25}. Because limb asymmetry can be sensitive to mid-transit time \cite{murphy24}, we use these additional observations to more tightly constrain the mid-transit time ($T_0$) than the NIRISS/SOSS observations alone (Table \ref{tab:wlc_combineall3}). We analyze the NIRSpec/G395H data for both the ``NRS1" and ``NRS2" detectors using the standard FIREFLy pipeline v1.0.0 \cite{sing24}. We measure the white light curve of WASP-94A~b from these observations between 2.75 to 3.66~{$\upmu$}m for NRS1 and 3.86 to 5.11~{$\upmu$}m for NRS2. Then, we fitted the transit light curves for each detector with the Catwoman transit model following the same methodology as the white light curve fitting for the NIRISS/SOSS observations. Fig.~\ref{wlc_all3}A-C show the resulting white light curves. Fig.~\ref{wlc_all3}D shows the measured $T_0$ from each of the three transits with the Catwoman transit model, along with the combined $T_0$ and its uncertainty. The transit observed with NIRSpec happened 32 transits before the transit observed with NIRISS/SOSS. After calculating the $T_0$ of the NIRSpec transit in the epoch of the NIRISS/SOSS measurement, we combine the three $T_0$ measurements using inverse-variance weighted averaging and compute the uncertainty from the variance of the weighted average. The three $T_0$ measurements are consistent with each other and combining them leads to a 21.9\% improvement on the $T_0$ precision. Table \ref{tab:wlc_combineall3} lists the constraints on $T_0$ from each white light curve and the combined $T_0$ value and its uncertainty, which we adopt for subsequent analysis.

\subsubsection{Fitting the spectroscopic light curves}

\begin{figure}
\centering
\includegraphics[width=1.0\textwidth]{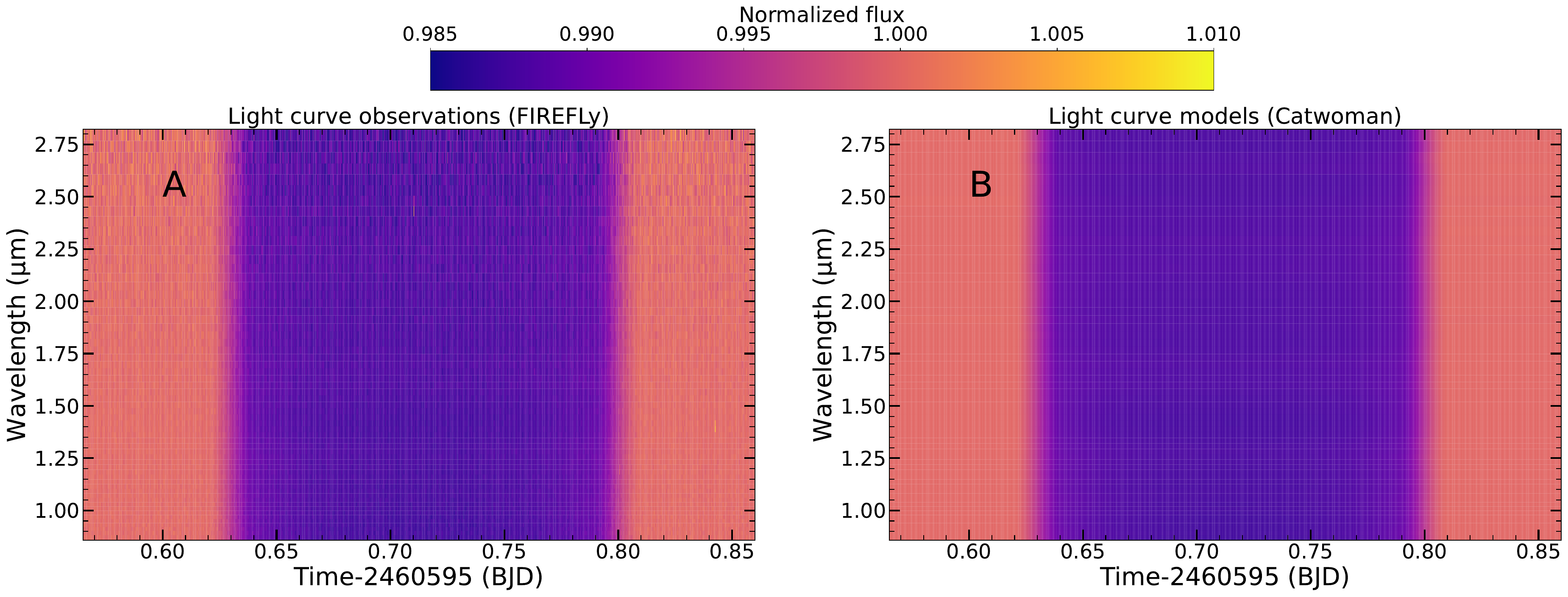}
\caption{{\bf Observed and modeled spectroscopic light curves}. ({\bf A}) the observed spectroscopic light curves as a heat map, from the FIREFLy data reduction. ({\bf B}) same as panel A, but for the Catwoman model fitted to these data.}\label{LCcolorplot}
\end{figure}

\begin{figure}[h]
\centering
\includegraphics[width=1.0\textwidth]{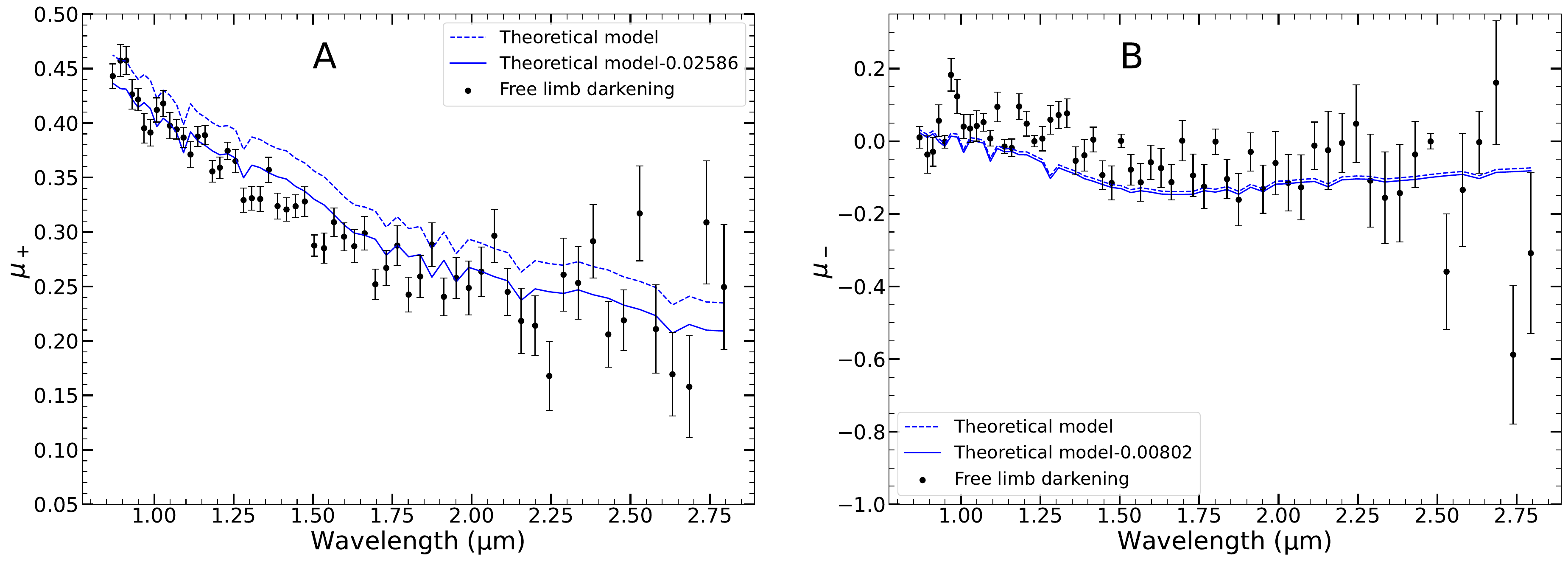}
\includegraphics[width=1.0\textwidth]{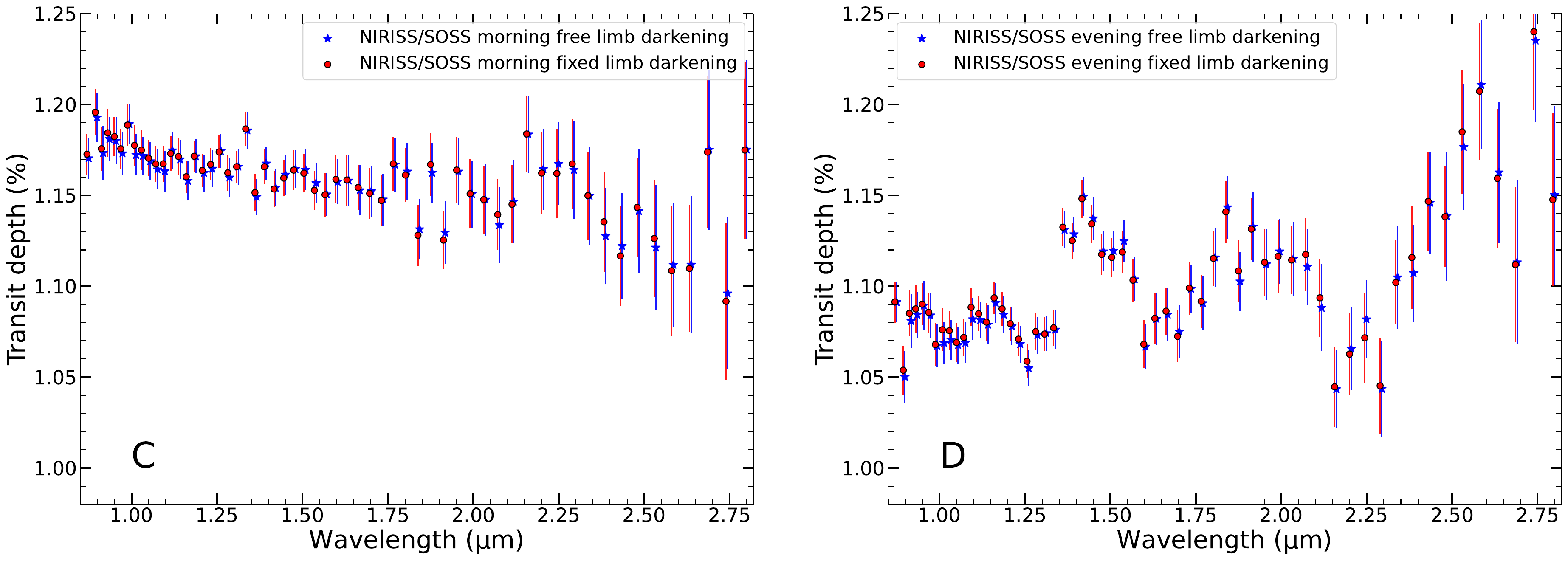}
\caption{{\bf Effect of stellar limb darkening on the transmission spectrum}. ({\bf A}) the $u_+$ parameter when fitted to each spectroscopic channel (black points). The $u_+$ parameter predicted from theoretical models  with and without offset (see text) are shown with blue lines. ({\bf B}) Same as panel A, but for the $u_-$ parameter. ({\bf C})-({\bf D}) the measured transmission spectrum of the morning and evening limbs, respectively. The blue points are the spectrum when limb darkening is fitted freely and the red points are the spectrum when limb darkening parameters are fixed to the 3D stellar model+offset. The spectrum with the freely fitted limb darkening are shown with an added offset of 0.005~{$\upmu$}m to wavelengths for visual clarity. All error bars show 1$\sigma$ uncertainty.}\label{LDfreefix}
\end{figure}

While fitting the spectroscopic light curves with Catwoman, we fix the $T_0$ (to the combined value in Table \ref{tab:wlc_combineall3}), $a/R_{\rm S}$, period, and $b$ to the median values determined from the white light curves. We assume that the planet's orbit is circular and its spin axis is perfectly aligned with the orbital axis of the planet  (spin-orbit angle $\phi$= 90$^{\circ}$), as expected for a tidally locked planet at such close proximity to its host star \cite{fabrycky2007,hut1981,peale1999}. We bin the spectroscopic light curves to a spectral resolving power $R\sim50$ before fitting them with a transit model. In addition to the morning/evening transit depths and the limb darkening coefficients, we also fitted a systematics model that varies linearly with time to the light curve from each spectroscopic channel. We use the \texttt{emcee} sampler for fitting these light curves. Fig.~\ref{LCcolorplot}B shows the wavelength vs. time 2D map of the normalized best-fitting model light curves for each spectroscopic channel. Fig.~\ref{LDfreefix}A-B shows a comparison between $u_+$ and $u_-$ constrained from fitting the spectroscopic light curves with theoretically predicted values for WASP-94A \cite{magic15}. The theoretical predictions match the wavelength-dependence of both limb darkening parameters, but show a constant offset from the measured values from the light curves. Therefore, we apply an offset of -0.02586 to the model predictions for $u_+$ and -0.008 to the model predictions of $u_-$ to match the measured coefficient values. The stellar models with the applied offsets are also shown  in Fig.~\ref{LDfreefix}A-B. We fitted the spectroscopic light curves again by fixing the limb darkening parameters to this modified stellar model prediction, to measure our final transmission spectra for the morning and the evening limbs. Fig.~\ref{LDfreefix}C-D show a comparison of the transmission spectrum derived for each limb with free limb darkening coefficients and the fixed limb darkening coefficients. We find that the features in the spectrum of each limb change very little in each case. We use the spectrum measured with the fixed limb darkening coefficients for our subsequent analysis. 

\begin{figure}[h]
\centering
\includegraphics[width=1.0\textwidth]{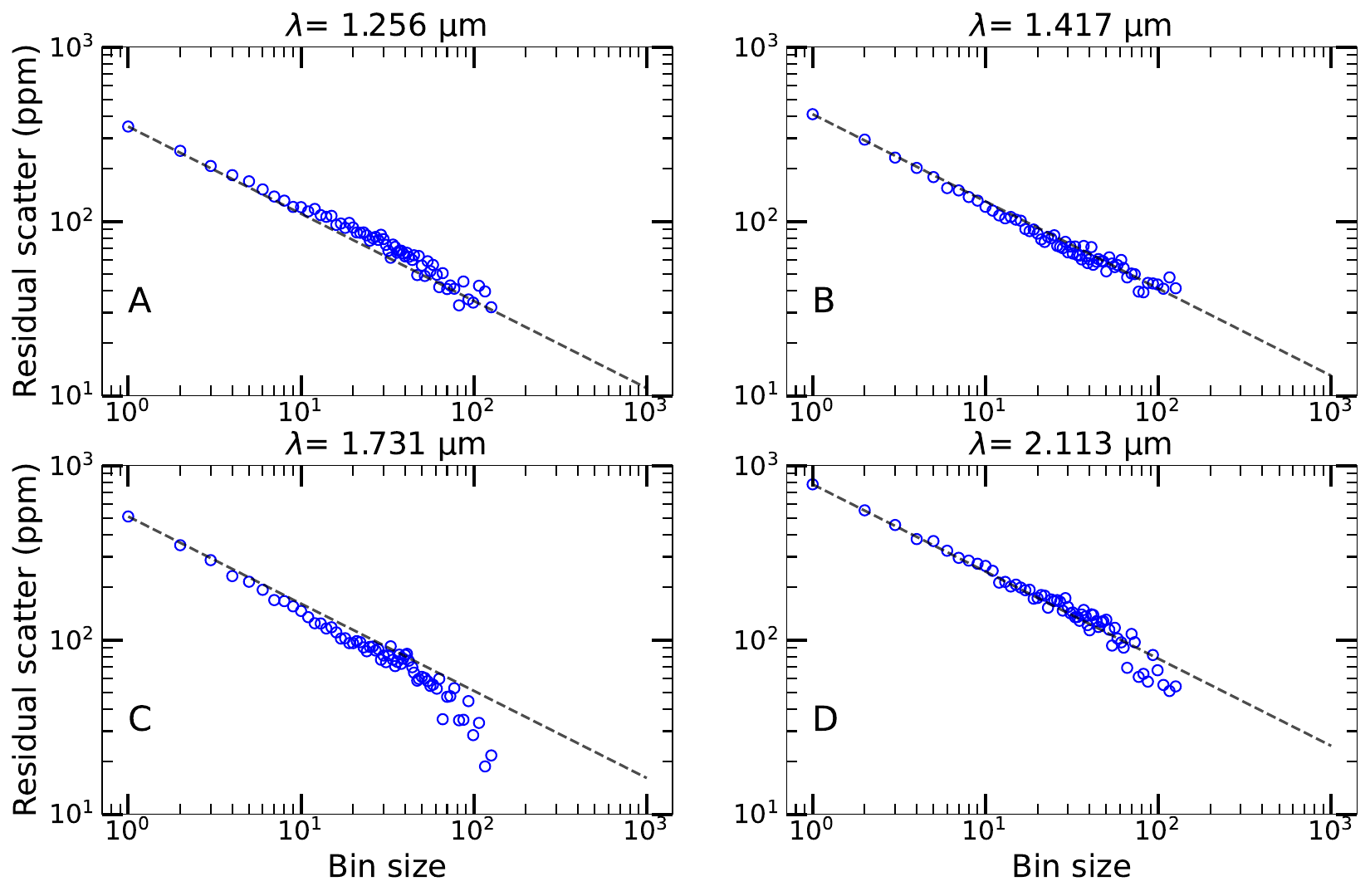}
\caption{{\bf Correlated noise in the spectroscopic light curves}. ({\bf A})-({\bf D}) the standard deviation of residuals as a function of bin size for four randomly selected spectroscopic channels at different wavelengths (blue points). The expected white noise behaviour is also shown (black dashed lines). In each case, we conclude there is negligible correlated noise is present in the spectroscopic light curves. }\label{rednoise}
\end{figure}

Fig.~\ref{rednoise}A-D show plots of the standard deviation of residuals vs. bin size for four randomly selected spectroscopic channels. We find that the spectroscopic light curves have negligible correlated noise. The residuals follow a 1/$\sqrt{N}$ behavior, where $N$ is the bin size. 

\subsubsection{Dependence of spectra on $T_0$}\label{sec:t0}

\begin{figure}[h]
\centering
\includegraphics[width=1.0\textwidth]{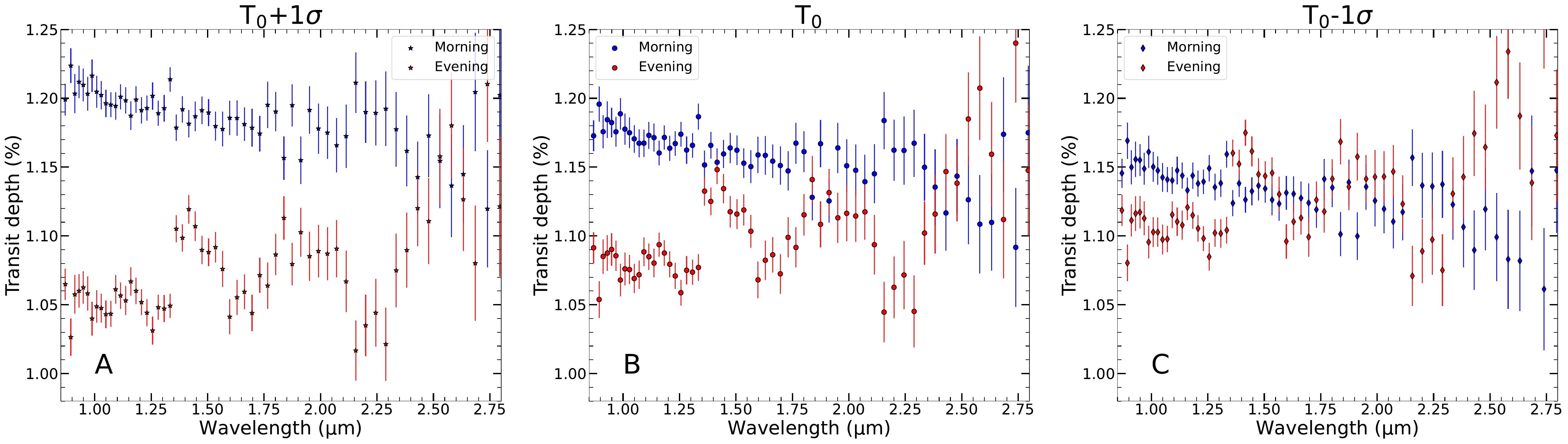}
\includegraphics[width=1.0\textwidth]{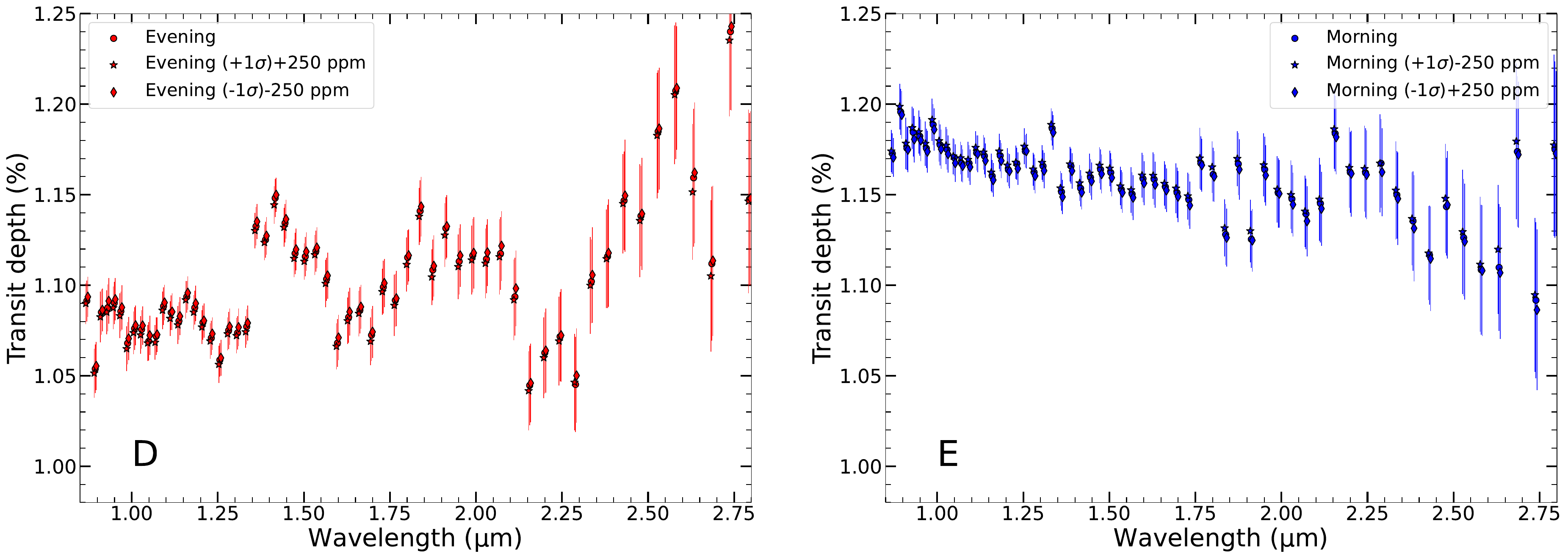}
\caption{{\bf Effect of mid-transit time uncertainty on the measured transmission spectrum for the morning and evening limb}. ({\bf A}) the morning (blue points) and evening limb spectrum (red points) if the mid-transit time is fixed to $T_0$+1$\sigma$, where $T_0$ is the error-weighted mid-transit time. ({\bf B})-({\bf C}) Same as panel A, but when the mid-transit time is fixed to $T_0$ and $T_0$-$1\sigma$, respectively. ({\bf D}) the three evening limb spectra from panels A-C, when an offset of +250~ppm and -250~ppm is applied to the evening limb spectrum from the $T_0$+$1\sigma$ and the $T_0$-$1\sigma$ cases, respectively. The three spectra are shown with different symbols: spectra corresponding to $T_0$ with circles, $T_0+1\sigma$ with stars, and $T_0-1\sigma$ with diamonds. A offset of $\pm$0.003~{$\upmu$}m has been applied to the wavelengths of the spectra corresponding to $T_0\pm1\sigma$ for visual clarity. ({\bf E}) same as panel D, but for the morning limb spectrum. All error bars represent 1$\sigma$ uncertainty.}\label{t0vsoffsetspec}
\end{figure}

The mid-transit time $T_0$ has a substantial effect on the measured transmission spectrum for the morning and evening limbs of WASP-94A~b. 
Following previous work \cite{murphy24}, we calculate a timing uncertainty of 7.1~seconds is needed to probe a limb asymmetry feature that spans 1 atmospheric pressure scale height, so the 1.4~{$\upmu$}m H$_2$O feature of 3.7 scale heights observed for WASP-94Ab corresponds to 26.3 second precision. Fig.~\ref{t0vsoffsetspec}A-C show how fixing the mid-transit time to the combined value $T_0$, $T_0$-1$\sigma$, and $T_0$+1$\sigma$ (Table \ref{tab:wlc_combineall3}) changes the resulting morning and evening limb spectrum. Fig.~\ref{t0vsoffsetspec}D-E show the morning and evening spectra measured by setting the mid-transit time to $T_0\pm1\sigma$, when a constant offset of $\pm$ 250~ppm is added to them. The offset causes them to overlap with the morning/evening limb spectrum derived by setting the mid-transit time to the combined $T_0$. The $1\sigma$ uncertainty on the combined mid-transit time corresponds to moving the morning limb spectrum relative to the evening spectrum by $\pm$250~ppm in transit depth. However, the shape of the spectrum and the spectral features for the morning and evening limb are not affected by the mid-transit time uncertainty. Therefore, we incorporate this additive shift within our modeling framework, described below (\S\ref{sec:retrieval}). Following an analytical formulation \cite{murphy24}, our 12~seconds uncertainty on the mid-transit time provides sensitivity to any limb-asymmetry signal $>$2 atmospheric pressure scale heights between the morning and evening limbs of WASP-94A~b. This is related to the 250~ppm absolute transit depth uncertainty of the morning and evening limb spectra, not to the relative strengths of spectral features. 

\subsubsection{Extracting 1D spectrum assuming spherical planet}

\begin{figure}[h]
\centering
\includegraphics[width=0.8\textwidth]{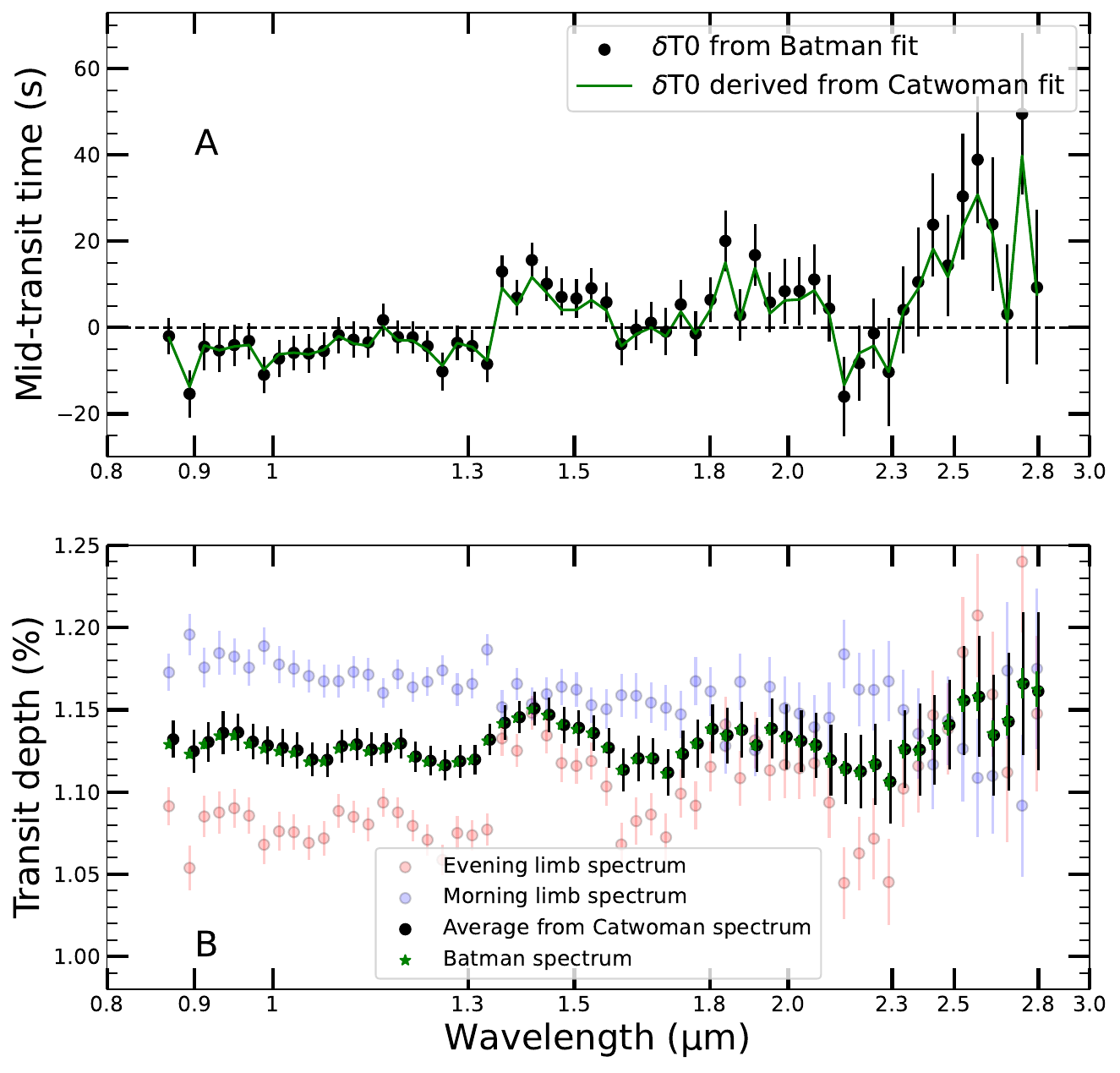}
\caption{{\bf Additional evidence of limb asymmetry in the spectroscopic light curves.} ({\bf A}) the mid-transit time ($T_0$) as a function of wavelength (black points) for all the spectroscopic light curves fitted with the spherical planet model. This allows an independent $T_0$ for each channel, which can compensate for the limb asymmetry of the planet. We find peaks that match the wavelengths of the H$_2$O absorption bands, implying that they are stronger in one limb than the other. The $T_0$ spectrum calculated from the separate spectra of the two limbs is shown for comparison (green line). ({\bf B}) the average planet spectrum (black points) derived from the spectrum of the two limbs (blue and red points), compared to the spherical planet spectrum extracted from the full transit (black points). A 0.005~{$\upmu$}m offset has been added to the wavelengths of the spectrum shown with black points for visual clarity. All error bars represent 1$\sigma$ uncertainty.}\label{T0vscatwoman}
\end{figure}

We also derive a transmission spectrum of the planet by fitting the spectroscopic light curves with the Batman transit software \cite{batman15}, which assumes that the planet is a perfect sphere. We fitted the white light curve from the NIRISS/SOSS observations with Batman, which finds a mid-transit time $T_0$=2460595.714929 ($\pm$1.7$\times{10^{-5}}$)~BJD. The uncertainty on this mid-transit time from the spherical planet model is an order of magnitude smaller than we derived using Catwoman (Table \ref{tab:wlcfits}) and differs by $\sim$1.4$\sigma$ from the limb-asymmetric planet model.
Fig.~\ref{compare_spectra}E-F shows the spectrum extracted with the Batman transit model. We search for limb--to--limb asymmetry by fitting the mid-transit time ($T_0$) for each spectroscopic channel with the Batman transit model \cite{espinoza24}. The asymmetry in the spectroscopic light curves due to the asymmetric limbs then appears as a shifting $T_0$ at wavelengths corresponding to absorption features, such as those of H$_2$O. Fig.~\ref{T0vscatwoman}A shows the $T_0$ from Batman for each spectroscopic light curve: the derived value of $T_0$ changes in the region of the H$_2$O absorption band between 1.35 to 1.5~{$\upmu$}m. This is consistent with the $T_0$ signature calculated analytically \cite{murphy24} from the transmission spectrum of the morning and the evening limbs measured with the Catwoman transit model. Fig.~\ref{T0vscatwoman}B also shows a comparison between the 1D spherical spectrum calculated by averaging the spectrum of the two limbs, with the 1D spherical spectrum derived from fitting the light curves with the Batman spherical transit model. The two spectra are consistent but have different uncertainties due to difference in the number of parameters involved in fitting the light curves. We adopt the spherical spectrum measured with a fixed $T_0$ in the subsequent analysis. 

\subsection{Fu Pipeline}

The data reduction steps with the Fu pipeline \cite{fu24,guangweifu_2026_tswift} were consistent with previous studies using SOSS \cite{fu24}. In stage 1, the data were processed by the JWST pipeline \cite{Bushouse2024} up to the \texttt{dark\_current\_step}, followed by group-level background subtraction using the \texttt{dark\_current.fits} files \cite{fu24}. The spectra PSF was masked and a column-by-column median subtraction was performed using the unmasked regions. Subsequently, the datasets were processed through the \texttt{jump\_step} and \texttt{ramp\_fit\_step}, yielding \texttt{rampfitstep.fits} files. 

Next, we performed spectral extraction by identifying the vertical location of the spectral trace in each column through a column-by-column cross-correlation with the spectral PSF. Spectra were extracted using a fixed, 30-pixel-wide aperture centered around this identified spectral trace. Each spectroscopic light curve was cleaned by removing outliers using a rolling median.
 
Similar to FIREFLy, we next fitted the white light curve with Catwoman model \cite{espinoza21} and the \texttt{emcee} sampler \cite{emcee13}, incorporating nine free parameters: visit-long slope, constant offset, mid-transit time, mean planet radius $R_p$, deviation from the mean planet radius $\Delta R_p$, inclination, $a/R_{S}$, $u_1$, and $u_2$. To enhance sampling efficiency and alleviate degeneracy between the limbs radii, the default $R_{\rm M}$ and $R_{\rm E}$ parameters are reparameterized as $R_{\rm E}$ = $R_p$ - $\Delta{R}_p$/2 and $R_{\rm M}$ = $R_p$ + $\Delta{R}_p$/2, where $R_{\rm M}$ and $R_{\rm E}$ are the radius of each limb. The best-fitting white light curve parameters are listed in Table \ref{tab:wlcfits}.

Similar to \texttt{FIREFLy}, we used the best-fitting parameters from the white light curve (mid-transit time, inclination, and $a/R_{\rm S}$) as fixed values in the subsequent spectroscopic channel analyses. Model for each channel included four free parameters: visit-long slope, constant offset, $R_p$, and $\Delta R_p$. The two quadratic limb darkening parameters were fixed to the  theoretical values with the offsets \cite{magic15} discussed above. After fitting, $R_p$, and $\Delta{R}_p$ were converted back to $R_{\rm E}$ and $R_{\rm M}$. The final morning and evening limb transit spectra are shown in Fig. \ref{compare_spectra}A-B. We discuss the differences below (\S\ref{sec:difference_pipe}).

\subsection{Eureka!}

We modified the \eureka~pipeline \cite{Bell2022} to account for the NIRISS/SOSS data. 
In Stage 1, we made no changes to the \eureka~wrapper around the default JWST pipeline, version 1.15.1 \cite{Bushouse2024}. Unlike the FIREFLy and Fu reductions, we did not correct for $1/f$ noise or apply group-level background subtraction at this stage.  The resulting \texttt{rateints} files were found to be already sufficiently clean and the final \eureka~ transmission spectrum matches the other reductions; therefore, we decided that these additional steps were unnecessary for this dataset.  In Stage 2, we skipped the \texttt{photom} step, as recommended for time-series observations \cite{Bouwman23}.  

In Stage 3, we supplied the \texttt{PASTASOSS} software \cite{baines23,baines232} with the NIRISS pupil position reported in the header to compute the trace and 1D wavelength solution for each spectral order. We then performed a double-iteration, $4\sigma$ outlier rejection of the sky background region along the time axis and computed a median flux frame that is free of bad pixels.
Before straightening the trace for a given order, we first masked the flux from the other order using the trace computed above and applied the spectral extraction half-width.  We then shifted each column by an integer number of pixels to produce a roughly-straightened 2D spectrum for each order.  Each order's spectrum was centered at the location specified in the \eureka~control file (ECF).
The column-by-column background subtraction routine operates on each spectral order.  For our reduction, we used a background half-width of 22 pixels and a zeroth-order polynomial.
We performed both standard and optimal spectral extraction for each order, using a spectral half-width of 17 pixels and the median frame as our optimal weighting function.  In Stage 4, we computed white and spectroscopic light curves for each spectral order.

We fitted the first order white light curve (0.86 to 2.80~{$\upmu$}m) using the Catwoman transit model \cite{espinoza21} with free quadratic limb darkening parameters, a quadratic baseline in time, and a scatter multiplier term.  We estimate uncertainties using \texttt{emcee} \cite{emcee13} and list the results in Table \ref{tab:wlcfits}.  
For the first order spectroscopic fits, we fix the transit time, inclination, and semi-major axis parameters to the best-fitting white light curve values.  We also apply a fixed limb darkening model using ExoTiC-LD \cite{Grant2024exotic} and the MPS2 grid \cite{kostogryz2022stellar}. For the spectroscopic light curves, we fitted a linear trend in time.  We tested several resolutions and settled on 59 channels at constant resolving power ($R\sim50$).

\subsection{Comparison of spectrum between independent pipelines}\label{sec:difference_pipe}

\begin{figure}[h]
\centering
\includegraphics[width=1.0\textwidth]{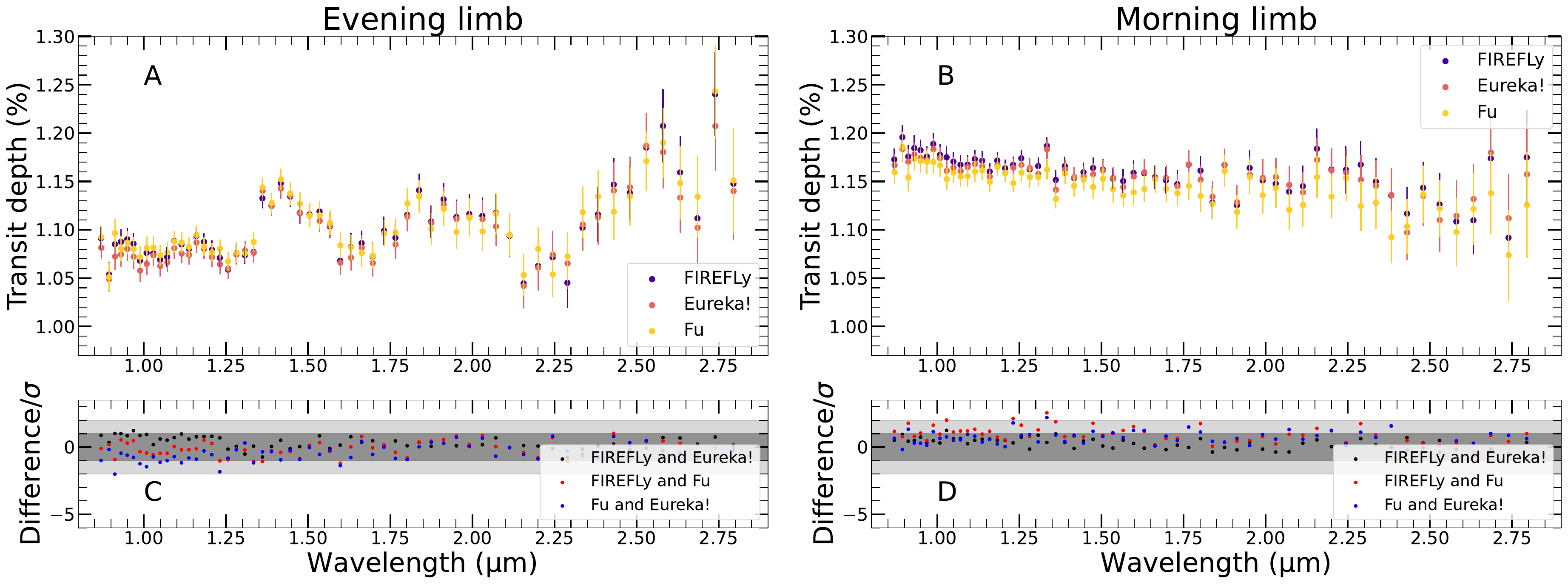}
\includegraphics[width=1.0\textwidth]{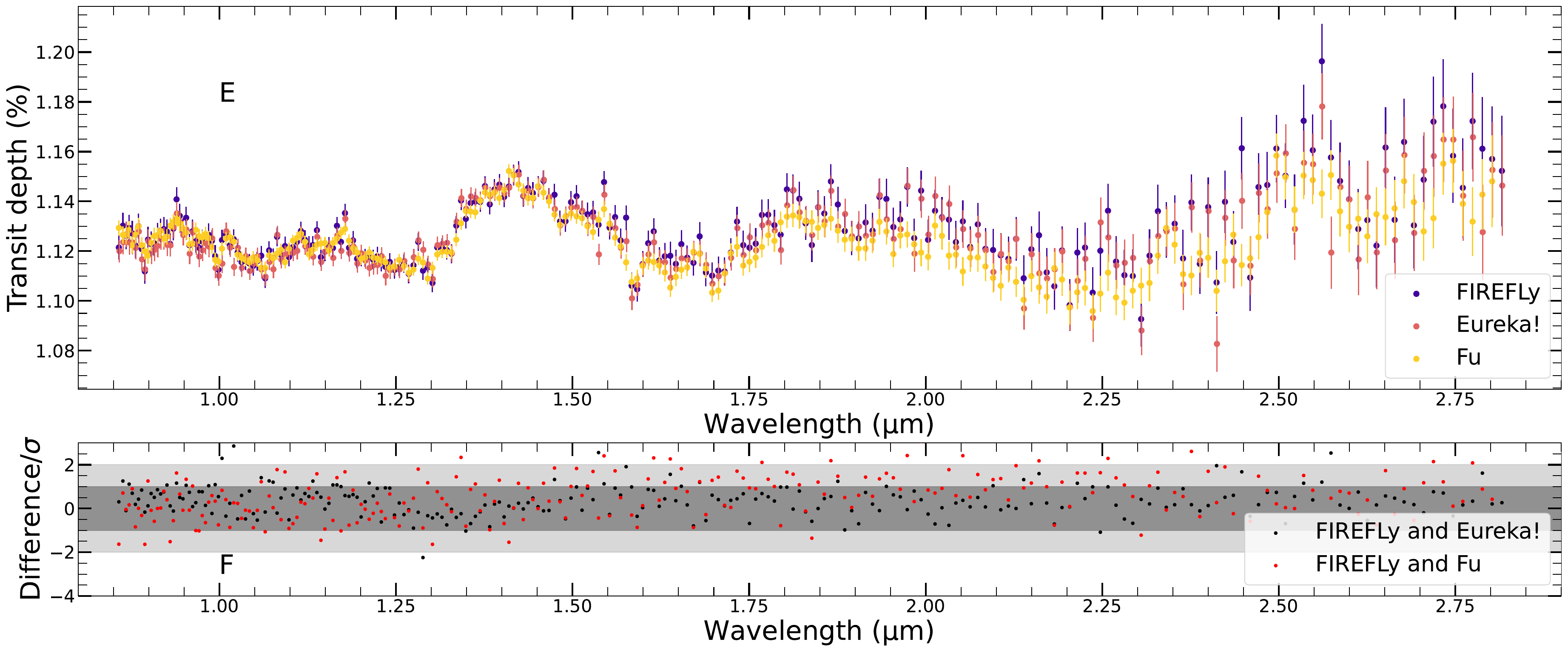}
\caption{{\bf Comparison between different data reduction pipelines.} ({\bf A}) the evening limb spectrum measured from three different reduction pipelines -- FIREFLy, Eureka!, and Fu (see legend). ({\bf B}) same for the morning limb spectrum. ({\bf C})-({\bf D}) the residuals between the spectrum divided by the uncertainties ($\sigma$) derived from each pipeline for the evening and morning limbs, respectively. The range where the residuals between pipelines are within $1\sigma$ and $2\sigma$ are shown with the dark and light gray regions, respectively. ({\bf E}) comparison between the spectrum from the three pipelines assuming a spherical planet model. ({\bf F}) Same as panels C-D, but for the spherical planet spectrum. All error bars represent $1\sigma$ uncertainty. }\label{compare_spectra}
\end{figure}

Fig.~\ref{compare_spectra}A-D shows a comparison of the transmission spectrum of the evening and morning limbs measured with the three pipelines described above. The shape of the spectrum for each limb is consistent between the three pipelines. However, there are small offsets in transit depths between each of them. We find that this offset is driven by differences in the $T_0$ constraints from the white light curve by each pipeline, as discussed in \S\ref{sec:t0}. The FIREFLy pipeline uses the NIRSpec/G395H white light curve along with the NIRISS/SOSS white light curve to determine the mid-transit time. The Fu and the Eureka! pipelines use only the white light curve from the NIRISS/SOSS observation to measure $T_0$. Although the derived $T_0$ differ by less than their reported 1$\sigma$ uncertainty, they still affect the offset between the morning and evening spectrum because $T_0$ is used as a fixed input for fitting the spectroscopic light curves in each pipeline. Fig.~\ref{t0_vs_diff} shows the difference in transit depth at 1.4~{$\upmu$}m between the morning and evening limbs from each pipeline, which depends on the mid-transit time $T_0$ obtained by that pipeline. This effect leads to about 100~ppm offset differences between the spectra derived by the three pipelines. This is less than the $\pm$250~ppm offset between the morning limb and evening limb spectra caused by the 12 s uncertainty (1$\sigma$) on $T_0$ (see \S\ref{sec:t0}). To account for both offsets, we allow the morning spectrum to shift with a variable offset within $\pm$250~ppm relative to the evening spectrum in our modeling and Bayesian retrieval analysis below. Fig.~\ref{compare_spectra}E-F shows a comparison of the spherical spectrum measured with each pipeline, which are consistent with each other.

\section{Atmospheric Modeling}

\subsection{Atmospheric Retrievals}\label{sec:retrieval}

\subsubsection{Retrievals on morning and evening spectra}

We use the \texttt{PICASO}  atmospheric model \cite{Mukherjee22,batalha19,mukherjee20} and the \texttt{VIRGA} cloud model \cite{ackerman2001cloud,rooney21} to perform a 1.5D Bayesian retrieval on the transmission spectrum of the morning and evening limbs of the planet simultaneously. Our retrieval of both limbs uses a previous $T(P)$ parametrization \cite{guillot10}, which includes 5 free parameters: equilibrium temperature $T_{\rm eq}$, opacity in the infrared wavelengths $\log(\kappa_{\rm IR})$, opacity in the visible wavelengths $\log(\gamma)$, fraction of the visible stream in the two stream approximation $\alpha$, and intrinsic temperature $T_{\rm int}$. We assume that the $T_{\rm int}$ parameter is the same for the morning and the evening limbs, such that the $T(P)$ profiles of both limbs converge to the same temperatures in the deep atmosphere. We also assume that the $\log(\kappa_{\rm IR})$, $\log(\gamma)$ and $\alpha$ parameters remain the same between both the limbs,  due to the lack of absorption features in the spectrum of the morning limb. These $T(P)$ profile parameters control the shape of the $T(P)$ profile, so fixing them forces the morning limb to have the same $T(P)$ shape as the evening limb in the upper atmosphere. 
We allow the morning and evening limbs to have different $T_{\rm eq}$ values to account for the temperature difference between the limbs. 

\begin{figure}[h]
\centering
\includegraphics[width=0.5\textwidth]{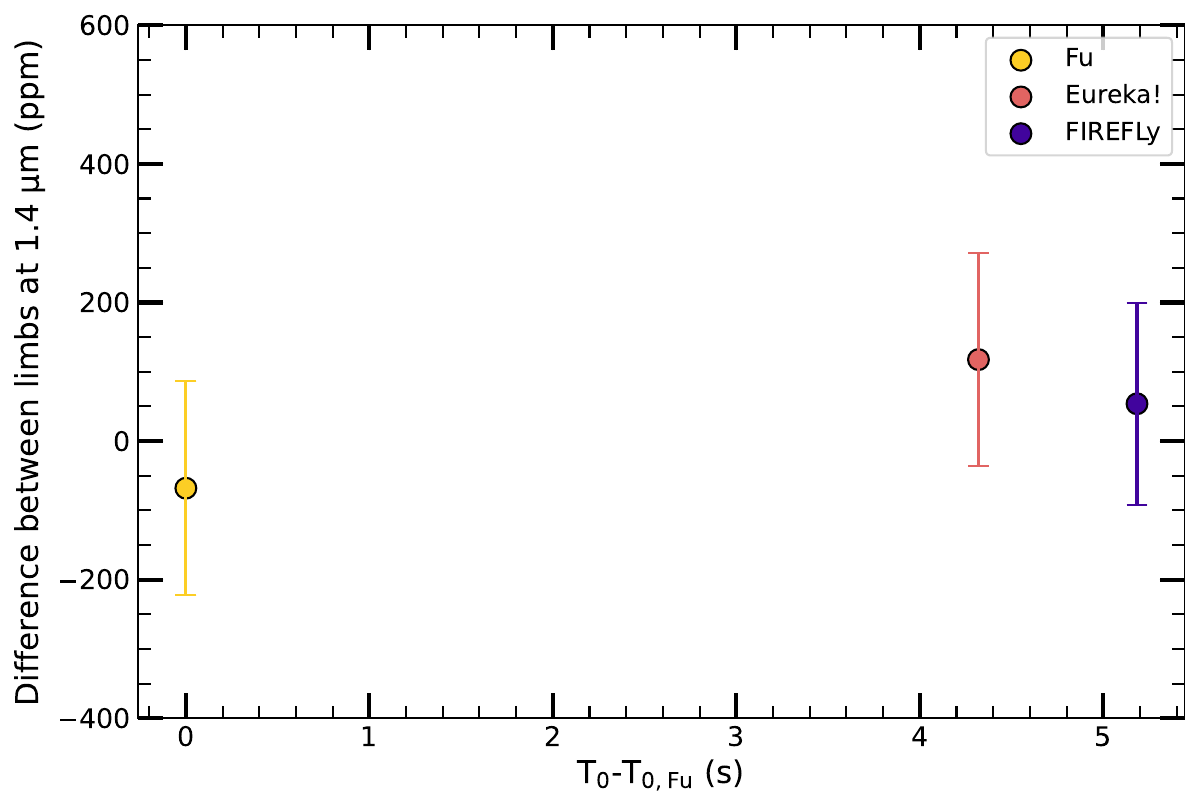}
\caption{{\bf Comparison between mid-transit time and the difference in transit depths between limbs.} The difference in the transit depths between the morning and evening limbs at 1.4~{$\upmu$}m from the three data reduction pipelines are plotted as a function of the constrained mid-transit time from the white light curves, expressed as offsets from the Fu pipeline value. The uncertainty on $T_0$ from each pipeline is larger than the difference in $T_0$ between the pipelines. The error bars show $1\sigma$ uncertainty.}\label{t0_vs_diff}
\end{figure}

We simulate the clouds in the morning and evening limbs of the planet's atmosphere using the \texttt{VIRGA} model \cite{rooney21,ackerman2001cloud}. The \texttt{VIRGA} cloud model simulates the cloud structure of the atmosphere using a balance between turbulent mixing and sedimentation of cloud droplets. The model includes two parameters: the cloud sedimentation factor $f_{\rm sed}$ and the vertical eddy diffusion coefficient $K_{\rm zz}$. The $f_{\rm sed}$ parameter is defined as the ratio between the sedimentation velocity of the cloud droplets and the velocity of convective mixing. Lower values of $f_{\rm sed}$ produce more vertically extended cloud decks \cite{rooney21,ackerman2001cloud}. We assume both the morning and the evening limbs have the same $K_{\rm zz}$ and $f_{\rm sed}$ parameter, so the vertical dynamics of cloud droplets between the two limbs are not different. We include three cloud forming species that are potentially relevant to the temperature regime of this planet: MgSiO$_3$, Fe, and MnS; their optical constants are adopted from previous work\cite{scott96,huffman67,natasha_batalha_2020_5179187}. We assume thermochemical equilibrium for both limbs, because disequilibrium chemistry is negligible at WASP-94A~b’s equilibrium temperature \cite{mukherjee25}. We vary the metallicity ([M/H]) and carbon-to-oxygen ratio (C/O) to calculate the abundances of gases in each limb based on the $T(P)$ profile. The transmission spectrum mostly shows H$_2$O absorption features on the spectrum of the evening limb. The H$_2$O abundance can be affected by processes like vertical mixing or photochemistry, but its abundance is almost insensitive to the strength of mixing at $T_{\rm eq}\ge 1000$K \cite{mukherjee25}. The fractional change in H$_2$O abundance due to photochemical processes is also negligible especially at pressures $\sim$ 1~mbar \cite{mukherjee25}. We find no observational evidence of C- or N- bearing gases including CH$_4$, CO, and NH$_3$. However, our retrieval includes opacities for H$_2$, H$_2$O, CO, CO$_2$, CH$_4$, H$_2$S, HCN, N$_2$, N$_2$O, and NH$_3$ \cite{HITRAN2016,Polyansky2018H2O,HITEMP2010,li15rovibrational,HUANG2014reliable,yurchenko13vibrational, yurchenko_2014,azzam16exomol,Harris2006hcn,Barber2014HCN,hitran2020,hitran2012,yurchenko11vibrationally,Wilzewski16}. We also include collision-induced opacities for H$_2$-H$_2$, He-H$_2$, and H$_2$-N$_2$ collisions \cite{Saumon12,Lenzuni1991h2h2}. We include a free parameter $xR_p$, which varies the planet's radius at 1~mbar as $R=R_p(1+xR_p)$, where $R_p$ is the measured planet radius from previous transit observations. We fit its value for both limbs simultaneously. Due to the dependence of the offset between the morning and evening limb spectra on the mid-transit time $T_0$ (Fig.~\ref{t0vsoffsetspec} and ~\ref{t0_vs_diff}), we allow an additive offset that varies between $\pm$250~ppm for the morning limb spectrum within our retrieval. Therefore, our retrieval model has 12 free parameters: 6 parameters for $T(P)$ profiles of the limbs, 2 chemical composition related parameters, 2 cloud parameters, 1 radius scaling parameter, and 1 offset parameter.

\begin{figure}[h]
\centering
\includegraphics[width=1.0\textwidth]{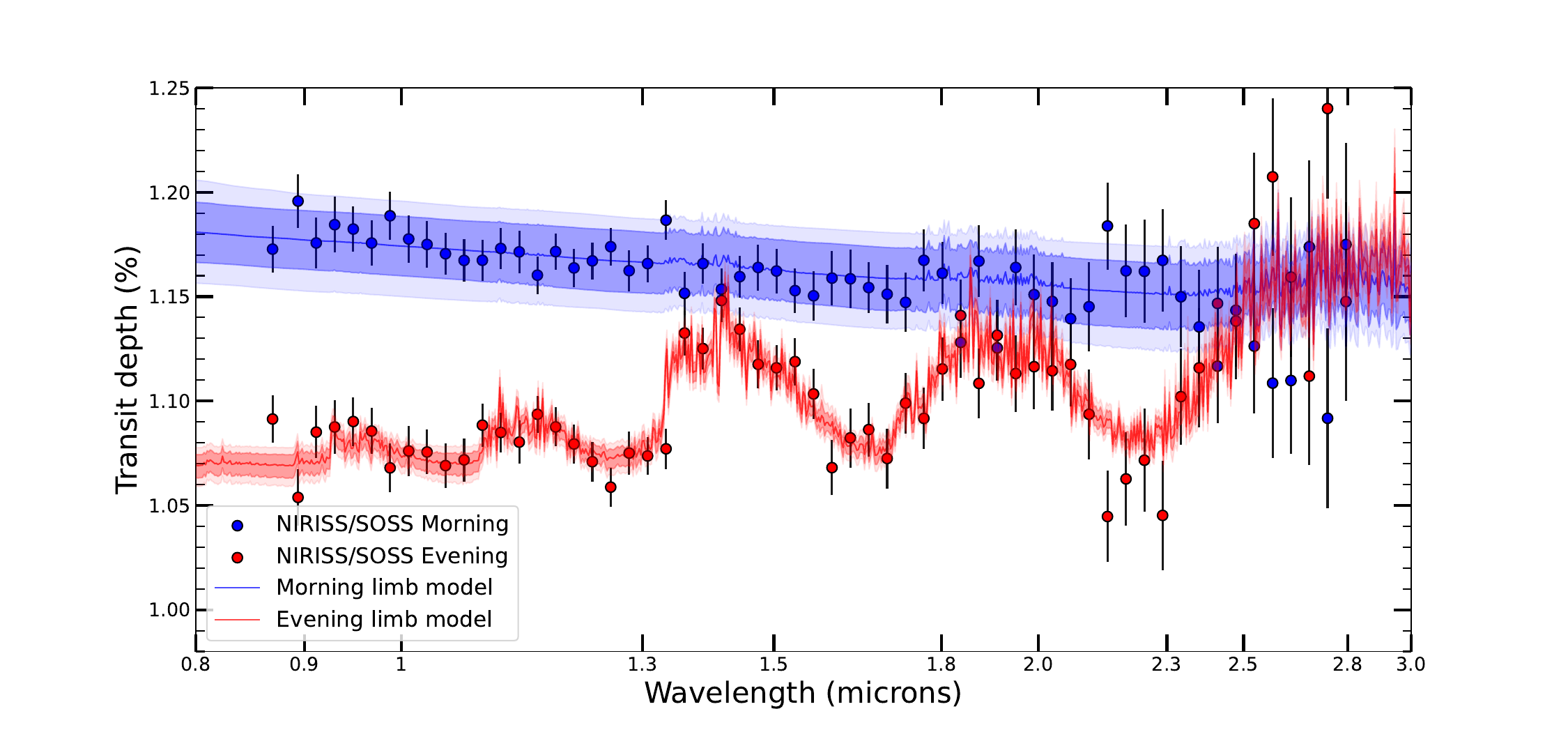}
\caption{{\bf Retrieved spectrum for each planet limb}. The retrieved median spectrum for the evening and morning limbs along with their 1$\sigma$ and 2$\sigma$ envelopes are shown in red and blue, respectively. The observed evening and morning spectrum are also shown with red and blue points, respectively. Error bars show $1\sigma$ uncertainty.}\label{limblimbret}
\end{figure}

\begin{figure}[h]
\centering
\includegraphics[width=1.0\textwidth]{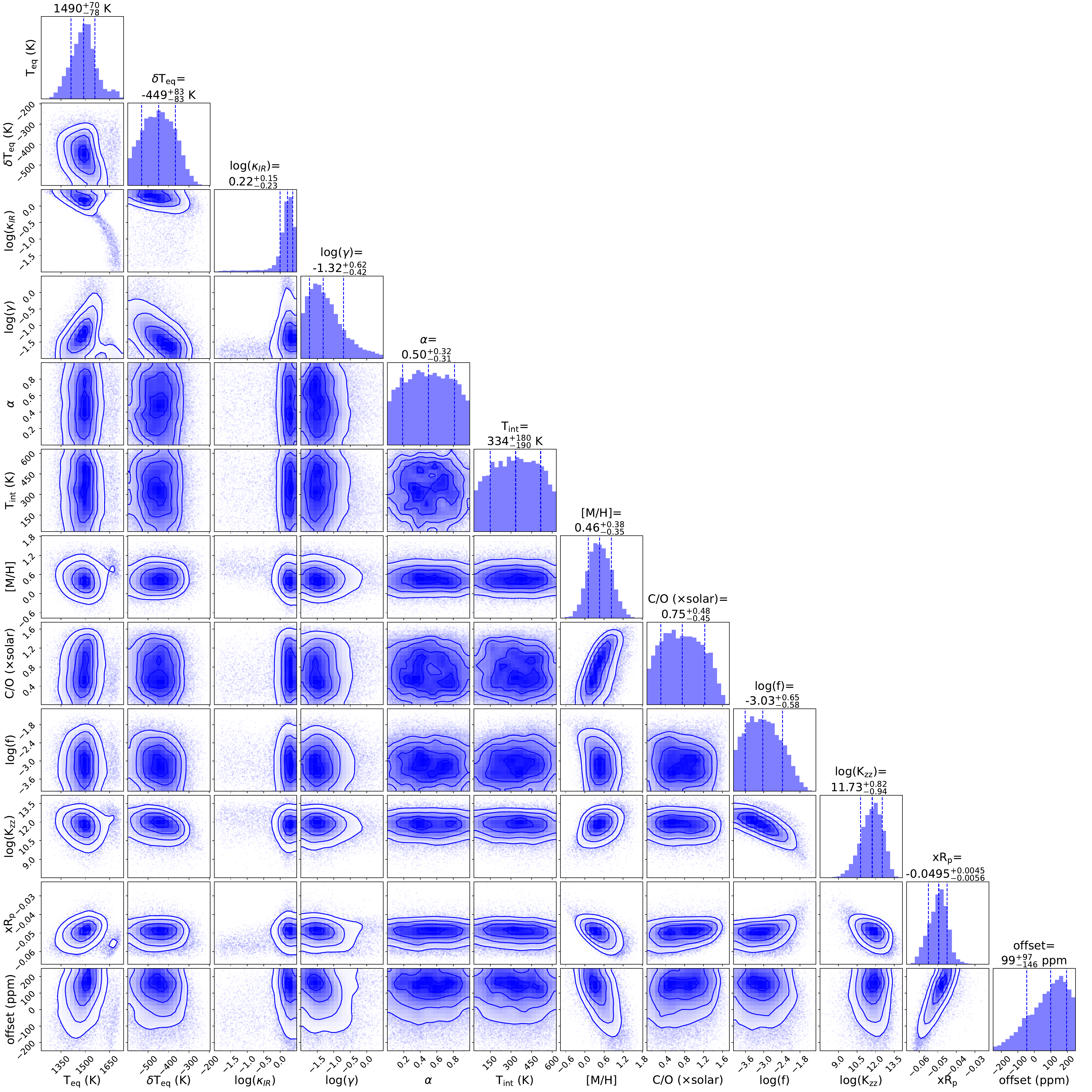}
\caption{{\bf Corner plot for simultaneous retrieval on both limbs}. The posterior probability distribution for each parameter are plotted in the diagonal panels. The vertical dashed lines show the median and $\pm1\sigma$ constraint for each parameter (also listed in Table~\ref{tab:retrieval}). The off-diagonal panels show the joint posterior probability distributions between each pair of parameters as a 2D heat map, showing correlations between fitted parameters. The contours on the heat map denote regions corresponding to $1\sigma$, $2\sigma$, and $3\sigma$ constraints, respectively. Units of parameters that have been fitted in logarithmic space are listed in Table~\ref{tab:retrieval}.}\label{corner}
\end{figure}

We use the pyMultiNest sampler \cite{feroz} for parameter estimation from the observed spectra. We define the log-likehood function, $\ln(\mathcal{L})$, in our retrieval as 
\begin{equation}
   \ln(\mathcal{L})=  -\frac{1}{2}\sum_i\left(\dfrac{y^{\rm morning}_i+O-f^{\rm morning}_i}{\sigma^{\rm morning}_i}\right)^2 -\frac{1}{2}\sum_i\left(\dfrac{y^{\rm evening}_i-f^{\rm evening}_i}{\sigma^{\rm evening}_i}\right)^2
\end{equation}
where $y^{\rm morning}_{i}$ and $y^{\rm evening}_{i}$ are the observed transit depth in the $i^{\rm th}$ spectral channel and $\sigma_{\rm morning/evening}$ denote their corresponding uncertainties. The $i$ parameter denotes each individual spectral channel. The $f^{\rm evening}_i$ and $f^{\rm morning}_i$ are the model transit depths from our retrieval model and $O$ is the offset parameter. We apply uniform priors on all 12 parameters, which is summarized in Table \ref{tab:retrieval}. We perform retrievals with 2000 live points and use evidence tolerance of 0.5 for convergence of the pyMultiNest routine. Fig.~\ref{limblimbret} shows the median retrieved spectrum for each limb along with $\pm1\sigma$ and $\pm2\sigma$ envelopes. Fig.~\ref{corner} shows the posterior probability distribution of the parameters and the correlations between them. The median and 1$\sigma$ constraint on each parameter are listed in Table \ref{tab:retrieval}. The $T(P)$ profile and the cloud properties of each limb are shown in Fig.~\ref{figtpclouds}. We cannot determine the specific species responsible for the clouds from our observed spectrum. Silicate clouds might be composed of other Si- containing minerals such as fosterite (Mg$_2$SiO$_4$) or quartz (SiO$_2$) rather than MgSiO$_3$, but this does not affect our conclusions.

Fig.~\ref{contribution}A-B show the atmospheric pressures that contribute to the observed spectra as a function of wavelength for the evening and morning limb of the planet. These contribution functions were calculated from the derivative of the wavelength and pressure dependent atmospheric transmittance, with respect to pressure in the best-fitting model from the 1.5D retrieval following previous work \cite{rustamkulov23}. The contribution function measures the relative contribution of a pressure layer to the transmission spectrum for a given wavelength. We find that the evening limb spectrum is sensitive to pressure ranges between $\sim$0.01 to 4~mbar whereas the morning limb spectrum is sensitive to $\sim$0.001 to 0.01~mbar. 

\begin{figure}[h]
\centering
\includegraphics[width=1.0\textwidth]{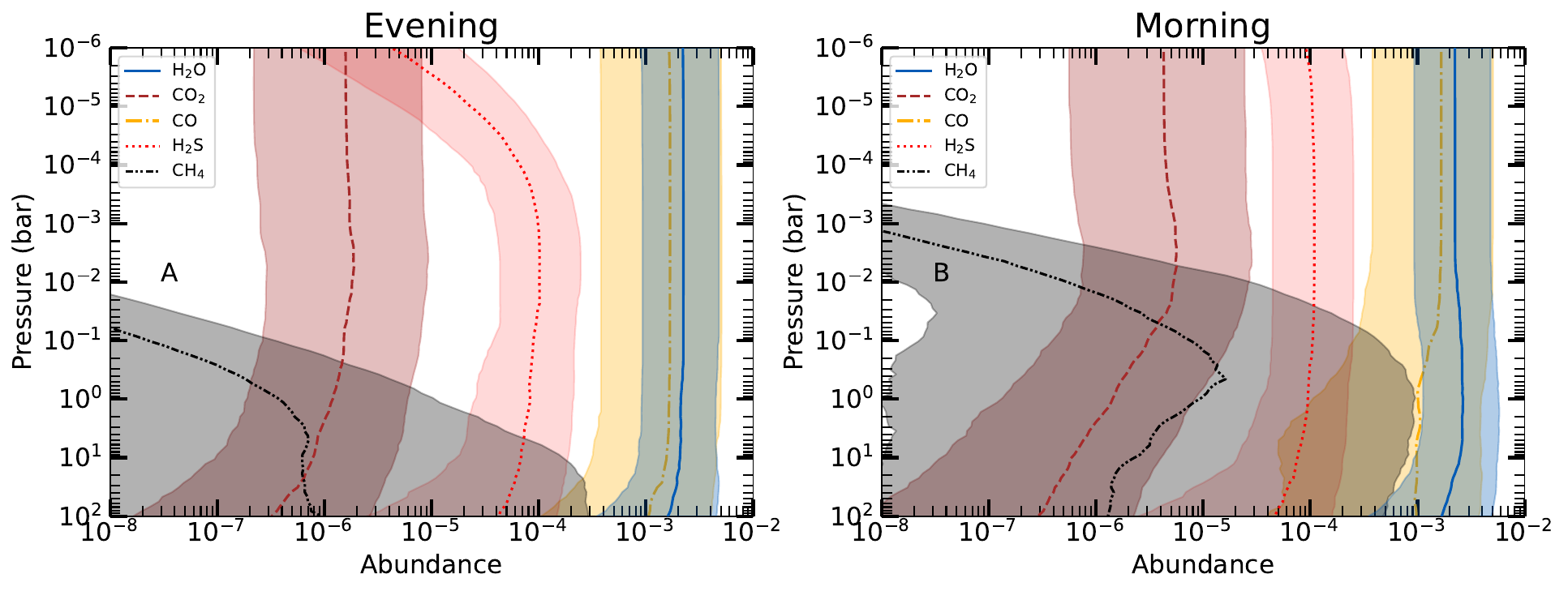}
\caption{{\bf Constraints on molecular abundances}. Abundances of potential atmospheric molecular gases on ({\bf A}) the  evening limb and ({\bf B}) the morning limb. Lines show the median volume mixing ratio and the shaded regions are their $1\sigma$ envelope. Only H$_2$O is detected in the spectrum but by assuming chemical equilibrium these calculations predict the abundances of other gases.}\label{chemistry}
\end{figure}

Fig.~\ref{chemistry} shows gas abundance constraints obtained from the 1.5D retrievals on each planet limb. Only H$_2$O features were identified in the transmission spectrum, but our retrieval analysis constrains other gases too. Both limbs of the planet have the same metallicity and C/O constraints, but the allowed abundances of individual gases are different between the morning and evening limbs due to the different $T(P)$ profiles. For example, CH$_4$ could be more abundant in the colder morning limb than the hotter evening limb (Fig.~\ref{chemistry}). The H$_2$O abundance constraints for each limb are consistent with each other (Fig.~\ref{chemistry}).

We perform two additional retrievals without H$_2$O and without clouds, which we use to determine the detection significance of each.   
Table \ref{tab:bayes_evi} compares the Bayesian evidence of each retrieval with the nominal 1.5D retrieval. We conclude that H$_2$O is detected at 10$\sigma$ and clouds at 9.7$\sigma$.

\subsubsection{Retrievals on the spherical spectrum}

\begin{figure}[h]
\centering
\includegraphics[width=1.0\textwidth]{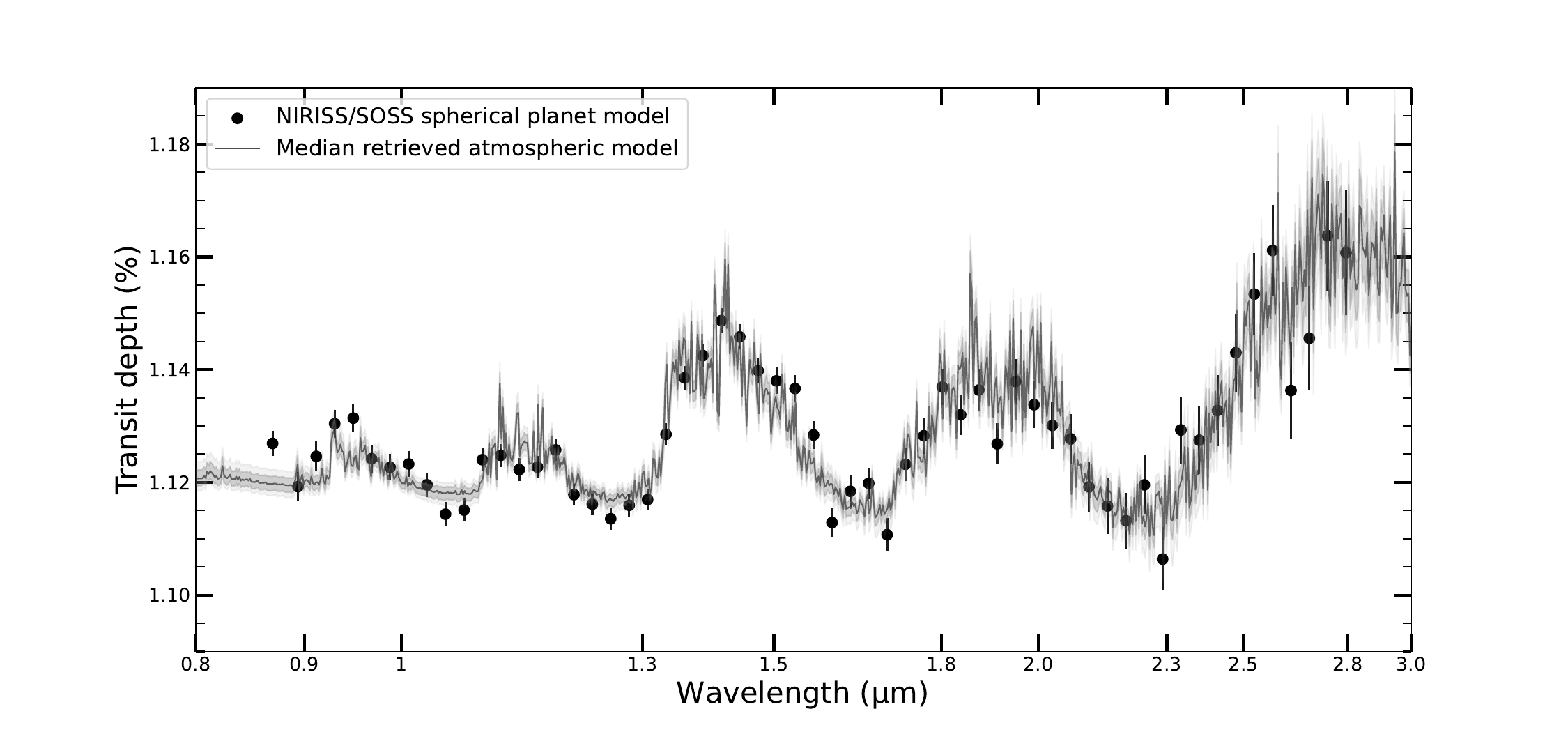}
\caption{{\bf Retrieved spectrum for the spherical planet spectrum}. The retrieved median spherical planet spectrum (gray line) along with its 1$\sigma$ (dark shading) and 2$\sigma$ (light shading) envelopes, compared to the measured spherical planet spectrum (black points). Error bars show $1\sigma$ uncertainties.}\label{1limbret}
\end{figure}

\begin{figure}[h]
\centering
\includegraphics[width=1.0\textwidth]{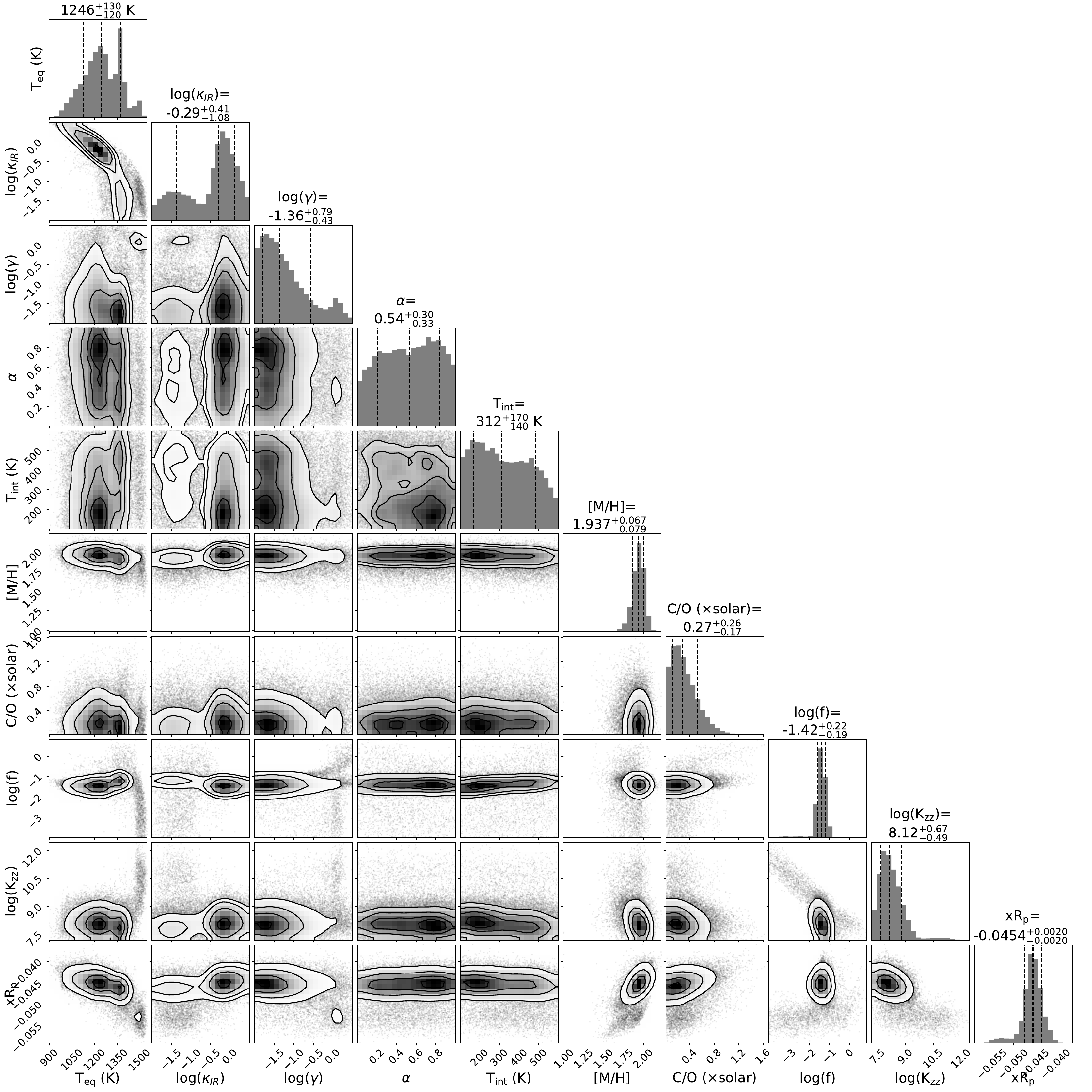}
\caption{{\bf Corner plot for 1D retrieval on the spherical spectrum of the planet}. Same as Fig.~\ref{corner}, but for the 1D \texttt{PICASO} retrieval on the spherical spectrum of the planet. Units of parameters that have been fitted in logarithmic space are listed in Table~\ref{tab:retrieval_1d}.}\label{1limbcorner}
\end{figure}

We also use \texttt{PICASO} to perform 1D atmospheric retrievals on the spherically symmetric spectrum. The forward model is the same as for the simultaneous morning/evening limb model, except with only one limb. The only free parameters and priors that differ from the morning/evening limb model are the removal of ${\delta}T_{\rm eq}$ and the offset between the morning and evening limb spectra. Fig.~\ref{1limbret} and Fig.~\ref{1limbcorner} show the retrieved transmission spectrum and the corner plot of the posterior probability distributions, respectively. The priors and constraints on each parameter are listed in Table \ref{tab:retrieval_1d}.

\begin{figure}[h]
\centering
\includegraphics[width=1.0\textwidth]{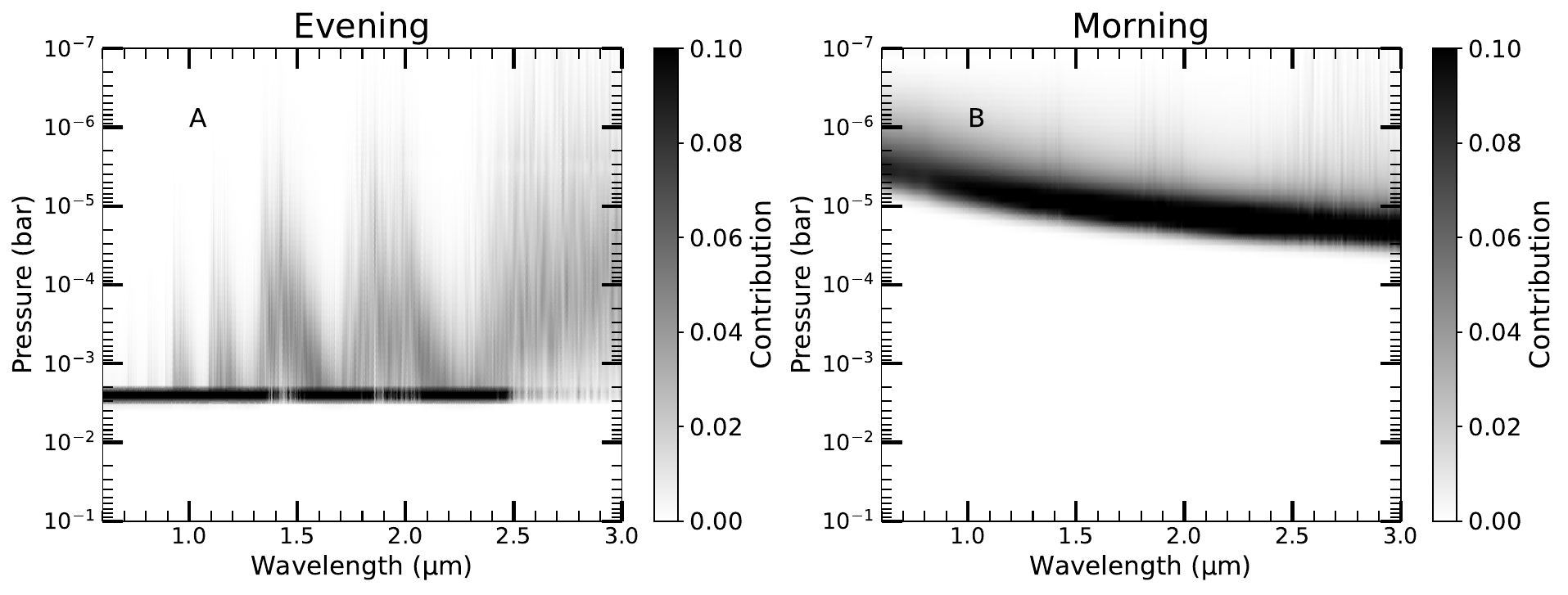}
\caption{{\bf Contribution function for the evening and morning limb spectrum}. ({\bf A}) the contribution function of the evening limb is plotted as a heat map for each pressure and wavelength. The contribution function measures the relative contribution of each pressure layer to the transmission spectrum for each wavelength and has been plotted here for the best-fitting 1.5D retrieval model for the evening limb. The contribution function was derived from the derivative of the transmittance of the atmosphere with respect to pressure for each wavelength. A darker gray color corresponds to higher contribution. ({\bf B}) same as panel A, but for the morning limb spectrum. }\label{contribution}
\end{figure}

\begin{figure}[h]
\centering
\includegraphics[width=1.0\textwidth]{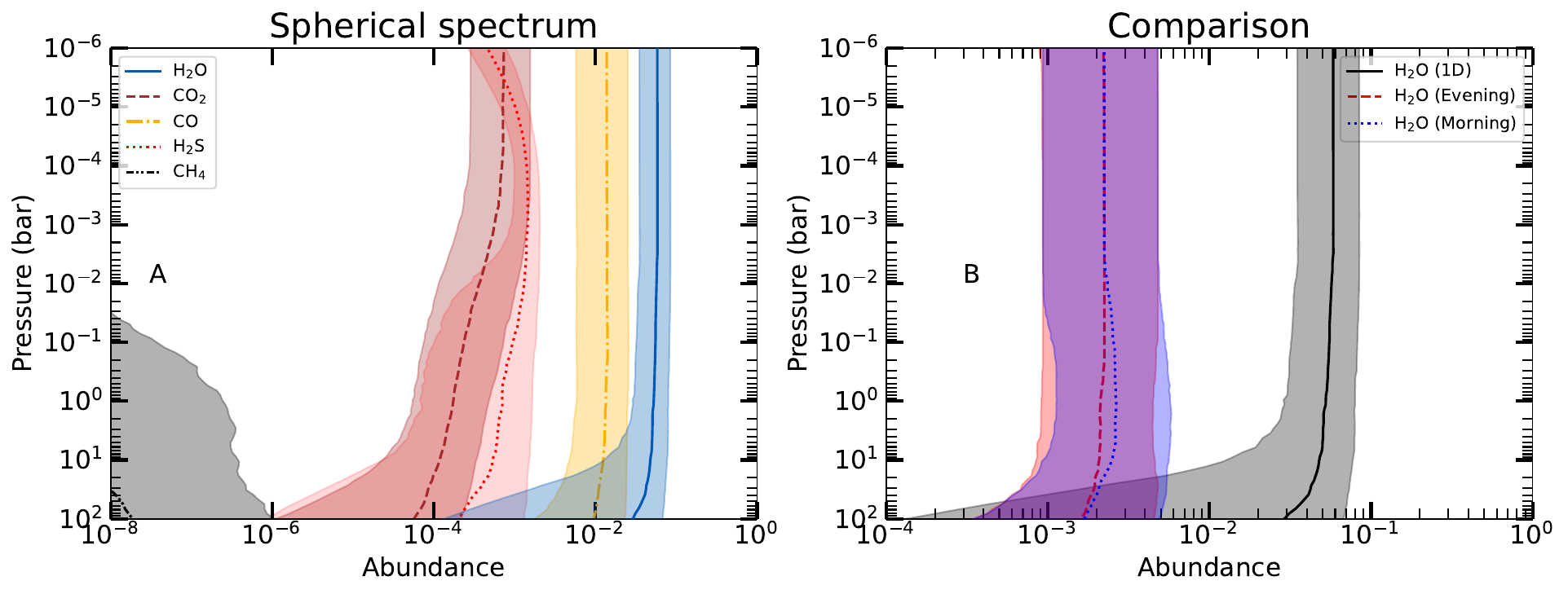}
\includegraphics[width=0.5\textwidth]{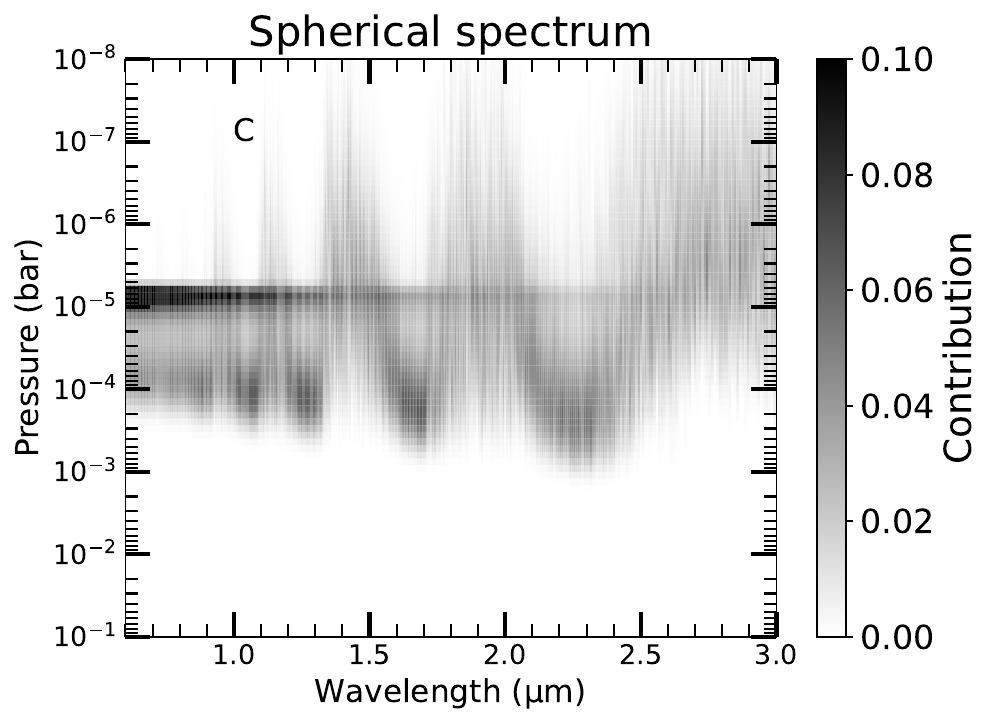}
\caption{{\bf Abundance constraints from the 1D retrieval on the spherical spectrum}. ({\bf A}) the biased abundance constraints from the 1D retrieval on the spherical transmission spectrum. Plotting symbols are the same as Fig.~\ref{chemistry}. ({\bf B}) comparison between the constraints on H$_2$O abundance from the 1.5D retrieval on the morning and evening spectrum with the 1D retrieval on the spherical spectrum. The blue and red lines show the median retrieved H$_2$O abundance profile from the morning and evening limb, respectively. The black line shows the H$_2$O abundance retrieved from the spherical planet spectrum. The shaded regions show $1\sigma$ envelopes on the abundances. ({\bf C}) the contribution function (as in Fig.~\ref{contribution}) of the 1D retrieval on the spherical spectrum.}\label{1limbchemcont}
\end{figure}

Fig.~\ref{1limbchemcont}A shows the abundance constraints for various gases from the 1D retrieval on the spherical spectrum. Fig.~\ref{1limbchemcont}B compares the H$_2$O abundances from the 1D retrieval on the spherical spectrum and the 1.5D retrievals on the morning and evening limb spectrum. We find that the spherical retrieval has 20 times higher H$_2$O abundance than either the morning or evening limbs, which we ascribe to bias in the spherical measurement. Fig.~\ref{1limbchemcont}C shows the contribution function for the transmission spectrum from the 1D retrieval; comparing this to the contribution functions for the two limbs (Fig.~\ref{contribution}) shows that the spherical case is affected more by lower pressures (higher altitudes).

We ascribe this bias to the use of a 1D retrieval model with global cloud cover to make inferences about an atmosphere that has patchy cloud coverage between the two limbs. The spherical transmission spectrum is an average between the morning and evening limb spectrum that effectively dilutes the strength of the gas absorption features (Fig.~\ref{figmhcto}). In such cases, ignoring the cloud coverage causes the retrieval framework to decrease the atmospheric pressure scale height to match the absorption feature sizes \cite{line16}. This can be achieved either by decreasing the temperature or increasing the mean molecular weight of the atmosphere. However, changing the temperature also affects the shape of the features because gaseous opacities are highly temperature dependent. Therefore, the model tends to enhance the atmospheric mean molecular weight by increasing the atmospheric metallicity. This bias does not arise from assumptions about chemistry (e.g., thermochemical equilibrium) because those were the same in our 1.5D and 1D retrievals. 

\subsubsection{Patchy cloud retrievals on spherical spectrum}

We also use the \texttt{POSEIDON} Python package \cite{MacDonald2017,MacDonald2023} to perform an atmosphere retrieval on our WASP-94A~b spherical transmission spectrum with 1D and patchy 2D cloud models. This atmosphere model assumes background gas consisting of H$_2$ and He in the ratio He/H$_2$ = 0.17 \cite{asplund09}.
We construct atmospheres with 150 layers uniformly spaced in log pressure, ranging from 10$^{-7}$ to 100~bar, assuming hydrostatic equilibrium.
We impose the boundary condition that the atmosphere reaches a pressure of 10$^{-3}$ bar at the planet's reference radius, which we fit for.
We freely retrieve the $T(P)$ profile following published methods \cite{Madhusudhan2009}. This model considers the following trace species: CO, CO$_2$, CH$_4$, H$_2$O, SO$_2$, H$_2$S, HCN, NH$_3$, K, and OCS.
We assume equilibrium chemistry and retrieve the atmospheric metallicity and C/O ratio.
For clouds and aerosols, we use the inhomogeneous cloud and haze parameterization  \cite{MacDonald2017} which introduces several additional parameters: a Rayleigh enhancement factor $\log a$ for the Rayleigh scattering cross-section of H$_2$, a scattering slope $\gamma$, a cloud-top pressure $\log P_{\text{cloud}}$, and the fraction of limb cloud coverage $\phi_{\text{cloud}}$. This cloud model is substantially different from that used in both the 1.5D and 1D retrievals with \texttt{PICASO}, because it does not assume specific aerosol optical properties to simulate clouds and hazes.
For the 1D cloud model we assume a cloud coverage fraction $\phi_{\text{cloud}} = 1$, while for the 2D patchy cloud model we allow it to vary as a free parameter.
In total, the 1D cloud model has 12 free parameters: 6 $T(P)$ profile parameters, 3 cloud/haze parameters, and the reference radius, atmospheric metallicity, and C/O ratio. The patchy 2D cloud model has 13 free parameters: the same as the 1D cloud model, plus the cloud coverage fraction $\phi$.

We calculate model spectra in the \texttt{POSEIDON} retrievals by solving the equation of radiative transfer in a cylindrical coordinate system for 100 incident stellar rays, which are attenuated according to the atmospheric opacity along line-of-sight.
We pre-compute our opacities at high resolving power, $R=60,000$, across a grid of temperatures and pressures using the \texttt{Cthulhu} Python package \cite{Agrawal2024}.
We derive the opacities using the following spectroscopic line lists or measured cross sections: CO \cite{li15rovibrational}, CO$_2$  \cite{yurchenko20}, CH$_4$ \cite{Yurchenko2024}, H$_2$O \cite{Polyansky2018H2O}, SO$_2$  \cite{Underwood2016}, H$_2$S \cite{azzam16exomol}, HCN \cite{Barber2014HCN}, NH$_3$ \cite{coles19}, K \cite{Ryabchikova2015}, and OCS \cite{Owens2024}.
We also include continuum collision-induced absorption from H$_2$-H$_2$ and H$_2$-He pairs \cite{Karman2019} and Rayleigh scattering for all gases \cite{MacDonald2022}. 
We calculate model transmission spectra using opacity sampling on a wavelength grid from 0.8 to 2.9~{$\upmu$}m,  and we use nested sampling via the \texttt{PyMultiNest} \cite{feroz,Buchner2014} Python package with 1000 live points to explore the parameter space for both of our retrievals.

\begin{figure}[h]
\centering
\includegraphics[width=1.0\textwidth]{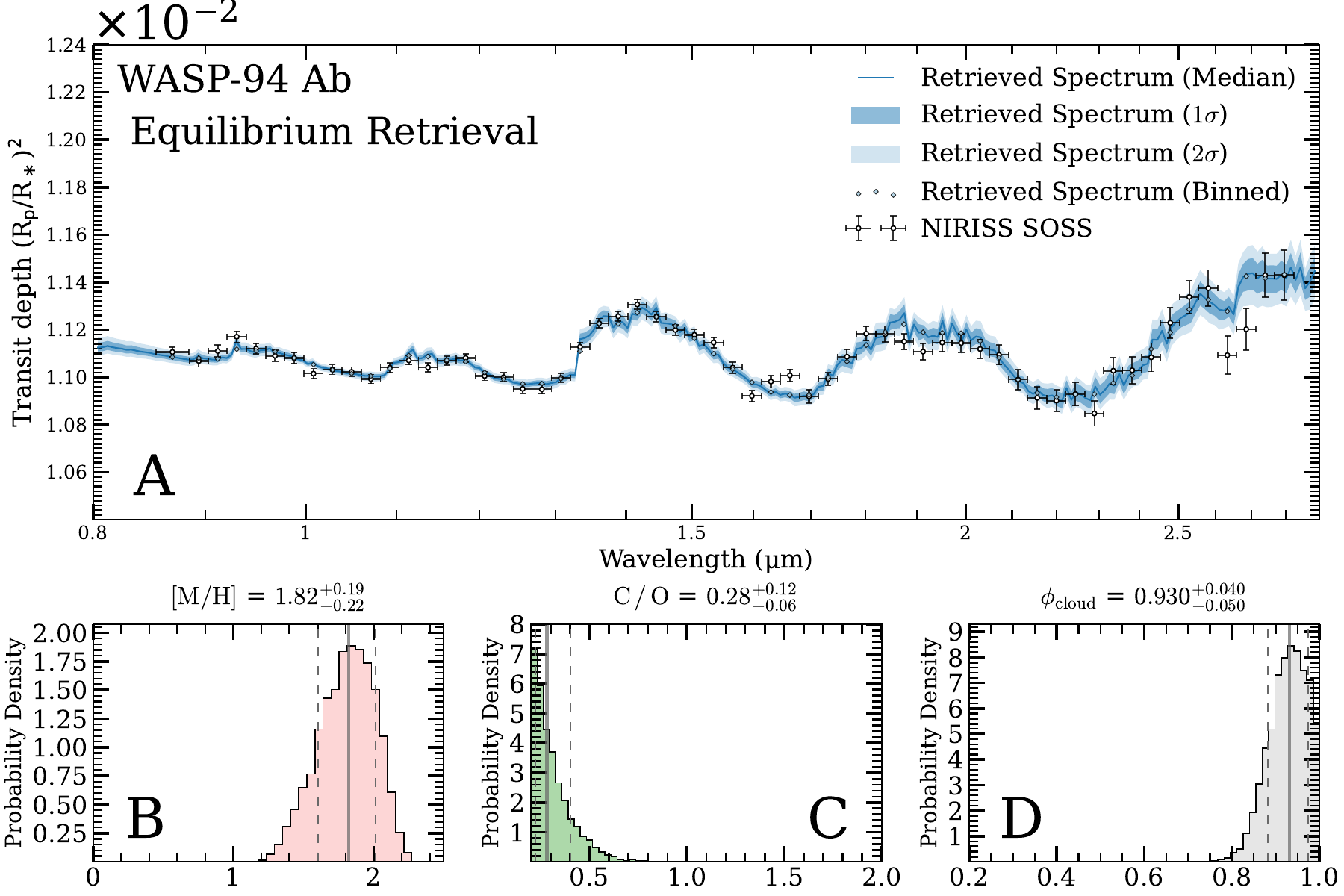}
\caption{{\bf Results of the patchy 2D cloud retrieval with POSEIDON}. ({\bf A}) the retrieved median spectrum (blue line) with the 1$\sigma$ (dark shading) and 2$\sigma$ (light shading) regions, along with the measured spherical planet spectrum (unfilled black points). The median model spectrum binned to the resolving power of the data are shown with the blue filled diamond points. All error bars represent $1\sigma$ uncertainties. Histograms of the posterior probability distributions for ({\bf B}) atmospheric metallicity, ({\bf C}) C/O ratio, and ({\bf D}) cloud coverage fraction. The gray solid vertical lines in ({\bf B})-({\bf D}) show the median values of each posterior probability distribution. The gray dashed lines show the $\pm1\sigma$ values for the posterior probability distributions.}\label{poseidon2Dretrieval}
\end{figure}

\texttt{POSEIDON} retrievals indicate an atmospheric metallicity ([M/H]=+1.91$\pm$0.15) and C/O ratio (C/O=0.27$\pm$0.07) that are severely inconsistent with our 1.5D \texttt{PICASO} simultaneous morninga and evening limb retrieval. However, the constraints on metallicity ([M/H]=+1.82$\pm$0.20) and C/O (C/O=0.28$\pm$0.09) obtained with the patchy cloud \texttt{POSEIDON} retrieval on the spherical spectrum are consistent with the constraints from the \texttt{PICASO} 1D retrieval on the spherical spectrum. The fitted patchy cloud models are shown in Fig.~\ref{poseidon2Dretrieval}A. The retrieved probability distributions for the metallicity and C/O ratio are shown in Fig.~\ref{poseidon2Dretrieval}B-D. 
The retrieved cloud parameter values are $\log~a=3.87^{+0.44}_{-0.42}$, $\gamma=-3.10^{+0.79}_{-0.95}$, and $\log~P_{\rm cloud}=-3.45^{+2.95}_{-0.55}$. The cloud coverage fraction $\phi_{\text{cloud}}=$$0.93^{+0.04}_{-0.05}$ is consistent with a fully cloudy atmosphere, conflicting with the minimal cloud contribution to the evening limb spectrum in our limb-resolved observations. The posterior probability distribution for the cloud coverage fraction is shown in Fig.~\ref{poseidon2Dretrieval}D. We interpret this inconsistency as further evidence of a bias when using spherical transmission spectra, even if the cloud coverage is allowed to be patchy. 

\subsection{3D Global Circulation Models}
\label{UM_GCM}

We use the Met Office \textsc{Unified Model} (\textsc{UM}) GCM adopting the same fundamental setup as previous work \cite{drummond20,zamyatina23,zamyatina24} to simulate the 3D atmospheric structure of WASP-94A~b. The dynamical core of the \textsc{UM}, \textsc{ENDGame} \cite{wood14,mayne13,mayne14b} solves the deep-atmosphere non-hydrostatic equations of motion using a semi-implicit semi-Lagrangian scheme on a constant angular grid. The radiative transfer is computed using SOCRATES\cite{edwards96,edwards_slingo_96,amundsen14,amundsen16,amundsen17} which solves the two-stream radiative transfer equations, handling opacities using the equivalent extinction and correlated-k methods. Opacity sources contained in the radiative transfer scheme are absorption due to CO, H$_2$O, CO$_2$, HCN, CH$_4$, NH$_3$, Li, Na, K, Rb, Cs and collision-induced absorption due to H$_2$-H$_2$ and H$_2$-He, together with Rayleigh scattering due to H$_2$ and He [\cite{zamyatina23}, their Appendix A]. \textsc{UM} computes the chemical structure and evolution of hot Jupiter atmospheres, using an ``equilibrium'' scheme.

We adopt the \textsc{PHOENIX} BT-Settl stellar spectrum \cite{rajpurohit13} with the closest match to the stellar parameters of WASP-94A \cite{ahrer24}. The \textsc{UM} simulations were initialized with a day-side mean pressure-temperature profile between $10^{-5}$ to ${10^{2}}$~bar from the 1D radiative-convective-chemistry model \textsc{ATMO} \cite{tremblin15,drummond16,drummond19} assuming chemical equilibrium and a chemical network \cite{venot12}. We applied an adiabatic correction to the input pressure-temperature profile between pressures of $10^{-1}$ to $10^{1}$~bar, which was smoothed using a Bézier curve to an isotherm of temperature $3500$~K between $10^{1}$ to $10^{2}$~bar. The adjusted profile corrects for fundamentally unphysical deep atmosphere thermal inversions due to initial conditions \cite{drummond20}. We performed one simulation at solar metallicity \cite{asplund09} with adjusted abundances for C, O, N, and K \cite{caffau_solar_2011}. The gas phase chemistry is treated by an equilibrium chemical scheme which calculated the chemical abundances using (a) a Gibbs minimization scheme \cite{Gordon94} for the C-H-O-N chemical species present in the reduced chemical network \cite{venot19}, and (b) the smoothing of the transformation boundaries for the alkali metals \cite{burrows99} .
The \text{UM} employs a height based grid structure. Due to the temperature contrast between the two sides of the atmosphere, the pressure, for a given height, on the day-side and night-side varies by orders of magnitude. Therefore, as material advects between the day-side and night-side, gravity waves are induced, which cascade through the model domain causing numerical instabilities. To maintain model stability, we limit the minimum pressure to $10^{-5}$~bar.
The \textsc{UM} simulation was evolved for $1000$ Earth days until the upper atmosphere ($10^{-5}$ to $10^{0}$ bar) reached pseudo-equilibrium, determined from the settling of maximum wind velocities, total outgoing top of atmosphere flux, and atmospheric kinetic energy. The input parameters for our simulations of WASP-94Ab are listed in Table \ref{tab:3d_gcm_parameters}.

\begin{figure}[h]
\centering
\includegraphics[width=0.9\textwidth]{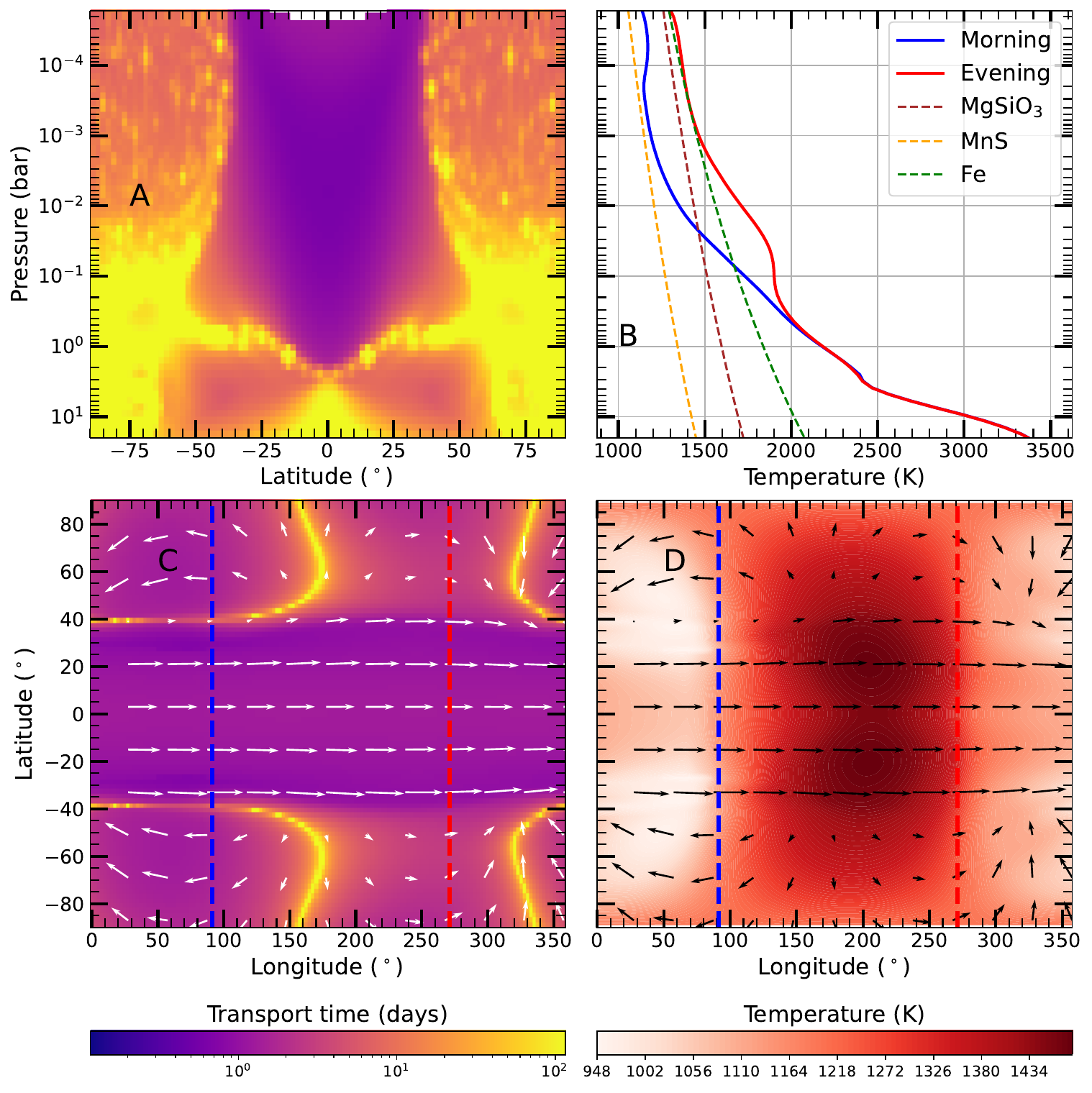}
\caption{{\bf 3D atmospheric structure from the GCM simulations of WASP-94A~b.} All panels show final day of the simulation. ({\bf A}) zonal-mean zonal advective timescale (color bar) as a function of pressure and latitude. ({\bf B}) $T(P)$ profiles, at latitude $0^{\circ}$, for the morning (blue) and evening (red) limbs. Dashed lines are condensation curves for MgSiO$_3$ (brown), Fe (green), and MnS (yellow), as in Fig.~\ref{figtpclouds}. ({\bf C}) the zonal advective timescale at $0.01$~mbar (color bar) as a function of longitude and latitude. The wind direction is marked with white arrows. The red and blue dashed lines show the locations of the evening and morning limb, respectively. ({\bf D}) the temperature (colour bar) for an isobaric surface at $0.01$~mbar. Horizontal wind vectors are over-plotted with black arrows.}\label{fig:GCM_timescales}
\end{figure}

Fig.~\ref{fig:GCM_timescales}A shows the resulting zonal-mean zonal advection timescale as a function of pressure. We find an equatorial jet extends between the latitudes of $\pm 40^{\circ}$, to a maximum depth of $1$~bar. This region has a minimum transport timescale of $1.15$ days. Surrounding the equatorial jet, a retrograde flow develops with a typical transport timescale an order of magnitude slower than that within the jet. Fig.~\ref{fig:GCM_timescales}C shows the zonal advection timescales at 0.01 mbar across different longitudes and latitudes, indicating the circulation pattern in the model. Fig.~\ref{fig:GCM_timescales}B shows the $T(P)$ profile for the two limbs at mid-latitudes of the GCM. A temperature difference of $300$~K is found between the morning and evening limbs of WASP-94A~b at $1$~mbar, similar to the value we inferred from the Bayesian retrievals on the limb asymmetric spectrum (449$\pm$ 83 K). The temperature difference in the GCM averaged across all latitudes between the morning and evening limb is slightly lower $\sim$265 K. Fig.~\ref{fig:GCM_timescales}D shows the temperature at 0.01~mbar across different latitudes and longitudes of the model; we find that the evening limb is hotter than the morning limb for most of the mid-latitudes surrounding the equatorial region of the planet.

\begin{figure}[h]
\centering
\includegraphics[width=1\textwidth]{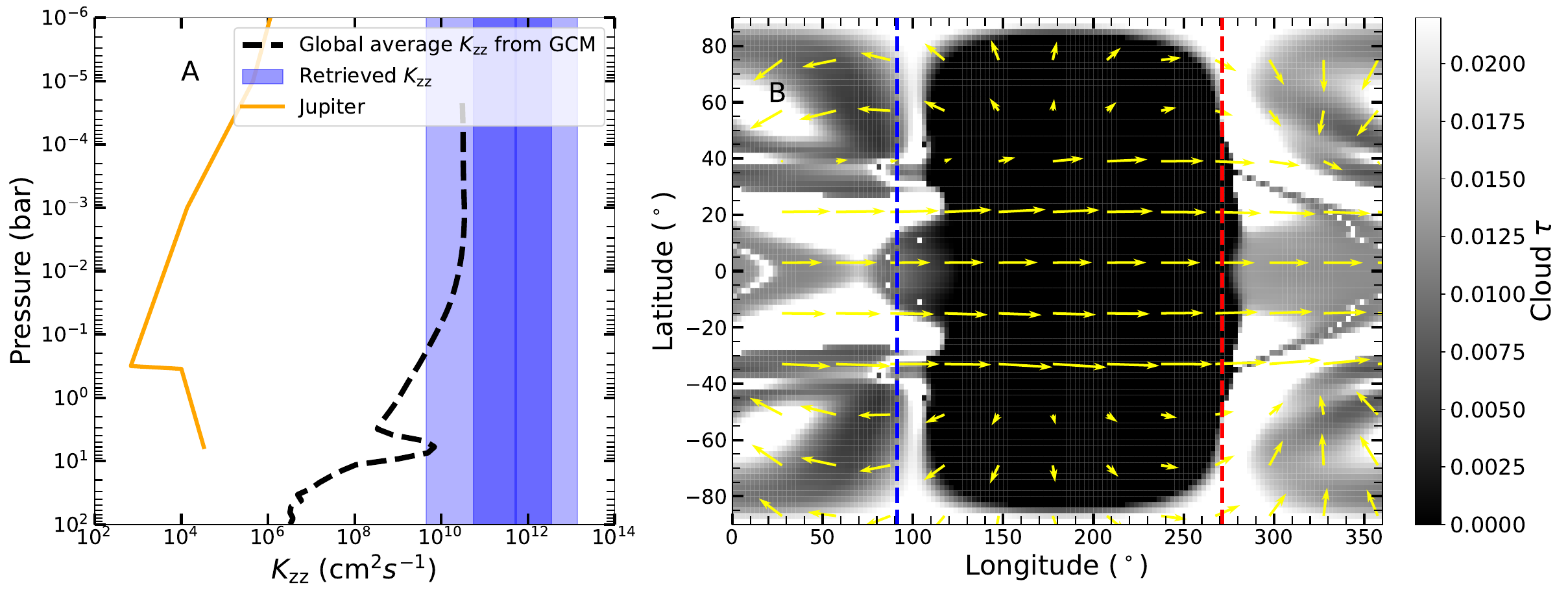}
\caption{{\bf Vertical mixing and cloud properties from GCM simulations of WASP-94A~b.} ({\bf A}) the global mean $K_{\rm zz}$ profile from the GCM (black dashed line) compared to constraints on $K_{\rm zz}$ from the limb asymmetry observations (blue shaded region). The median (solid blue line), $1\sigma$ (dark shading) and $2\sigma$ (light shading) constraints are shown. The measured $K_{\rm zz}$ profile for Jupiter \cite{moses05} is shown with the orange line. ({\bf B}) the cloud optical depth (grayscale) at $0.01$~mbar obtained by post-processing the GCM with \texttt{VIRGA} to model MgSiO$_3$, Fe, and MnS clouds. The red and blue dashed lines show the locations of the evening and morning limb, respectively. The morning limb appears cloudy while the evening limb is relatively clear. Wind vectors are over-plotted with yellow arrows.}\label{fig:Kzz}
\end{figure}

To investigate the cloud distribution at the morning and evening limbs, we use the cloud-free GCM outputs as inputs for the \texttt{VIRGA} cloud model \cite{batalha20,rooney21}. The $T(P)$ and $K_{\rm zz}$ profiles at each latitude-longitude pair of the GCM were used as inputs to the \texttt{VIRGA} model to calculate the cloud distribution. A map of the cloud optical depth at $\sim$0.01 mbar as a function of latitude and longitude is shown in Fig.~\ref{fig:Kzz}B. The colder $T(P)$ profile of the morning limb (Fig.~\ref{fig:GCM_timescales}B) allows formation of Fe, MnS, and MgSiO$_3$ clouds. Enhanced night-side cloud formation appears as increased opacity in Fig.~\ref{fig:Kzz}B. Condensates in the mid-latitudes of the night-side are transported by the equatorial jet to the morning limb, resulting in a cloudy morning limb. However, Fig.~\ref{fig:GCM_timescales}B shows that the evening limb is too hot for these condensates. As a result, the cloud droplets are evaporated before they reach the evening limb, resulting in lower cloud opacity at the limb (Fig.~\ref{fig:Kzz}B).

Fig.~\ref{fig:Kzz}A shows the globally averaged $K_{\rm zz}$ profile estimated using the root mean square (RMS) of the vertical velocity along isobars and scale heights of the atmosphere. The $K_{\rm zz}$ profile from our GCM simulation is consistent with the constraints from the retrievals from $1$~bar to $0.01$~mbar, differing by only 1.3$\sigma$. $K_{\rm zz}$ is generally unknown in the atmospheres of exoplanets \cite{Zahnle14,moses11,Mukherjee22a,mukherjee24}. $K_{\rm zz}$ calculations using the RMS vertical velocity along isobars or the upward diffusive flux matched to the averaged vertical flux from the dynamics can differ by orders of magnitude \cite{parmentier13}. Therefore the value of $K_{\rm zz}$ derived from our GCM could be an overestimation by several orders of magnitude. We use an alternative analytical prescription  \cite{parmentier13} to estimate the $K_{\rm zz}$ parameter. We find that to keep cloud droplets with radii between 0.01 to 1~{$\upmu$}m aloft near a pressure of 0.01\,mbar at a temperature of 1500$\pm$100\,K, $K_{\rm zz}$ should be within $10^{9}$ to ${10^{11}}$ cm$^2$~s$^{-1}$. This is consistent with both the cloud model and the GCM. Therefore, the vertical mixing at the morning limb is strong enough to sustain small condensate droplets at lower pressures, as probed by our observations of the morning limb.

\subsection{Self-consistent climate models}

To simulate the variation of the 1.4~{$\upmu$}m H$_2$O feature amplitude with $T_{\rm eq}$ of exoplanets (Fig.~\ref{figmhcto}B), we use the 1D radiative-convective equilibrium climate modeling suite in \texttt{PICASO} \cite{Mukherjee22}. This climate model self-consistently solves the thermal and chemical structure of exoplanet atmospheres assuming radiative-convective equilibrium and thermochemical equilibrium. We simulated 1D cloud-free atmospheric models across an equilibrium temperature ($T_{\rm eq}$) range of 500 to 2500~K in 100 K increments, assuming solar metallicity and C/O ratios. We computed the transmission spectrum from these models and calculated the predicted amplitude of the 1.4~{$\upmu$}m H$_2$O feature (Fig.~\ref{figmhcto}B, red line), following previous work \cite{fu17}. To simulate the muting of the H$_2$O feature amplitude due to gray clouds, we post-process these clear atmosphere models with clouds, assuming a wavelength-independent cloud optical depth profile 
\begin{equation}
    \tau_{\rm cld}= \kappa_{\rm cld}\dfrac{\delta{P}}{g}
\end{equation}
following previous work \cite{rustamkulov23}. Here, $\kappa_{\rm cld}$ is the cloud opacity, $\delta{P}$ is the pressure difference between atmospheric layers, and $g$ is the gravity of the planet. We assume a relatively high cloud opacity $\kappa_{\rm cld}=10^{-1}$ cm$^2$~g$^{-1}$ and generate the transmission spectra from these cloudy atmospheres. Following the same methodology as the clear atmosphere models, we then compute the H$_2$O feature amplitude from the cloudy atmosphere models (Fig.~\ref{figmhcto}B, gray line). To simulate very cloudy conditions as seen in the morning limb of WASP-94A~b, we assume a model transmission spectrum with the same slope as measured from the WASP-94A~b's morning limb spectrum (\S\ref{sec:slope}). The H$_2$O feature amplitude was then measured from these model spectra (Fig.~\ref{figmhcto}B, blue line). We compute the H$_2$O feature amplitude of an atmosphere with limb-asymmetry by linearly combining the transmission spectra of the clear (red line in Fig.~\ref{figmhcto}B) and the very cloudy atmosphere models (blue line in Fig.~\ref{figmhcto}B) with a weight of 0.5 for each. This weighting assumes one limb of the model planet is fully clear and the other is cloudy. The H$_2$O feature amplitude of the resulting transmission spectra was then measured (gray dashed line in Fig.~\ref{figmhcto}B). This H$_2$O feature amplitude is expected from the spherical transmission spectrum of a planet with limb asymmetry similar to WASP-94A~b.

\subsection{Slope of Morning Limb Transmission Spectrum}\label{sec:slope}

\begin{figure}[h]
\centering
\includegraphics[width=1\textwidth]{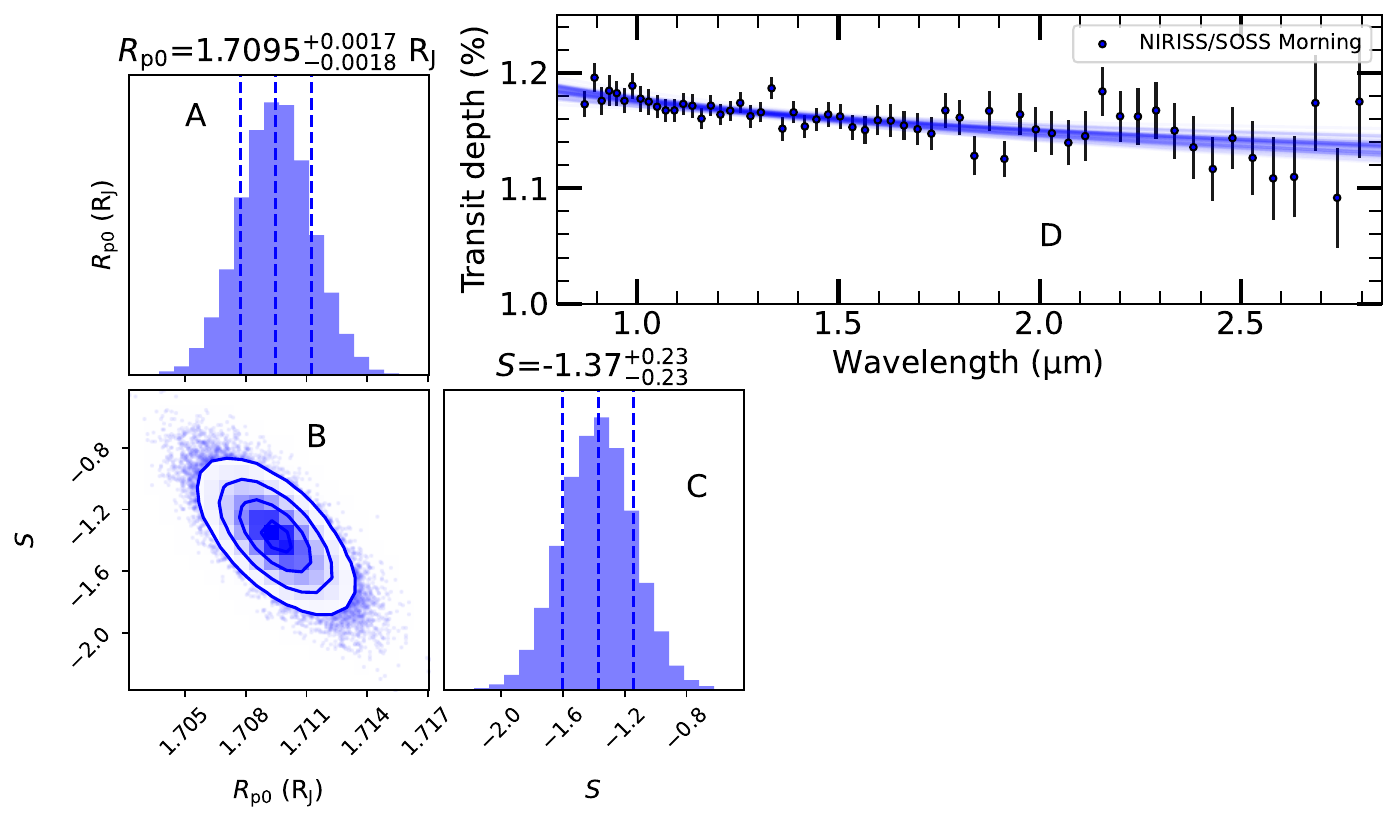}
\caption{{\bf Slope of morning limb spectrum.} Constraints on ({\bf A}) planet radius, and ({\bf C}) slope $S$ of the morning limb spectrum. The left, middle, and right dashed vertical lines indicate the $-1\sigma$, median, and $1\sigma$ values in both panels. ({\bf B}) shows their joint posterior probability distribution. The contours represent $1\sigma$, $2\sigma$, and $3\sigma$ envelopes of the joint posterior probability distribution. ({\bf D}) the morning limb spectrum (blue points) and 500 samples of the fitted parametric model (blue lines). All error bars show $1\sigma$ uncertainties. }\label{fig:slope}
\end{figure}

We use a parametric model to measure the scattering slope of the morning limb transmission spectrum. Following previous work \cite{pont13,ohno20}, we define the wavelength-dependant planet radius,
\begin{equation}
    R_p (\lambda)= R_{p_0} + H{\times}S{\times}\ln\left(\dfrac{\lambda}{1 {\mu}m}\right)
\end{equation}
where $H$ is the atmospheric pressure scale height, $R_{p_0}$ is the baseline planet radius, and $S$ is the slope of the transmission spectrum. We fitted this parametrization to the morning limb spectrum of WASP-94A~b, assuming $H$=1460~km, which is the pressure scale height at WASP-94A~b's equilibrium temperature. Fig.~\ref{fig:slope} shows the posterior probability distributions, which constrain $S$=-1.37$\pm$0.23. This is inconsistent with the slope expected for Rayleigh scattering, which is $S$=-4 \cite{pont13}. Instead, the slope is close to the value expected from  grains in the settling regime with a constant mass fraction distribution of sizes, which is $S=-1$ \cite{pont13}. 

\subsection{Do hazes fit better than clouds?}



We interpret the observations of WASP-94A~b as due to condensate clouds that form on the colder planetary night-side and then cycle between a cooler and cloudier morning limb and a clearer and hotter evening limb. However, we do not detect any specific cloud absorption feature in our observations. Therefore, we also test whether clouds or hazes provide a better fit to the morning limb spectrum of WASP-94A~b. We use the flexible cloud layer functionality of the \texttt{VIRGA} cloud model to simulate the effect of various aerosol species on the morning limb transmission spectrum from their refractive indices, following previous methods \cite{schlawin2024}. We populate the atmosphere with arbitrary aerosol species using their real and imaginary refractive indices. We assume a Gaussian distribution for the logarithm of aerosol droplet radii with two parameters: the mean droplet size ($r_{\rm mean}$), and the dimensionless standard deviation of the droplet size distribution ($\sigma_{\rm r}$). We fix  $\sigma_{\rm r}$ to 0.01 to determine whether the resulting predicted morning limb spectrum has potentially observable features caused by aerosols. If the width of the aerosol particle size distribution is allowed to be higher, the resulting broad particle size distribution tends to erase aerosol-driven features making it difficult to probe such features in the morning limb spectrum. 

The aerosol model we use for silicate clouds has three tunable inputs: the base pressure of the cloud deck ($P_{\rm base}$), a column density of all aerosol particles ($n$), and a scaling parameter for the layer-by-layer aerosol optical depth ($f_{\rm layer}$). The optical constants of the assumed cloud species (MgSiO$_3$) are used to compute the optical depth ($\tau$), asymmetry parameter ($g_0$), and the single scattering albedo ($w_0$) at the base pressure $P_{\rm base}$. The layer-by-layer cloud optical depth is set to 0 for pressures greater than $P_{\rm base}$. For lower pressures, it is scaled with $e^{-f_{\rm layer}z/H}$ for layers with pressures smaller than $P_{\rm base}$, where $z/H$ is the ratio of atmospheric altitude ($z$) and pressure scale height ($H$) that increases with decreasing pressure. This scaling factor ensures that the cloud optical depth decreases exponentially with increasing altitude above the aerosol base, as found in microphysical cloud simulations \cite{ackerman2001cloud,gao18,helling20}. Our aerosol model for clouds therefore has four tunable parameters: $n$, $f_{\rm layer}$, $P_{\rm base}$, and $r_{\rm mean}$ for a given aerosol species with input refractive indices. In addition to these aerosol parameters, we assume that the morning limb atmosphere has an isothermal thermal structure with temperature $T_{\rm iso}$. We allow a uniform prior on $T_{\rm iso}$ between 100 to 3000~K. We also determine an H$_2$O abundance while fitting the aerosol model to the morning limb spectrum. We allow the H$_2$O abundance $X_{\rm H{_2}O}$ to vary freely between $-1{\ge}\log_{10}(X_{\rm H{_2}O}){\ge}{-10}$. The H$_2$ and He abundances, $X_{\rm H_2}$ and $X_{\rm He}$, are then calculated using $X_{\rm H_2}+X_{\rm He}=1-X_{\rm H_2O}$ and $X_{\rm H_2}/X_{\rm He}=5.1349$ \cite{asplund09}. Therefore, the aerosol model for the morning limb has 6 free parameters: $n$, $f_{\rm layer}$, $P_{\rm base}$, $r_{\rm mean}$, $T_{\rm iso}$, and $\log_{10}(X_{\rm H_2O})$.

To simulate haze-like aerosols, we use a different optical depth profile than clouds. The dependence of haze optical depth on pressure is dictated by the variation of haze particle concentration and haze particle sizes with pressure. Because hazes are expected to be produced photochemically at low pressures, their concentration can decrease with increasing pressure \cite{arfaux24,ohno20}. Haze particle sizes are expected to increase with increasing pressure due to collisional growth of haze particles \cite{Arfaux22,arfaux24,ohno20}. The layer-by-layer optical depth profile of hazes could therefore increase, decrease, or remain constant with increasing pressure. We assume that the haze opacity $\kappa$ follows a power-law like pressure dependence $\kappa\propto{(P/P_{\rm top})^{-\beta}}$ \cite{ohno20}, where $P$ is atmospheric pressure, $P_{\rm top}$ is the pressure of the top of the haze layer, and $\beta$ is the dimensionless power-law parameter. The vertical haze optical depth ($\tau$) profile is $\tau\propto\kappa{dP}$,
which implies $\tau\propto{(P/P_{\rm top})^{1-\beta}}$ in this parametrization. If $0\le\beta<1$, the haze optical depth will increase with increasing pressure, similar to our cloud model. However, if $1\le\beta$, the haze optical depth remains constant ($\beta=1$) or decreases with increasing pressure. The haze optical depth is assumed to be 0 for $P<P_{\rm top}$ in this model. The optical constants of the assumed haze species are used to compute the haze optical depth ($\tau$), asymmetry parameter ($g_0$), and the single scattering albedo ($w_0$) at the top pressure $P_{\rm top}$, which are then scaled with the assumed pressure-dependence described above for $P>P_{\rm top}$. Therefore, our haze aerosol model also has 6 free parameters: $n$, $P_{\rm top}$, $\beta$, $r_{\rm mean}$, $T_{\rm iso}$, and $\log_{10}$($X_{\rm H_2O}$).

We fit the aerosol model to the morning limb transmission spectrum assuming 5 possible types of aerosols with different refractive indices. Four are used in the haze model: Titan tholins \cite{khare84}, soots \cite{lavvas17}, Saturn-like phosphorus hazes \cite{noy81,sromovsky20,fletcher23} (haze model), 400 K tholins \cite{He24} (haze model), and one in the cloud model: enstatite (MgSiO$_3$) \cite{scott96}. We use the same Bayesian sampling method as our retrievals to estimate the free parameters of our aerosol model for each type of aerosol. 

\begin{figure}[h]
\centering
\includegraphics[width=1\textwidth]{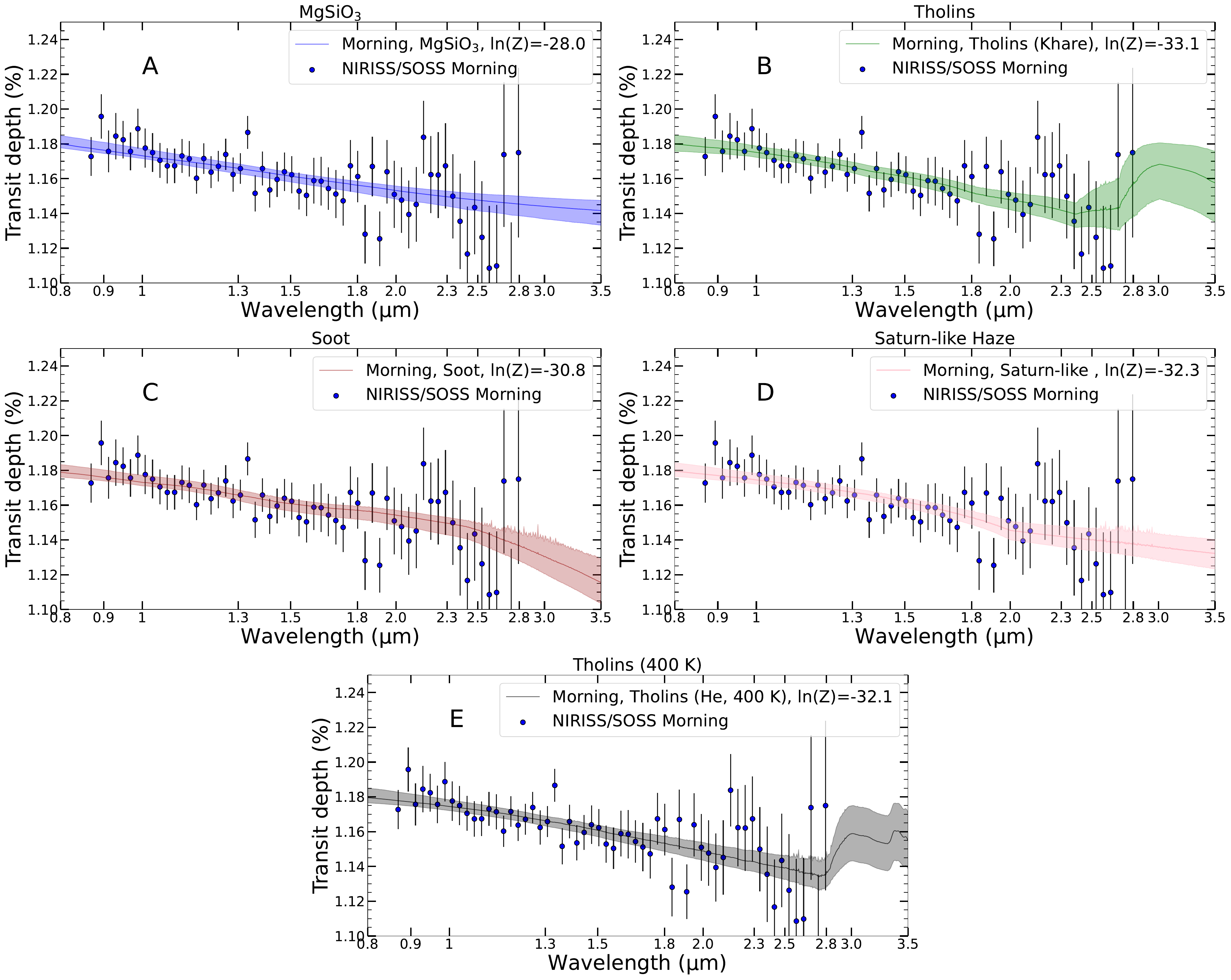}
\caption{{\bf The morning limb spectrum with best-fitting models of different aerosol species.} ({\bf A}) the median (line) and 1$\sigma$ envelope (shading) of the best-fitting MgSiO$_3$ cloud model compared to the morning limb spectrum (blue data points with $1\sigma$ error bars). The other panels show the same for haze models of ({\bf B}) Titan-like tholins, ({\bf C}) soot, ({\bf D}) Saturn-like hazes, and ({\bf E}) higher temperature (400 K) tholins. The model parameters are listed in Table~\ref{tab:hazecompare}. All error bars show $1\sigma$ uncertainties.}\label{fig:haze}
\end{figure}

Fig.~\ref{fig:haze} shows the best-fitting models fits for the morning limb spectrum from each of these haze/cloud species. The model with enstatite clouds has the highest Bayesian evidence $\ln(Z)=-28.0$. The highest evidence haze model is that of soot, which has $\ln(Z)=-30.8$. This corresponds to the soot model having a probability 0.06 times that of the enstatite model. The other haze species are more strongly disfavored, having probabilities 0.016 times that of the enstatite cloud model. The Bayesian evidence for each aerosol species along with constraints on the model parameters for them are presented in Table \ref{tab:hazecompare}.  We conclude there is moderate statistical preference for clouds over hazes as explanations for the morning limb of WASP-94A~b

\subsection{Stellar Activity}

Unocculted star spots can mimic the effect of aerosols on an exoplanet transmission spectrum by producing similar slopes at blue wavelengths \cite{narrett24,rackham19}. However, our analysis localizes the aerosols in the planet's atmosphere because they only appear prominently in one limb. Nevertheless star-spot or faculae crossing events during the transit can mimic planetary asymmetries in the transit light curve. 
We consider each of these scenarios in this section and quantify how they can affect our results.


\subsubsection{Occulted star spots or faculae}

\begin{figure}
\centering
\includegraphics[width=1.0\textwidth]{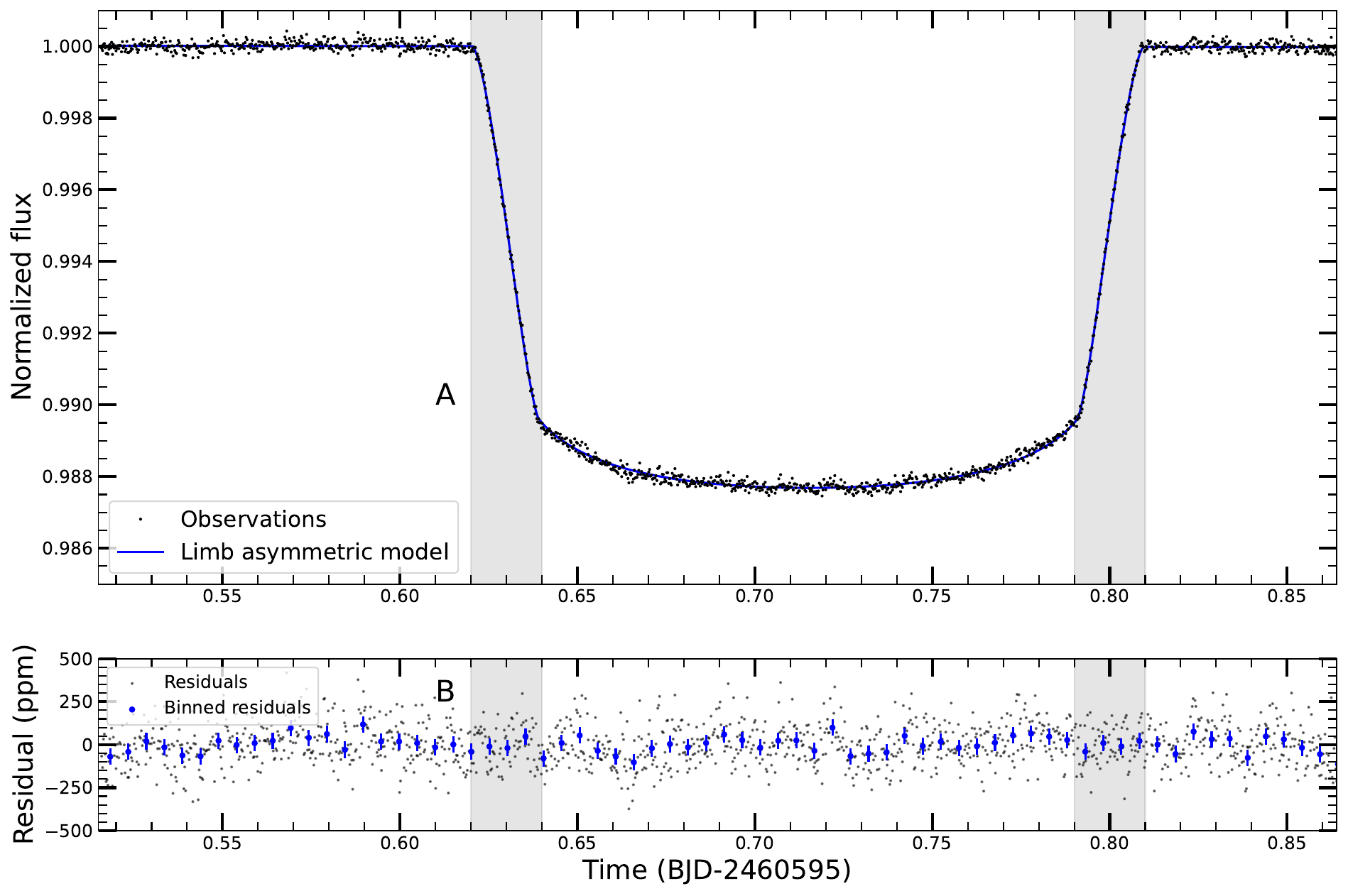}
\caption{{\bf Observed and modeled white light curve}. ({\bf A}) shows the observed white light curve of WASP-94A~b (black points) from the FIREFLy reduction and the best-fitting asymmetric limb model from Catwoman (blue line). The gray shaded regions mark the times of ingress and egress. ({\bf B}) shows the binned (blue points) and unbinned residuals (black points) between the model and the data, which have a MAD of 76~ppm. The binned residuals were calculated using bins containing 20 unbinned residuals. Error bars show 1$\sigma$ uncertainties.}\label{wlcwhole}
\end{figure}

\begin{figure}[h]
\centering
\includegraphics[width=1.0\textwidth]{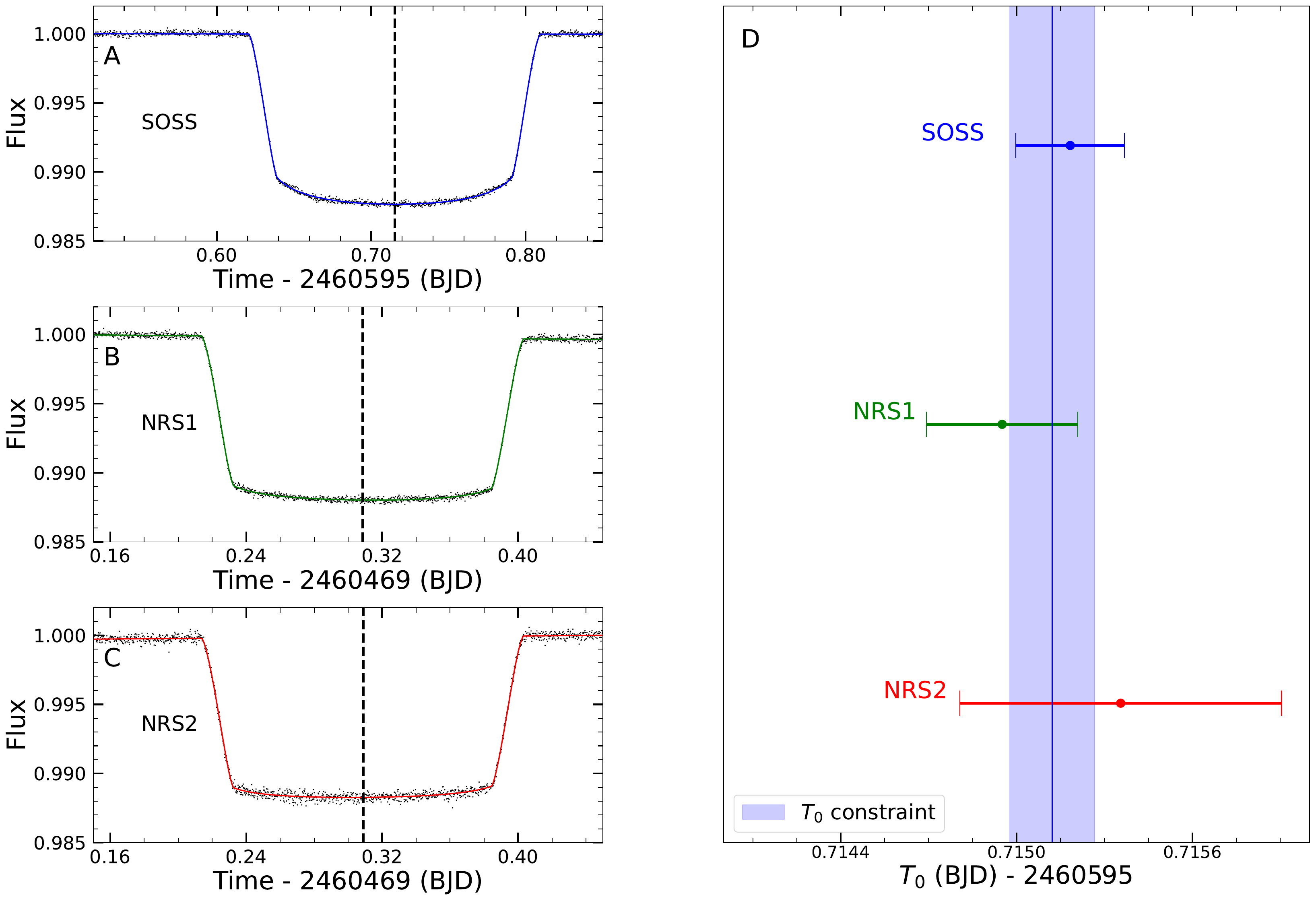}
\caption{{\bf White light curves of WASP-94A~b observed with NIRISS/SOSS and NIRSpec/G395H and measured mid-transit times}. ({\bf A}) the observed white light curve of WASP-94A~b (black points) with NIRISS/SOSS (same as Fig.~\ref{wlcwhole}) along with the best-fitting Catwoman transit model (blue line). The measured mid-transit time is shown (dashed black line). ({\bf B})-({\bf C}) Same as panel A, but for the transit observed with NIRSpec/G395H for the NRS1 and NRS2 detectors. ({\bf D}) Comparison of the measured mid-transit times. The measured mid-transit times from the NIRSpec observations have been projected forward by 32 transits, corresponding to the gap between the NIRISS and NIRSpec observations. The solid blue line and the shaded region show the weighted mid-transit time and its uncertainty calculated from these three measurements. All error bars shown here represent 1$\sigma$ uncertainty.}\label{wlc_all3}
\end{figure}

\begin{figure}[h]
\centering
\includegraphics[width=1\textwidth]{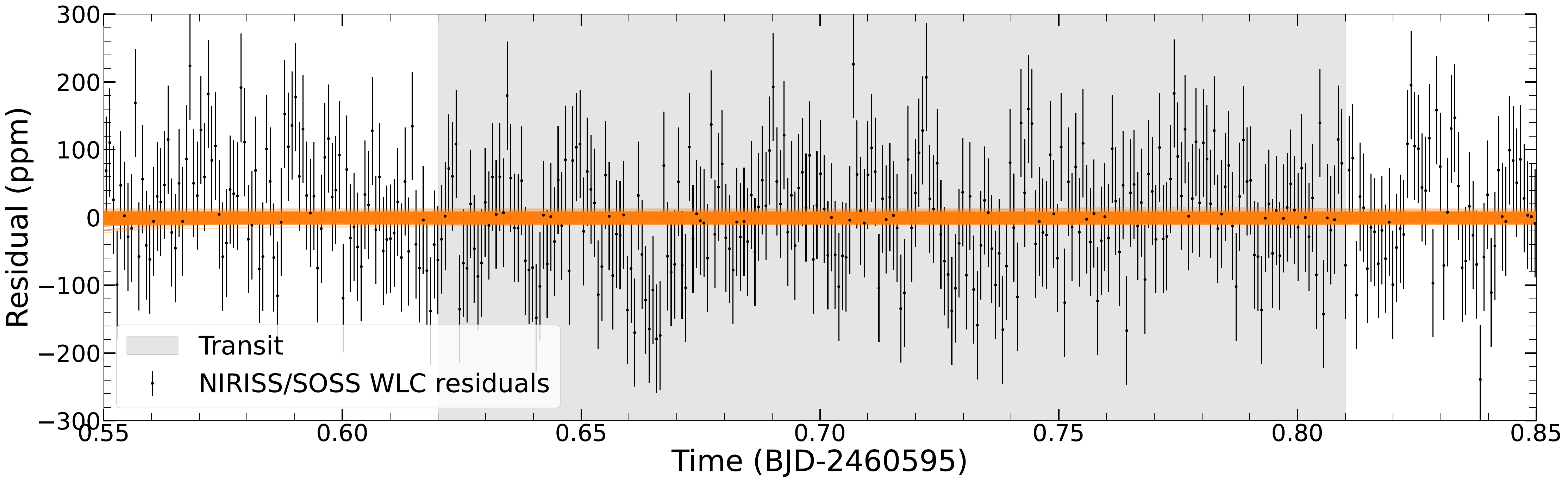}
\caption{{\bf Search for star spot-crossings.} The residuals between the white light curve observed with NIRISS/SOSS and the catwoman transit model (black data points with $1\sigma$ uncertainties). The orange lines are 1000 draws from a Bayesian sampling analysis which fits Gaussian models to the residuals. We do not identify any occulted spots or faculae during the transit (gray shading). }\label{fig:occulted}
\end{figure}

We analyze the residuals from the white light curve (Fig.~\ref{wlcwhole}) to search for spot or faculae crossings during the transit. We fitted the residuals with Gaussian functions that vary with time, to mimic the effects of spot or faculae crossings on the transit light curve. The Gaussian function has three free parameters: amplitude, mean time, and standard deviation. We use the \texttt{emcee} Bayesian sampler \cite{emcee13} to perform model fitting. We allow the amplitude to vary between -500 to 500~ppm and the mean time to vary within the duration of the transit. The standard deviation is allowed to vary from 8~s to the whole transit duration. Fig.~\ref{fig:occulted} shows the residuals from the white light curve along with 1000 models randomly drawn from the Bayesian sampler. None of the models show any Gaussian peaks or troughs, so we do not detect any spot or faculae crossing events during the observed transit of WASP-94A~b. They are therefore unlikely to affect our conclusions. 

\subsubsection{Unocculted Spots}

\begin{figure}[h]
\centering
\includegraphics[width=0.75\textwidth]{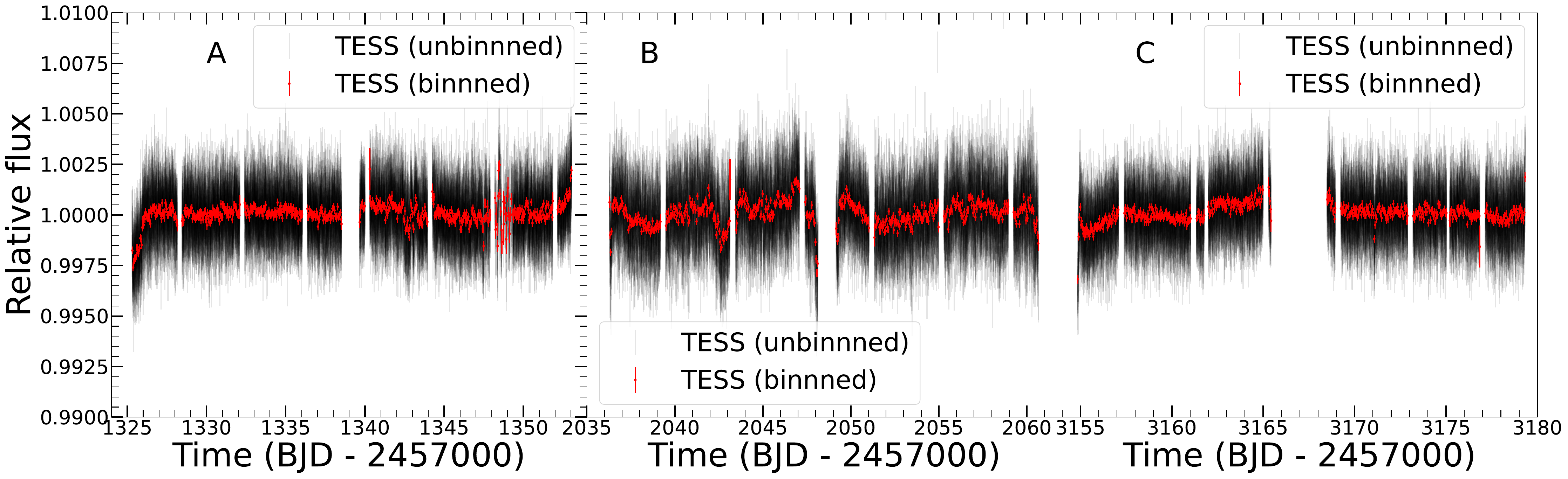}
\includegraphics[width=0.75\textwidth]{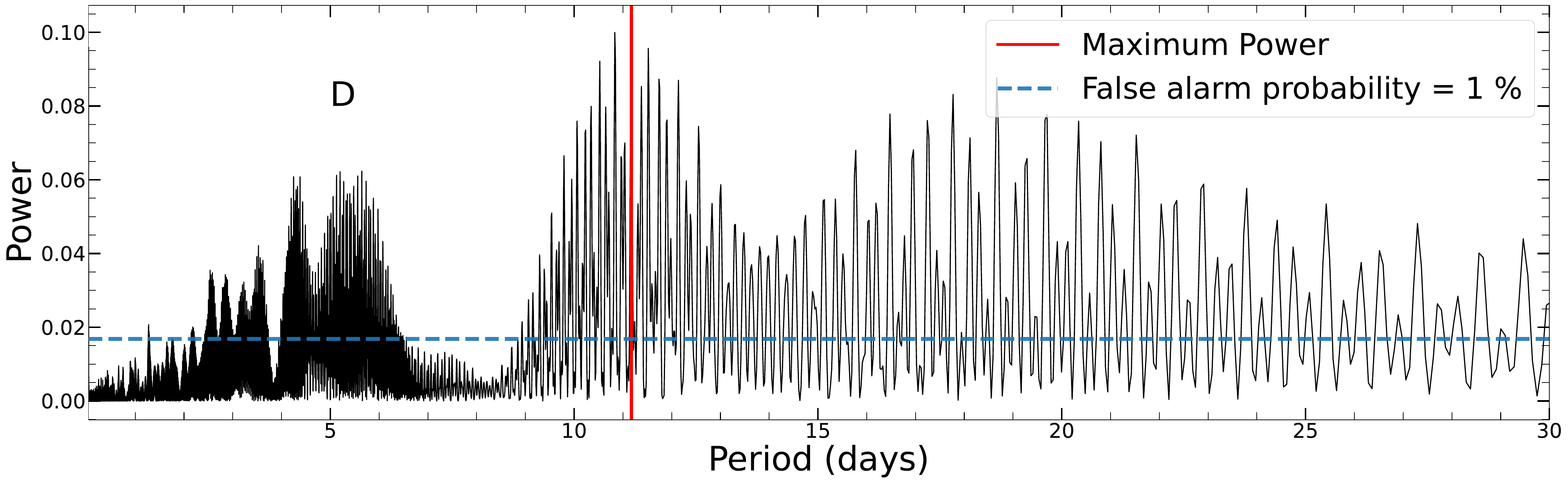}
\includegraphics[width=0.75\textwidth]{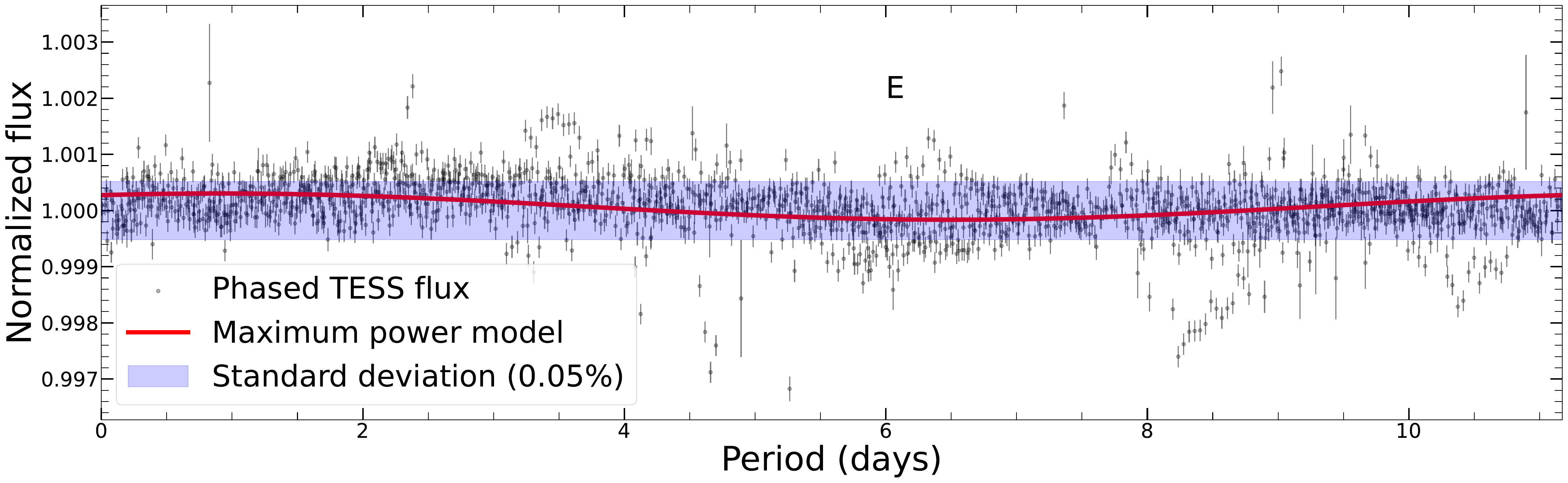}
\includegraphics[width=0.75\textwidth]{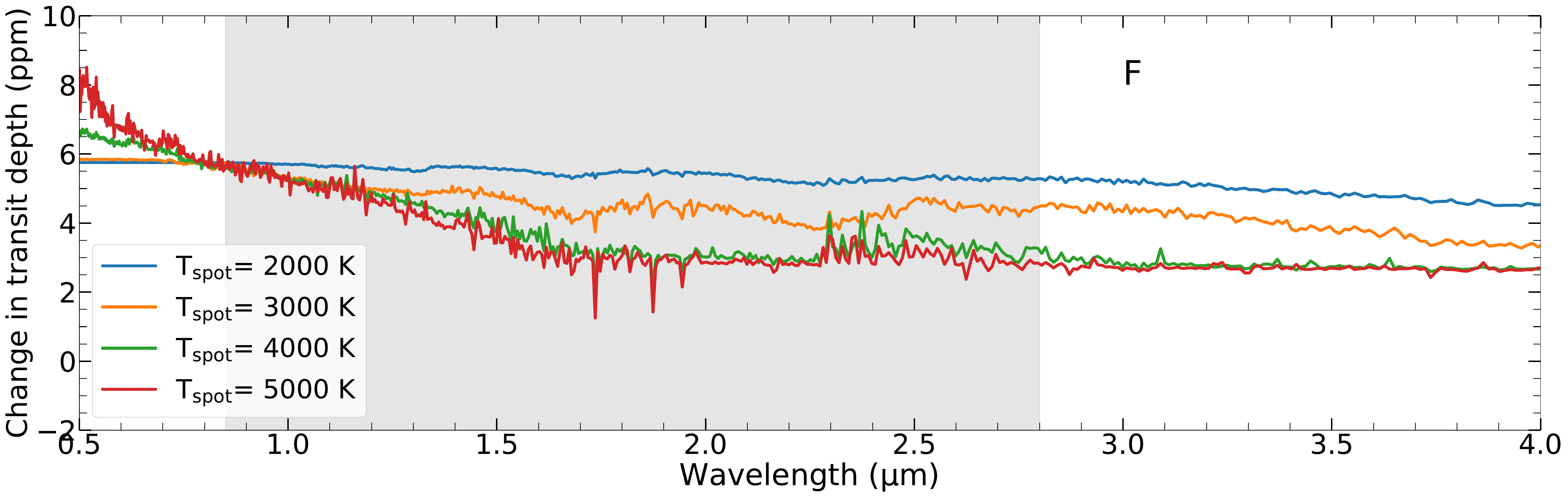}

\caption{{\bf Effects of unocculted star spots on the transmission spectrum}. ({\bf A})-({\bf C}) observed TESS light curve of WASP-94A at three separate epochs (black points) and in 1-hour bins (red points). Gaps are transits of WASP-94A~b that have been masked. ({\bf D}) Lomb-Scargle periodogram of the binned light curve. The period with the most power (11.173 days) is indicated by the red vertical line and the power corresponding to false alarm probability of 1\% is indicated with the dashed blue horizontal line. ({\bf E}) light curve phased with this period, overlain with the best-fitting sinusoid (red line). Blue shading indicates the $\pm$1 standard deviation of the binned flux of WASP-94A. We find stellar flux variability $\le$0.05\% in the TESS bandpass. ({\bf F}) potential change in transit depth due to unocculted spots with four spot temperatures across the wavelengths covered by our JWST observations (gray shading). The effect is $\le$6~ppm level.  }\label{fig:unocculted}
\end{figure}

Unocculted spots on the star can also contaminate the transmission spectra of exoplanets by altering the baseline stellar flux level in a wavelength dependant manner. We use TESS photometry of WASP-94A across three different epochs in 2018, 2020, and 2023 \cite{tessmukherjee2025-mastdoi} to quantify the extent to which variable unocculted spots might affect our results. Fig.~\ref{fig:unocculted}A-C  show the normalized flux of WASP-94A observed with TESS during these three epochs. We mask the transits of WASP-94A~b from the TESS light curve to leave only the host star brightness over time. Fig.~\ref{fig:unocculted}D shows the Lomb-Scargle periodogram of the binned TESS light curve. The periodogram has several weak but statistically significant peaks within the 10-22 days timescale. The strongest power appears at 11.1736 days. Fig.~\ref{fig:unocculted}E shows the phase-folded the binned light curve of WASP-94A using this time period. Fig.~\ref{fig:unocculted}E also shows the standard deviation of the entire binned TESS light curve. We find a variability of about $\sim$0.05\% with this period.

This period is much longer than the transit duration of WASP-94A~b, so any unocculted variable components can be assumed to be static during the transit. We estimate the change in transit depth caused by unocculted spots with various spot temperatures [\cite{sing11}; their Equation 4], matched to the 0.05\% level variability. Fig.~\ref{fig:unocculted}F shows the resulting change in transit depth as a function of wavelengths for various spot temperatures between 2000 and 5000 K; the nominal stellar effective temperature is assumed to be 6153 K \cite{ahrer24}. We find that if unocculted star spots are responsible for the $\sim$11~day timescale 0.05\% level variability in WASP-94A, then their effect on the measured transit depth is $\le$6~ppm at the wavelengths covered by our JWST observations. That is three orders of magnitude smaller than the transit depth we observe, and two orders of magnitude smaller than the limb asymmetry we measured. Therefore, unocculted star spots are unlikely to affect our conclusions.


\subsubsection{Atmospheric Mass Loss}

\begin{figure}[h]
\centering
\includegraphics[width=1.0\textwidth]{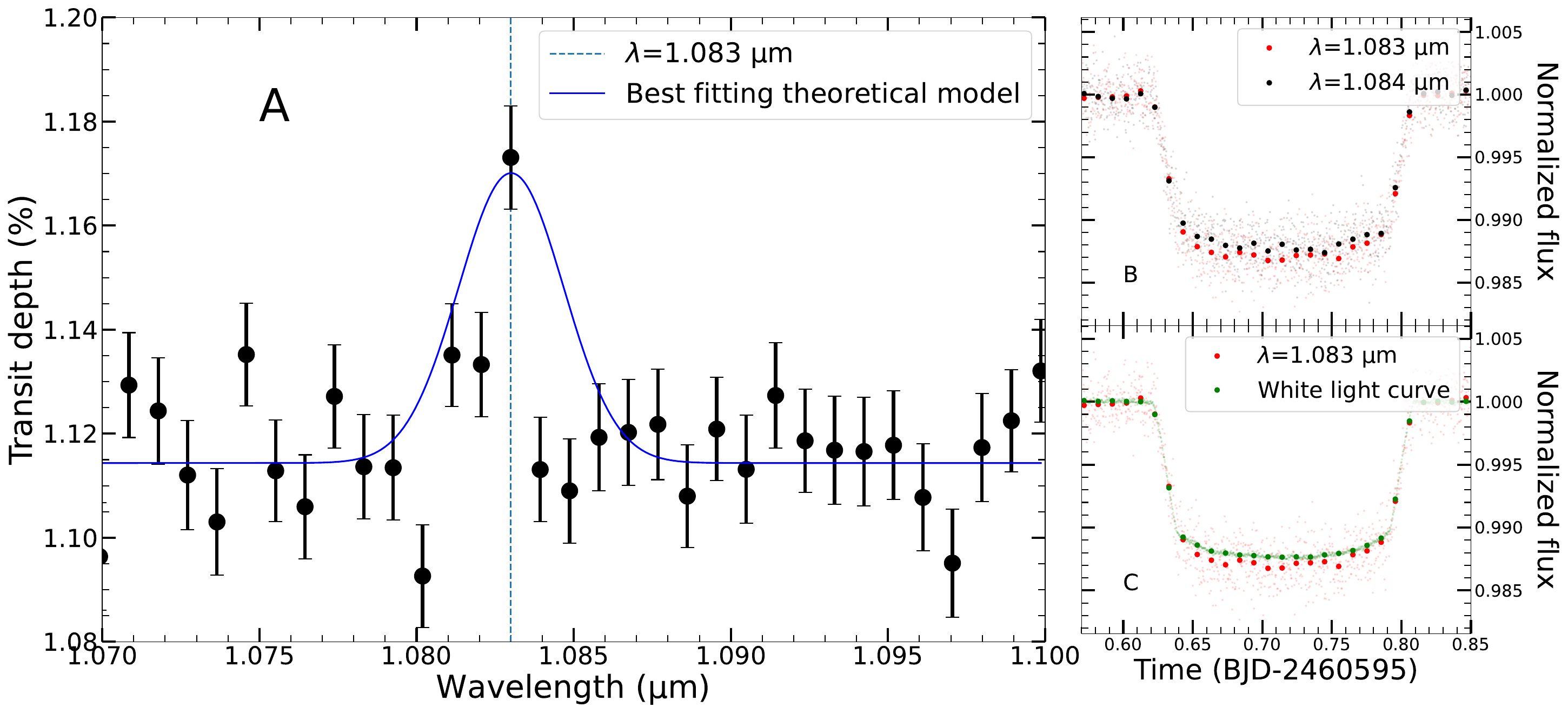}
\caption{{\bf Detection of Helium outflow from WASP-94A~b.} ({\bf A}) the excess transit depth of WASP-94A~b due to metastable He absorption at 1.083~{$\upmu$}m (black points). The best-fitting atmospheric mass-loss model for WASP-94A~b atmosphere is shown (blue solid line). The dashed blue vertical line indicates the wavelength of expected He absorption. ({\bf B}) comparison of the light curve within the He absorption band (1.083~{$\upmu$}m; red points)  with the light curve just outside it (1.084~{$\upmu$}m; black points), showing excess transit depth at a light curve level. Small points show the unbinned data, while large points show data binned into bins of 40 points each.  ({\bf C}) comparison of the light curve within the He absorption band (red points) with the white light curve of WASP-94A~b (green points; same as Fig.~\ref{wlcwhole}). Small and large points represent the same binning as panel B. Error bars represent $1\sigma$ uncertainty.}\label{helium}
\end{figure}

The unbinned 1D spherical transmission spectrum exhibits excess metastable 1.083~{$\upmu$}m helium absorption, though the line profile remains unresolved. Fig.~\ref{helium}A illustrates this absorption feature in the transmission spectrum and compares the light curve of the pixel containing the helium feature to a neighboring pixel and the white light curve in Fig.~\ref{helium}B-C. We constrain the helium transit depth to 1.17$\pm$0.01\% and compare it to the white-light depth of 1.12$\pm$0.02\%.

Fully resolved metastable helium absorption can be used to constrain atmospheric mass-loss parameters; however, the unresolved nature of the absorption profile from our observations complicates this process. To estimate the mass-loss rate of WASP-94A~b, we apply the one-dimensional, isothermal Parker-wind model \cite{parker58} using the p-winds code\cite{pwinds}. p-winds follows a mass-loss formulation from previous studies \cite{Oklopcic2018} with some modifications \cite{Lampon2021,Vissapragada2022b}. The model leverages planetary and stellar parameters to fit helium absorption features in the transmission spectrum and derive mass-loss properties, including the mass outflow rate and temperature. For this analysis, we adopt the high-energy spectral energy distribution (SED) of HD 149026 from the MUSCLES database as a proxy for the high-energy SED of WASP-94A \cite{musclesI,musclesII,musclesIII,musclesIV,musclesV}. However, given the unresolved absorption line profile in our observations, the mass-loss rate and outflow temperature remain degenerate with each other when fitting an isothermal Parker-wind model to the feature. Fig.~\ref{helium}A displays an example best-fit given the constraints placed on the mass-loss parameters discussed above. However, the degeneracy between the mass-loss rate and outflow temperature results in multiple best-fit solutions. The low resolving power of this observation, which does not fully resolve the helium triplet feature, leads to a broader line profile to account for the wide helium triplet absorption window \cite{fu22}. 

With an estimate of the mass-loss rate, the outflow temperature can be further constrained \cite{McCreery2025}. In the energy-limited mass-loss framework \cite{Caldiroli2022}, the mass-loss efficiency, $\varepsilon$, parameterizes the fraction of incident X-ray and ultraviolet (XUV) irradiation that powers atmospheric escape. The mass-loss rate ($\dot{M}$) can then be expressed in terms of the incident XUV flux ($F_{\rm XUV}$), the planetary density ($\rho_{\rm p}$), and the gravitational constant ($G$) as:

\begin{equation}
    \dot{M} = \frac{3 \varepsilon}{4G}\frac{F_{\rm XUV}}{\rho_{\rm p}}
    \label{eqn:elml}
\end{equation}

For a He/H ratio of 10/90, the efficiency is estimated to be $0.34 \pm 0.13$ prior to the predicted breakdown of energy-limited mass loss formulations ($F_{\rm XUV}/\rho_{\rm p} \sim 10^4$~cm$^3$~s$^{-3}$) and $0.10 \pm 0.06$ thereafter. Using Equation \ref{eqn:elml}, we constrain WASP-94A~b’s mass-loss rate to $7.3_{-2.6}^{+2.7} \times 10^{10}$~g~s$^{-1}$. We estimate an outflow temperature between 4000-6000 K by applying the p-winds model.


\begin{table}[]
\caption{{\bf Constraints on system and planetary parameters from different reduction pipelines}. In each case, the white light curve for WASP-94A~b was fitted with the Catwoman transit model. The first column lists the parameter followed by its units in the second column. The other three columns list the results from each data reduction pipeline. The values which are denoted as ``calculated" were not fitted in that reduction pipeline but were derived from the median parameter values of related parameters. This includes the impact parameter and inclination, with each computed using the other's median value. It also includes the limb darkening parameters $u_1$ and $u_2$ computed from $u_+$ and $u_-$, and vice versa.
}\label{tab:wlcfits}
\begin{tabular}{|l|l|l|l|l|}
\hline
\textbf{Parameter}      & \textbf{Unit}     & \textbf{FIREFLy}          & \textbf{Eureka!}          & \textbf{Fu} \\ \hline
($R_{\rm M}$/$R_{\rm S}$)$^2$   & \%        & 1.176$\pm$0.035       & $1.170\pm0.036$           &  $1.153\pm0.040$\\ \hline
($R_{\rm E}$/R$_{\rm S}$)$^2$   & \%        & 1.078$\pm$0.035       & $1.092\pm0.036$           &  $1.096\pm0.039$\\ \hline
Period                  & days              & 3.9502 (fixed)            & 3.950190                  &3.950190     \\ \hline
Transit Time ($T_0$)-2460595    & BJD   & 0.71518$\pm$0.00018 & $0.71511\pm0.00019$ & 0.71506$\pm$0.00021\\ \hline
$a/R_{\rm S}$                 & -                 & 7.312$\pm$0.019          & $7.313\pm0.016$           & 7.320$\pm$0.019          \\ \hline
Impact Parameter $b$    & -                 & 0.150$\pm$0.018           & 0.139 (calculated)                        &     0.141 (calculated)        \\ \hline
inclination $i$         & $^\circ$          & 88.82 (calculated)                         & $88.91\pm0.13$            & 88.89$\pm$0.17       \\ \hline
limb darkening $u_+$    & -                 & 0.3473$\pm$0.0071         & 0.3839 (calculated)                        &     0.351 (calculated)        \\ \hline
limb darkening $u_-$    & -                 & -0.004$\pm$0.015          & -0.1243 (calculated)                        &    -0.019 (calculated)         \\ \hline
limb darkening $u_1$    & -                 & 0.1716 (calculated)                        & 0.1298                    & 0.1660$\pm$0.0086             \\ \hline
limb darkening $u_2$    & -                 & 0.1756 (calculated)                         & 0.2541                    & 0.185$\pm$0.016          \\ \hline

Spin-orbit angle $\phi$ & $^{\circ}$             & 90 (fixed)      & 90 (fixed)        &   90 (fixed)          \\ \hline
\end{tabular}

\end{table}

\begin{table}[t]
\caption{{\bf Constraints on $T_0$ from fitting the observed white light curves of WASP-94A~b.} The $T_0$ measured from transits observed with NIRISS/SOSS and NIRSpec/G395H (NRS1 \& NRS2) are shown. The Catwoman transit model was used to constrain the $T_0$ from each light curve. The first column lists the instrument used for each transit observation. The second column shows the measured $T_0$ and the third column shows the $T_0$ in the epoch of the NIRISS/SOSS observations. The fourth and fifth columns show the uncertainty on $T_0$ in days and seconds, respectively. }

\label{tab:wlc_combineall3}
\centering
\small
\setlength{\tabcolsep}{5pt} 
\begin{tabular}{|l|l|l|l|l|}
\hline
\makecell{} &
\makecell{$T_0$\\ {(BJD)}} &
\makecell{$T_0$\\ (NIRISS/SOSS epoch)\\ {(BJD)}} &
\makecell{$T_0$ uncertainty\\ {(days)}} &
\makecell{$T_0$ uncertainty\\ {(s)}} \\
\hline
NIRISS/SOSS & 2460595.71518 & 2460595.71518 & 0.00018 & 15.5 \\
\hline
G395H/NRS1  & 2460469.30850 & 2460595.71495 & 0.00025 & 21.6 \\
\hline
G395H/NRS2  & 2460469.30890 & 2460595.71535 & 0.00054 & 46.6 \\
\hline
Combined    & --            & 2460595.71512 & 0.00014 & 12.1 \\
\hline
\end{tabular}

\end{table}

\begin{table}[]
\caption{{\bf Constraints from atmospheric retrieval on the morning and evening transmission spectra.} Fitted parameters in the \texttt{PICASO} atmospheric retrieval on the morning and evening limb transmission spectra are listed in the first column. The second column lists their units. The third column lists the applied prior to each parameter, where $\mathcal{U}$ denotes uniform prior probability distribution and the following numbers show the range of allowed values for each parameter in the retrieval. The median of the posterior probability distribution for each parameter is listed in the fourth column. The $1\sigma$ uncertainty on each parameter value is listed in the fifth column.}
\label{tab:retrieval}
\begin{tabular}{|l|l|l|l|l|}
\hline
\textbf{Parameter}     & \textbf{Units} & \textbf{Prior}           & \textbf{Median} & \textbf{$\pm1\sigma$} \\ \hline
$T_{\rm eq}$           & K              & $\mathcal{U}(800,2100)$  & 1490            & 74                    \\ \hline
${\delta}T_{\rm eq}$   & K              & $\mathcal{U}(0,-600)$    & -449            & 83                    \\ \hline
$\log\left(\dfrac{\kappa_{\rm IR}}{{\rm cm}^2{\rm g}^{-1}}\right)$ & -              & $\mathcal{U}(-2.0,+0.5)$ & 0.22            & 0.19                  \\ \hline
$\log\left(\dfrac{\gamma}{{\rm cm}^2{\rm g}^{-1}}\right)$          & -              & $\mathcal{U}(-2.0,+0.5)$ & -1.32           & 0.52                  \\ \hline
$\alpha$               & -              & $\mathcal{U}(0,1)$       & 0.5            & 0.32                  \\ \hline
$T_{\rm int}$          & K              & $\mathcal{U}(30,630)$    & 334             & 180                   \\ \hline
\mbox{[M/H]}               & -  & $\mathcal{U}(-1.0,+3.0)$ & +0.46           & 0.36                  \\ \hline
C/O                    & -  & $\mathcal{U}(0.01,2.0)$  & 0.75             & 0.46                  \\ \hline
$\log(f_{\rm sed})$     & -              & $\mathcal{U}(-4,+1.0)$   & -3.03            & 0.62                   \\ \hline
$\log\left(\dfrac{K_{\rm zz}}{{\rm cm}^2{\rm s}^{-1}}\right)$     & -       & $\mathcal{U}(7,14)$      & 11.73            & 0.875                   \\ \hline
$xR_p$                 & -              & $\mathcal{U}(-0.1,0)$    & -0.0495          & 0.005                 \\ \hline
offset ($O$)                 & ppm            & $\mathcal{U}(-250,250)$    & +100             & 120                   \\ \hline
\end{tabular}

\end{table}

\begin{table}[]
\caption{\textbf{Detection significance of H$_2$O and clouds.} The logarithm of the evidence (Z) of the nominal 1.5D retrieval done simultaneously on the morning and evening transmissions spectrum, the retrieval without any H$_2$O, and the retrieval without any clouds are listed in the first column. Logarithm of the Bayes factor (B) for the latter two retrievals relative to the nominal case is listed in the second column. The relative probabilities are listed in the third column along with their corresponding detection significances in the fourth column. We calculate the probabilities and detection significances from the Bayes factor following previous work \cite{thorngren2026}.}
\label{tab:bayes_evi}
\begin{tabular}{|l|l|l|l|l|}
\hline
Retrieval Type & ln(Z)  & ln(B) &  Relative probability & Detection Significance \\ \hline
Nominal        & -60.2  & -     &  1 & -  \\ \hline
No H$_2$O      & -114.6 & -54.4 &  $2.3\times{10^{-24}}$ &  $10\sigma$ \\ \hline
No Clouds      & -110.1 & -49.9 & $2.13\times{10^{-22}}$  &  $9.7\sigma$ \\ \hline
\end{tabular}

\end{table}

\begin{table}[]
\caption{{\bf Constraints from atmospheric retrieval on the spherical planet spectrum.} Same as Table~\ref{tab:retrieval}, but for the \texttt{PICASO} retrieval on the spherical planet spectrum.}
\label{tab:retrieval_1d}
\begin{tabular}{|l|l|l|l|l|}
\hline
\textbf{Parameter}     & \textbf{Units} & \textbf{Prior}           & \textbf{Median} & \textbf{$\pm1\sigma$} \\ \hline
$T_{\rm eq}$           & K              & $\mathcal{U}(800,2100)$  & 1246            & 130                    \\ \hline
$\log\left(\dfrac{\kappa_{\rm IR}}{{\rm cm}^2{\rm g}^{-1}}\right)$ & -              & $\mathcal{U}(-2.0,+0.5)$ & -0.29            & 0.74                 \\ \hline
$\log\left(\dfrac{\gamma}{{\rm cm}^2{\rm g}^{-1}}\right)$    & -              & $\mathcal{U}(-2.0,+0.5)$ & -1.36           & 0.61                  \\ \hline
$\alpha$               & -              & $\mathcal{U}(0,1)$       & 0.54            & 0.32                  \\ \hline
$T_{\rm int}$          & K              & $\mathcal{U}(30,630)$    & 312             & 160                   \\ \hline
\mbox{[M/H]}            & -  & $\mathcal{U}(-1.0,+3.0)$ & 1.937       & 0.073                   \\ \hline
C/O                    & - & $\mathcal{U}(0.01,2.0)$  & 0.27             & 0.21                  \\ \hline
$\log(f_{\rm sed})$     & -              & $\mathcal{U}(-4,+1.0)$   & -1.42            & 0.21                  \\ \hline
$\log\left(\dfrac{K_{\rm zz}}{{\rm cm}^2{\rm s}^{-1}}\right)$   & -      & $\mathcal{U}(7,14)$      & 8.12            & 0.58                   \\ \hline
$xR_p$                 & -              & $\mathcal{U}(-0.1,0)$    & -0.0454          & 0.0020                 \\ \hline
\end{tabular}

\end{table}

\begin{table}[]
\caption{\textbf{GCM input parameters.} Input parameters used in the \textsc{UM} general circulation model simulation for WASP-94A~b are listed in the first column followed by their adapted values in the second column. The third column lists the units of the parameters and the last column lists the references from which those value were adapted.}
\label{tab:3d_gcm_parameters}
\begin{tabular}{|l|l|l|l|}
\hline
\textbf{Parameter} & \textbf{Value}        & \textbf{Units} & \textbf{Reference}   \\ \hline
Inner radius       & {1.10$\times{10^8}$}          & m &  \cite{ahrer24}   \\ \hline
Domain height      & {3.67$\times{10^7}$}         & m & -     \\ \hline
Semi-major axis    & 0.0554                & AU   & \cite{ahrer24}             \\ \hline
Orbital period     & 3.95                  & Earth days & \cite{ahrer24}       \\ \hline
Rotation rate      & {1.84$\times{10^{-5}}$}       &  rad~s$^{-1}$  &  \cite{ahrer24}     \\ \hline
Surface gravity at inner boundary  & 4.74  & m~s$^{-2}$ &  \cite{ahrer24} \\ \hline
Specific gas constant     & 3517.24        & J~K$^{-1}$~kg$^{-1}$ &  - \\ \hline
Specific heat capacity    & $1.30\times{10^4}$   & J~K$^{-1}$~kg$^{-1}$ &  - \\ \hline
Stellar irradiance        & $1.46\times{10^6}$    &  W~m$^{-2}$ &  - \\ \hline
Stellar constant at 1 au  & 4494.55        & W~m$^{-2}$  &  -  \\ \hline
Intrinsic temperature     & 100            & K        &  -          \\ \hline
Longitude grid resolution & 2.5            & degrees &  -                       \\ \hline
Latitude grid resolution  & 2              & degrees  &  -                       \\ \hline
Number of vertical grid levels  & 87       & -    &  -                           \\ \hline
\end{tabular}

\end{table}

\begin{sidewaystable}[ht]
\caption{\textbf{Fitted parameters for different aerosol models.} The first column lists the aerosol type included in the model. The second to ninth column lists the constraints obtained on each parameter of the aerosol model. Logarithm of the Bayesian evidences (Z) for each model is listed in the tenth column, while the probability of the various haze models relative to the cloud model are listed in the eleventh column.}
\label{tab:hazecompare}
\centering
\begin{tabular}{|l|c|c|c|c|c|c|c|c|c|c|}
\hline
\begin{tabular}[c]{@{}c@{}} Aerosol\\ Type\end{tabular} & $\log\left(\dfrac{n}{\rm cm^{-2}}\right)$ & $\log\left(\dfrac{P_{\rm base}}{\rm bar}\right)$ & $\log\left(\dfrac{P_{\rm top}}{\rm bar}\right)$ & $\log(f_{\rm layer})$ & $\beta$ & $\log\left(\dfrac{r_{\rm mean}}{\rm cm}\right)$ & $T_{\rm iso}$ (K) & $\log(X_{\rm H_2O})$ & $\ln(Z)$ & \begin{tabular}[c]{@{}c@{}}Relative\\ probability\end{tabular} \\ \hline
\makecell[l]{Clouds \\ (MgSiO$_3$) \\ \cite{scott96}} & 14$\pm$1.9 & -2.34$\pm$0.86 & -- & 0.51$\pm$0.13 & -- & -5.06$\pm$0.17 & 650$\pm$130 & -8.0$\pm$2.0 & -28.0 & 1.0 \\ \hline
\makecell[l]{Tholin \\ \cite{khare84}} & 15$\pm$1.5 & -- & -8.22$\pm$0.64 & -- & 0.16$\pm$0.12 & -4.90$\pm$0.23 & 610$\pm$71 & -6.3$\pm$2.6 & -33.1 & 0.006 \\ \hline
\makecell[l]{Soot \\ \cite{lavvas17}} & 7.17$\pm$2.2 & -- & -5.1$\pm$1.7 & -- & 2.3$\pm$1.4 & -4.89$\pm$0.11 & 1500$\pm$810 & -9.1$\pm$2.2 & -30.8 & 0.06 \\ \hline
\makecell[l]{Tholin \\ \cite{he2022}} & 15$\pm$1.3 & -- & -7.2$\pm$0.78 & -- & 0.22$\pm$0.18 & -5.06$\pm$0.22 & 770$\pm$150 & -6.1$\pm$2.4 & -32.1 & 0.016 \\ \hline
\makecell[l]{Saturn-like\\ Phosphorus \\\cite{noy81,sromovsky20,fletcher23}} & 17$\pm$1.0 & -- & -7.6$\pm$1.2 & -- & 0.19$\pm$0.13 & -5.83$\pm$0.23 & 1100$\pm$230 & -6.2$\pm$1.9 & -32.3 & 0.013 \\ \hline
\end{tabular}
\end{sidewaystable}

\clearpage 





\end{document}